\documentclass[twocolumn]{aastex62}
\usepackage{amsmath}

\received{, 2023}
\revised{, 2023}
\accepted{, 2023}

\submitjournal{ApJS}

\shorttitle{SOC in GRB prompt phase}
\shortauthors{Li et al.}

\begin{document}

\title{EVIDENCE FOR SELF-ORGANIZED CRITICALITY PHENOMENA IN PROMPT PHASE OF SHORT GAMMA-RAY BURSTS}
\author[0000-0001-6469-8725]{Li, Xiu-Juan$^{\ddag}$}
\affiliation{School of Cyber Science and Engineering, Qufu Normal University, Qufu 273165, China}
\author[0000-0001-6469-8725]{Zhang, Wen-Long}
\affiliation{School of Physics and Physical Engineering, Qufu Normal University, Qufu 273165, China}
\author[0000-0001-6469-8725]{Yi, Shuang-Xi$^{\dag}$}
\affiliation{School of Physics and Physical Engineering, Qufu Normal University, Qufu 273165, China}
*\author[0000-0001-6469-8725]{Yang, Yu-Peng}
\affiliation{School of Physics and Physical Engineering, Qufu Normal University, Qufu 273165, China}
*\author[0000-0001-6469-8725]{Li, Jia-Lun}
\email{yisx2015@qfnu.edu.cn;lxj@qfnu.edu.cn}
\affiliation{School of Physics and Physical Engineering, Qufu Normal University, Qufu 273165, China}

\begin{abstract}
The prompt phase of gamma-ray burst (GRB) contains essential information regarding the physical nature and central engine,
which are as yet unknown. In this paper, we investigate the self-organized criticality (SOC) phenomena in GRB prompt phase as done in X-ray flares of GRBs.
We obtain the differential and cumulative distributions of 243 short GRB pulses, such as
peak flux,
FWHM, rise time, decay time, and peak time in the fourth BATSE TTE Catalog with the Markov Chain Monte Carlo (MCMC) technique. It is found that these distributions can be well described by
power-law models. In particular, comparisons are made
in 182 short GRB pulses in the third Swift GRB Catalog from 2004 December to 2019 July. The results are essentially consistent with those in BATSE ones. We notice that there is no obvious power-law index evolution across different energy bands for either BATSE or Swift sGRBs. The joint analysis suggests that GRB prompt phase can be explained by a Fractal-Diffusive, Self-Organized Criticality (FD-SOC) system with the spatial dimension $S = 3$ and the classical diffusion $\beta = 1$. Our findings show that GRB prompt phases and X-ray flares possess the very same magnetically dominated stochastic process and mechanism.
\end{abstract}

\keywords{
\href{http://astrothesaurus.org/uat/739}{High energy astrophysics (739)};
\href{http://astrothesaurus.org/uat/629}{Gamma-ray bursts (629)};
}

\section{Introduction} \label{sec:intro}
Gamma-ray burst (GRB) is a sudden release of gamma-ray emission, which lasts from milliseconds to thousands
of seconds. Despite many studies done on GRBs, the natures are still strongly debated. The lightcurves of GRB prompt phases are the imprints of the activities of central engine and contain the key information of internal energy dissipation and physical mechanisms. In general, it is difficult to accurately extract the properties of GRB lightcurves due to their notoriously complex and irregular structures. Fortunately, a fraction of GRBs are overwhelmingly single-peaked and double-peaked \cite[e.g.][]{Hakkila2018,Lxj1,Lxj2} providing valuable insights into the
physical processes by which GRBs release energy.

\cite{Bak1987} reported that a real, many-bodied, physical system in an external field assembles itself into a critical state which  can be triggered by a small perturbation and give rise to an avalanche-like chain reaction of any size due to some driving forces. This is known as the concept of self-organizing criticality (SOC) proposed as an
attempt to explain the existence of self-similarities over extended ranges of spatial and temporal scales in a wide variety
of systems.
It is worth noting that the SOC phenomena are commonly discussed in many astrophysical systems, such as solar flare, stellar flares, lunar craters, the asteroid belt, Saturn ring particles, magnetospheric substorms, radiation belt electrons, pulsar glitches, soft gamma-ray repeaters, black-hole objects, blazars, cosmic rays, X-ray bursts, and fast
radio bursts \citep[e.g.][]{Lu1991,Melatos2008,Aschwanden2014,Aschwanden2021,Aschwanden2022,Li2015,Wang2013,Yi2016,Yi2022,Liu2017,Du2021,Wang20171,Wang20172,Wang2021,Wei2021,Cheng2020,Zhang2022}, etc.

As a common astrophysical phenomena in the universe, X-ray flares are observed in
a good fraction of GRB afterglows \citep[e.g.][]{Chincarini2010,Margutti2010}. It is well known that GRB X-ray flares
are attributed to the erratic GRB central engine activities just as the GRB prompt emission components which also provide important clues to the nature of the central
engines \citep[e.g.][]{Burrows2005, Margutti2010, Abdo2011,Yi2015,Chang2021}. For example, the similar lag-luminosity relation provides strongly evidence for the direct link between GRB X-ray flares and prompt phases \citep{Margutti2010}.
\cite{Wang2013} investigated the statistical properties of X-ray flares of GRBs and found that both GRB X-ray flares and solar
flares might origin from the similar physical
mechanism, i.e., they might be produced by a magnetic
reconnection process and thus both can be well
explained by physical framework of a SOC system. Moreover, the further investigation based on a large sample of Swift GRB X-ray flares reported obtained the similar conclusions \cite[e.g.][]{Yi2016,Wei2022}.
Thus, it is necessary to determine whether the GRB prompt phases can also be explained by the SOC model and if so to probe the possible SOC behavior in GRB prompt phases, etc.

From a more physical point of view, \cite{Aschwanden2011,Aschwanden2012, Aschwanden2014} developed the definition of SOC systems and proposed an analytical macroscopic SOC model. Adopting the expectation criteria proposed by the above model, we can identify an SOC system by examining the power-law-like distributions of relevant observed parameters \citep[e.g.][]{Wang2013,Yi2016,Zhang2022}. Recently, \cite{Lv2020} investigated the properties of BATSE GRBs with multi-pulses in their prompt lightcurves reported by \cite{Hakkila2011}. By fitting the distributions of several observed pulses parameters, they tentatively suggested that SOC phenomena exist indeed in prompt phase of GRBs. However, the indisputable fact is that their investigations were
mainly performed for the bursts with at least three pulses and single-peaked and double-peaked bursts were not
included \cite[e.g.][]{Hakkila2018,Lxj1,Lxj2}. In this study, we systematically compile the temporal
properties of the single-peaked and double-peaked GRB pulses in the fourth BATSE TTE Catalog and the third Swift/BAT catalog given by our recent works \citep{Lxj1,Lxj2} and perform a analysis focusing on the differential and cumulative distributions of characterized pulse parameters to search for the evidences of the SOC system. Particularly, in order to avoid instrumental selection effect, we examine whether the distributions evolve among different energy channels for both satellite bursts. Sample selection and data analysis are presented in
Section 2. Section 3 displays our main results. We discuss some possible physical explanations of the results in Section 4. Finally, we summarize the results in Section 5.
\section{DATA and Method} \label{sec:observations}
Our recent works \citep{Lxj1,Lxj2} utilized the empirical ``KRL'' function proposed by \cite{Kocevski2003} to fit the lightcurves of short GRBs in the fourth BATSE TTE Catalog and the third Swift GRB Catalog from 2004 December
to 2019 July. For BATSE sGRBs, the photon counts are accumulated into four standard
energy channels, labeled as Ch1 (25-55 keV), Ch2 (55-110 keV), Ch3 (110-320 keV), and Ch4 ($\geq$320 keV). Similarly, the mask-weighted lightcurve data of Swift sGRBs are taken
from the Swift website \citep{Lien2016} for four energy
channels: Ch1 (15-25 keV), Ch2 (25-50 keV),
Ch3 (50-100 keV), and Ch4 (100-350 keV).
In total, 243 BATSE and 182 Swift pulses are obtained. Note that the pulse numbers in different energy channels are different due to either selection effect or low signal-to-noise ratio. Then, we extract the parameters of these pulses, including peak flux ($f_{\rm m}$), full width at half-maximum (FWHM), peak time ($t_ {\rm m}$), rise time\
($t_ {\rm r}$), and decay time ($t_ {\rm d}$). The detailed data processing refers to our previous works \citep{Lxj1,Lxj2}.

To identify the possible SOC features in GRB prompt phases, we study in detail the differential and cumulative distributions of these temporal parameters. Here, a empirical thresholded power-law distribution function is used to fit the differential occurrence frequency distributions of GRBs \citep{Aschwanden2015,Lv2020}, which can be written as,
\begin{equation} \label{eq1}
N_{\rm diff}=\frac{dN(x)}{dx}\propto(x_{0d}+x)^{-\alpha_{d}}, x_1 \leq x \leq x_2,
\end{equation}
where N is the number of events, $x_{\rm {0d}}$ is a constant by considering the threshold effects, $x_{\rm 1}$ and $x_{\rm 2}$ are the minimum and maximum values of scale-free range, and $\alpha_{\rm d}$ is the power-law index of differential distribution, respectively. This size function is identical to a ``Generalized
Pareto distribution'' \citep{Hosking1987}, the ``Generalized Pareto Type II distribution'' \citep[e.g.][]{Johnson1994, Arnold2015}, and the ``Lomax distribution'' \citep{Lomax1954}.
The uncertainty of the differential distribution is given by $\sigma _{diff,i}=\sqrt{N_{bin,i}}/\triangle x_i$, where $N_{bin,i}$ refers to the number of events of the $i$-bin, and $\triangle x_i$ is the bin size.

According to \cite{Aschwanden2015}, the cumulative distribution function of Equation \ref {eq1} can be obtained as,
\begin{equation} \label{eq2}
N_{\rm cum}(>x)=1+(N-1) \times \frac{(x_2+x_{0c})^{1-\alpha_c}-(x+x_{\rm 0c})^{1-\alpha_c}}{(x_2+x_{0c})^{1-\alpha_c }-(x_1+x_{0c})^{1-\alpha_{c}}},
\end{equation}
where $N$ is the total number of events, $x_{\rm {0c}}$ is a constant by considering the threshold effects, and $\alpha_{\rm c}$ is the power-law index of cumulative distribution. The uncertainty of the cumulative distribution in a given bin $i$ is estimated with $\sigma _{cum,i}=\sqrt{N_{i} }$, where $N_{\rm i}$ refers to the number of events of the bin.
We use the standard reduced chi-square ($\chi{_{\rm \nu} ^2}$) goodness to identify a best fit. The $\chi_{\rm \nu}$ can be written as
\begin{equation} \label{eq3}
\chi{_{\nu,diff}}=\sqrt{\frac{1}{(n_x-n_{par})} \sum \limits_{i=1}^{n_x} \frac{{[N_{fit,diff}(x_i)-N_{obs,diff}(x_i)]}^2}{\sigma_{diff,i}^2}}
\end{equation}
for the differential distribution function, and
\begin{equation} \label{eq3}
\chi{_{\nu,cum}}=\sqrt{\frac{1}{(n_x-n_{par})} \sum \limits_{i=1}^{n_x} \frac{{[N_{fit,cum}(x_i)-N_{obs,cum}(x_i)]}^2}{\sigma_{cum,i}^2}}
\end{equation}
for the cumulative distribution function \citep{Aschwanden2015},
where $n_{\rm x}$ is the number of logarithmic bins, $n_{\rm par}$ is the number of the free parameters, $N_{\rm diff,obs}(x_i)$ and $N_{\rm cum,obs}(x_i)$ are the observed values, $N_{\rm diff,fit}(x_i)$ and $N_{\rm cum,fit}(x_i)$ are the corresponding theoretical values for differential distribution and cumulative distribution, respectively. Note that the points below the threshold $x_0$ are just noise and do not contribute to the accuracy of the best-fit power-law index, thus they are ignored when the reduced chi-square is calculated.

Markov chain Monte Carlo (MCMC) method is used to fit GRB data for the self-organized criticality model with PYTHON package pymc\footnote{https://pypi.org/project/pymc/} and obtain the optimal distribution parameters and the 95\% confidence regions.

\section{Results} \label{sec:results}

\subsection{Distributions} \label{sec:results}
\begin{figure*}
\centering
\gridline{
\fig{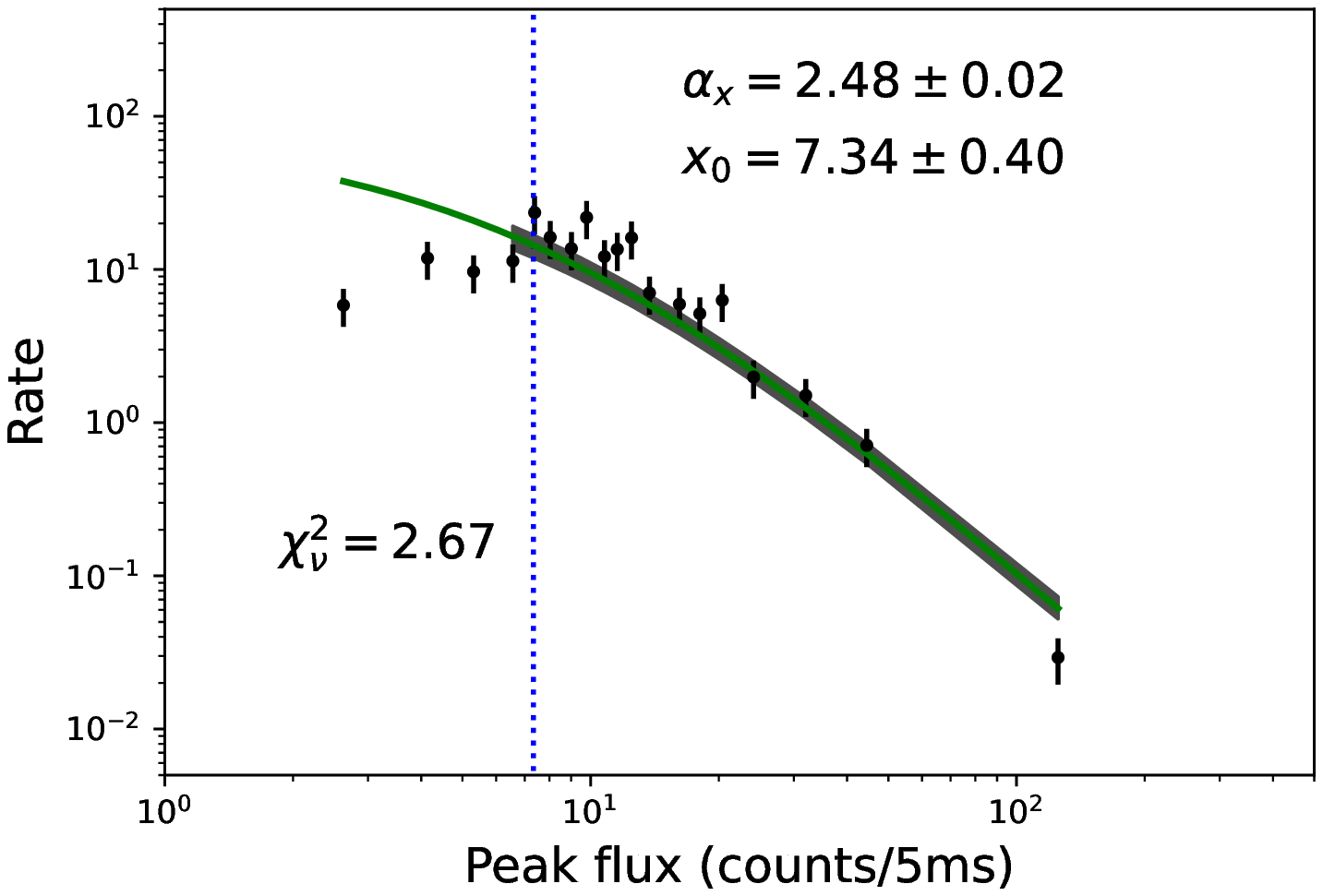}{0.5\linewidth}{(a)}
\fig{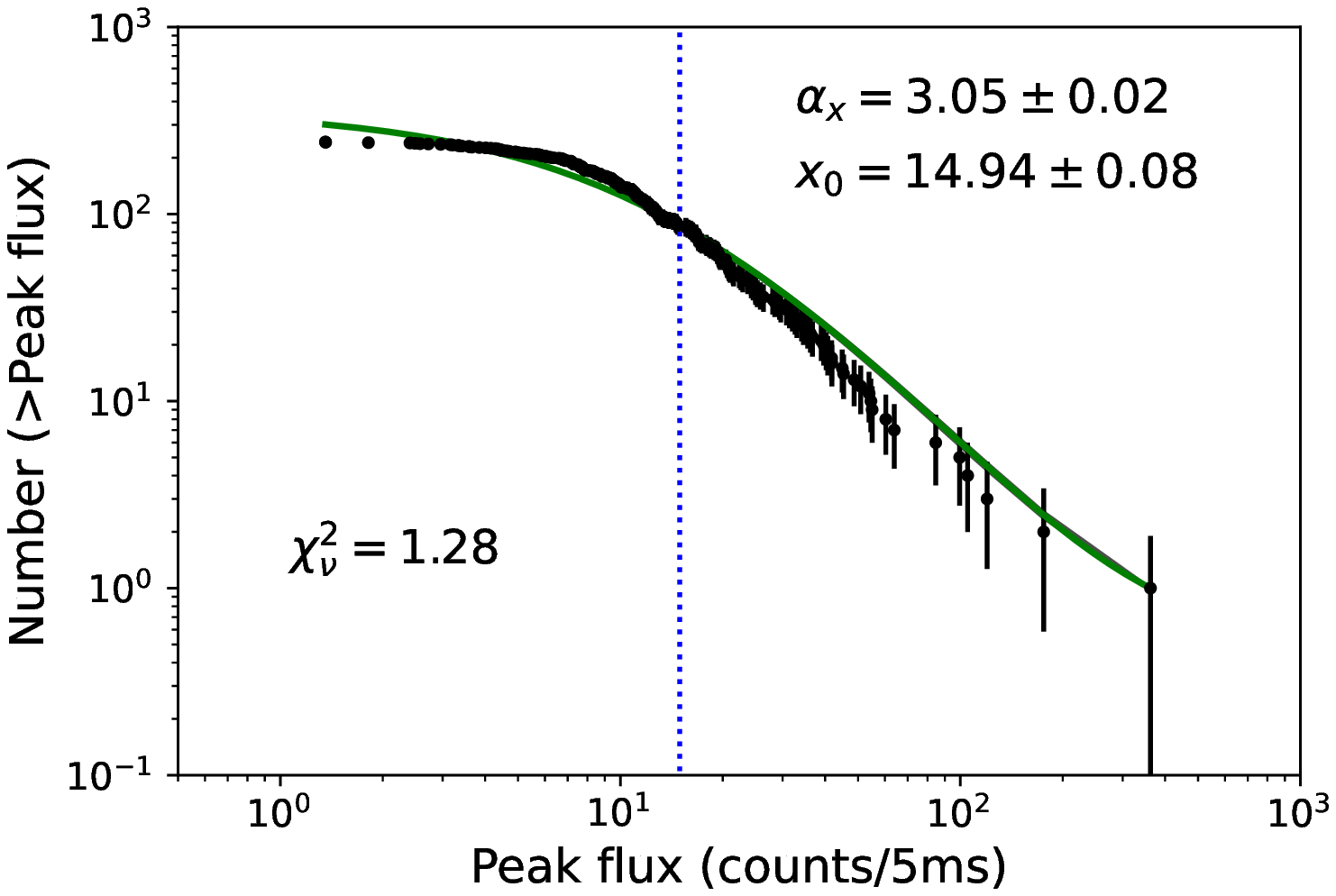}{0.5\linewidth}{(b)}
          }
 \gridline{
 \fig{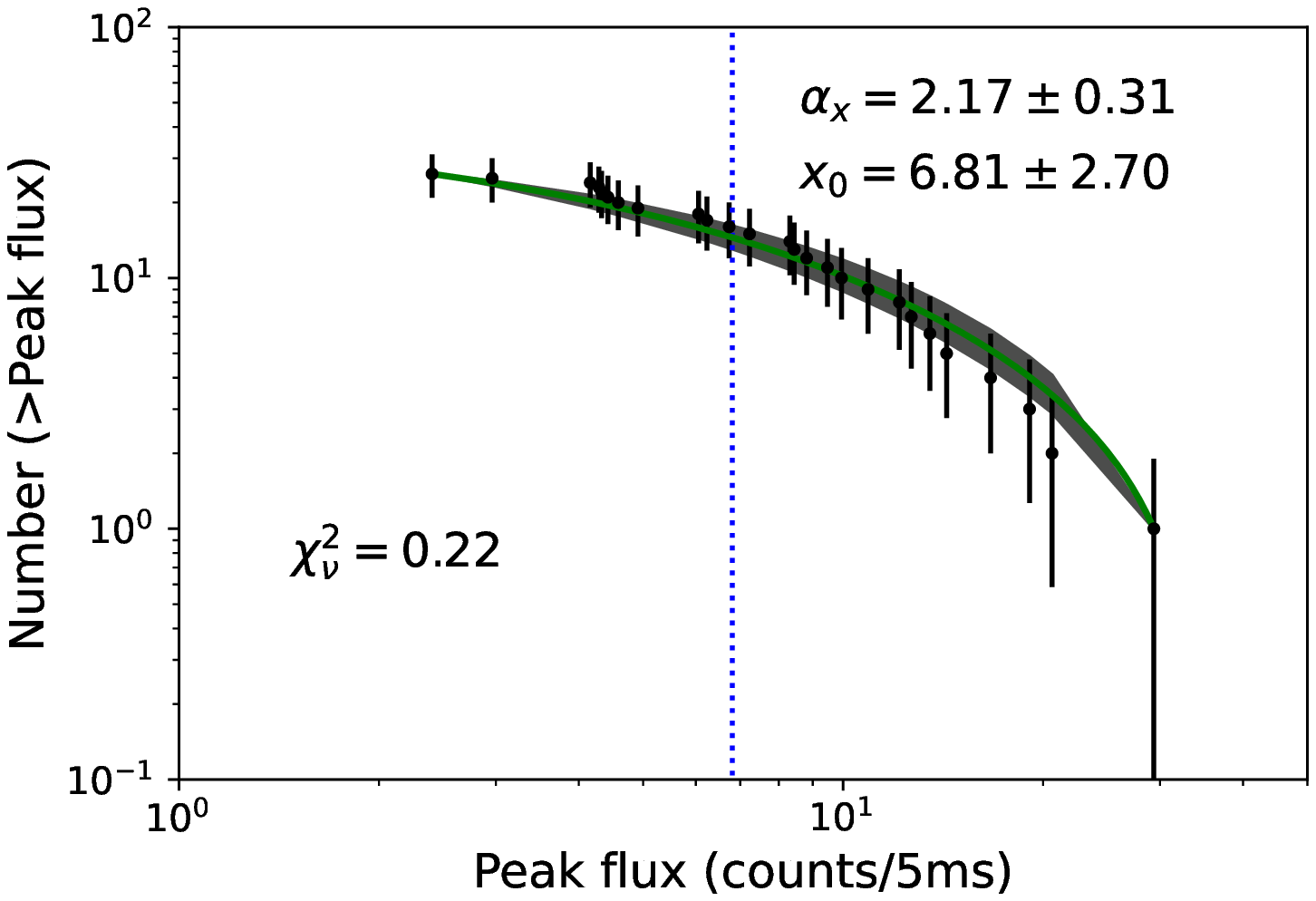}{0.5\textwidth}{(c)}
 \fig{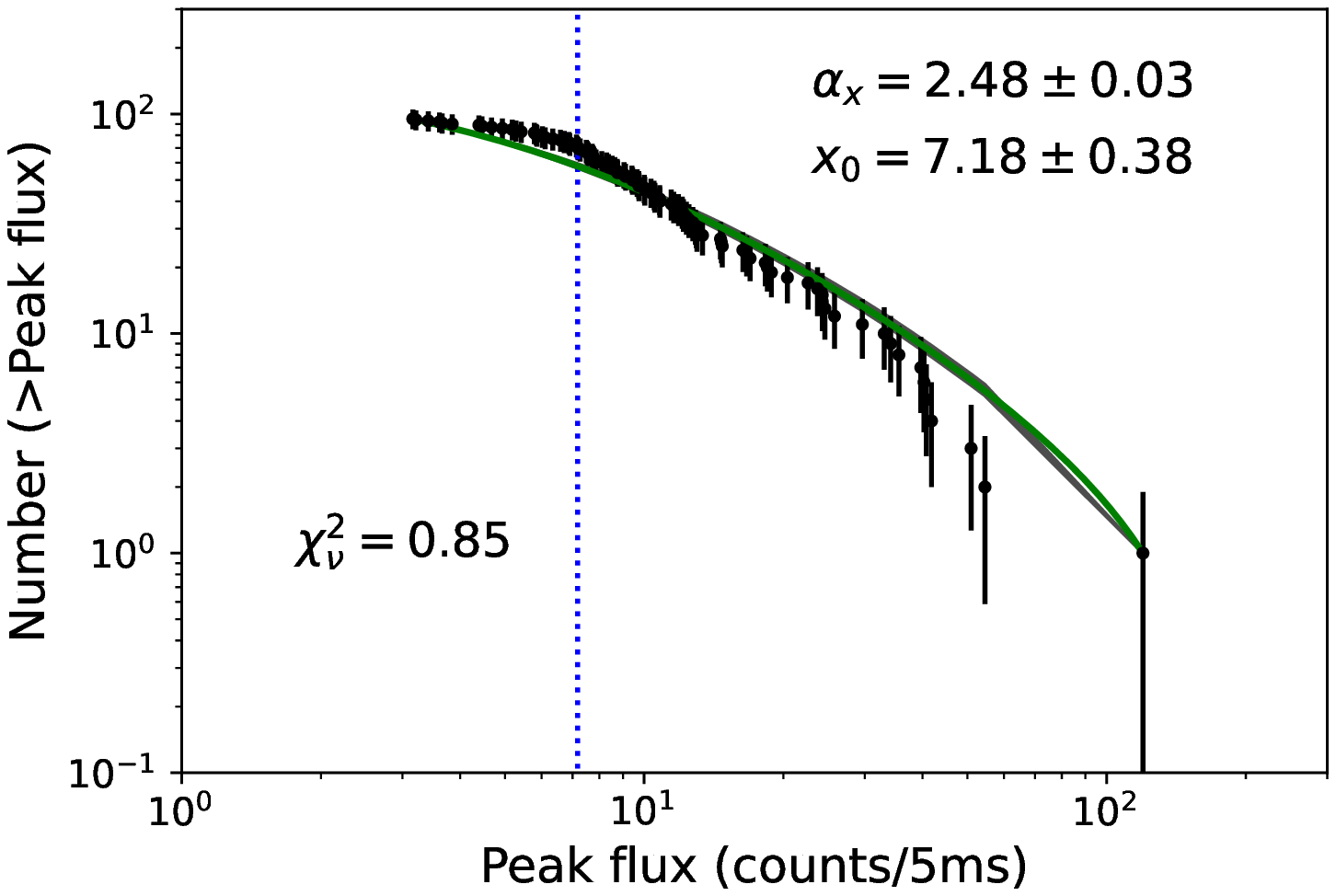}{0.5\textwidth}{(d)}
          }
 \gridline{
 \fig{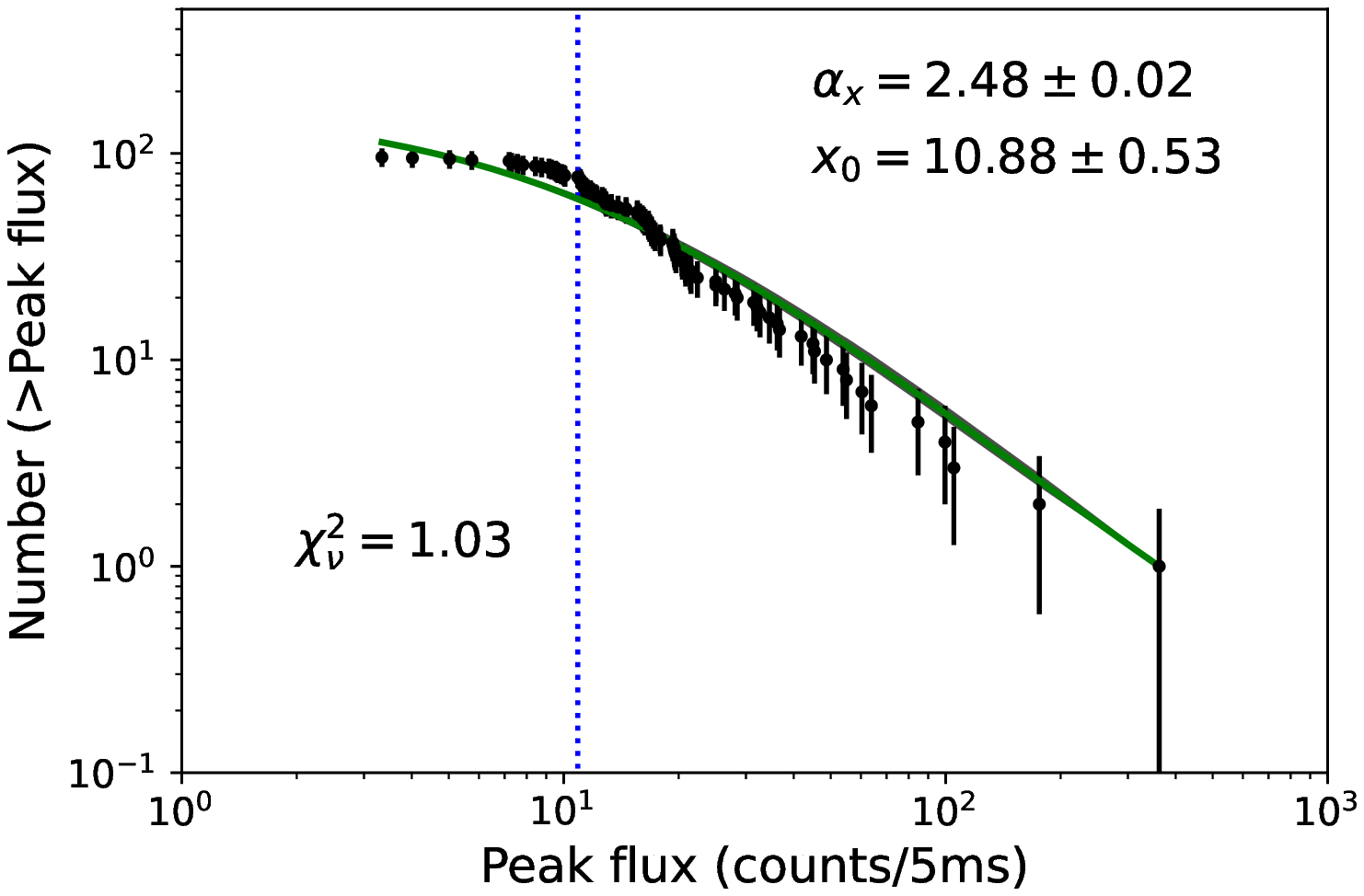}{0.5\textwidth}{(e)}
 \fig{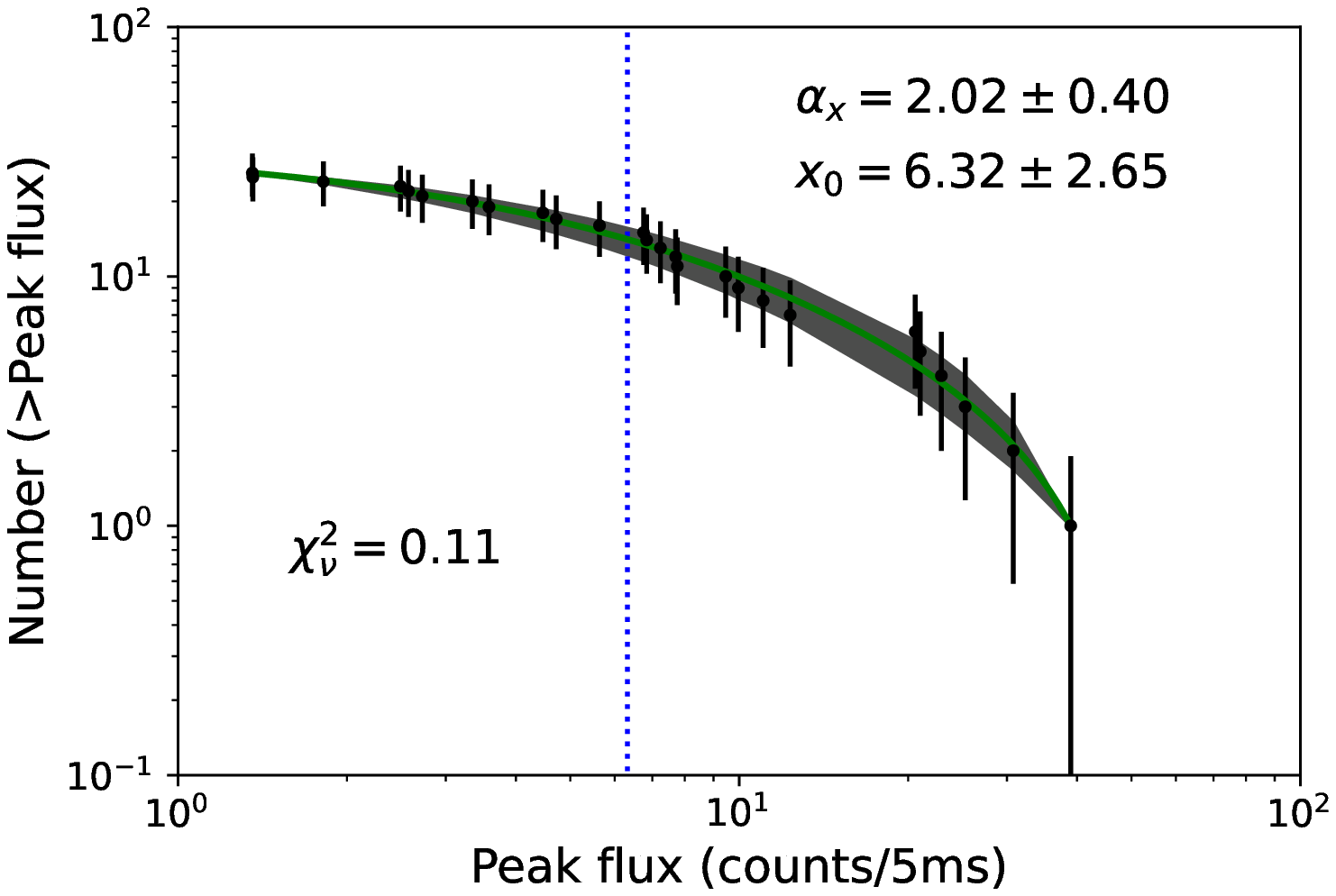}{0.5\textwidth}{(f)}
          }
\caption{The distributions of $f_{\rm m}$ for BATSE GRBs. (a) The differential distributions of $f_{\rm m}$ for the total BATSE sample. (b) The cumulative distributions of $f_{\rm m}$ for the total BATSE sample. (c) The cumulative distributions of $f_{\rm m}$ in Ch1. (d) The cumulative distributions of $f_{\rm m}$ in Ch2. (d) The cumulative distributions of $f_{\rm m}$ in Ch3. (e) The cumulative distributions of $f_{\rm m}$ in Ch4. The gray region represents the 95\% confidence level, the green solid line is the best fit, and the dotted line is marked as the threshold $x_{\rm 0}$. \label{fmdff}}
\end{figure*}

\begin{figure*}
\centering
\gridline{
\fig{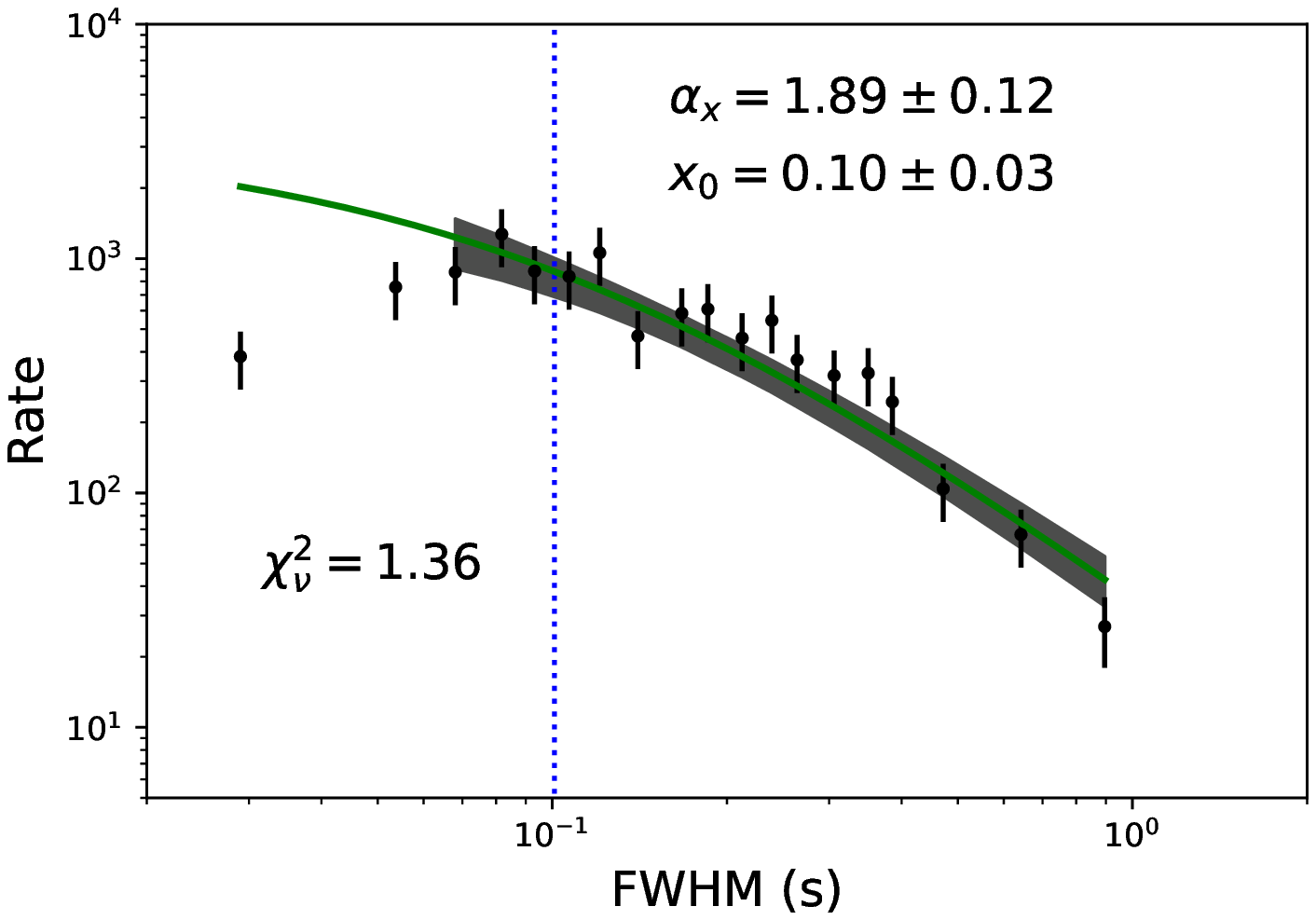}{0.5\textwidth}{(a)}
\fig{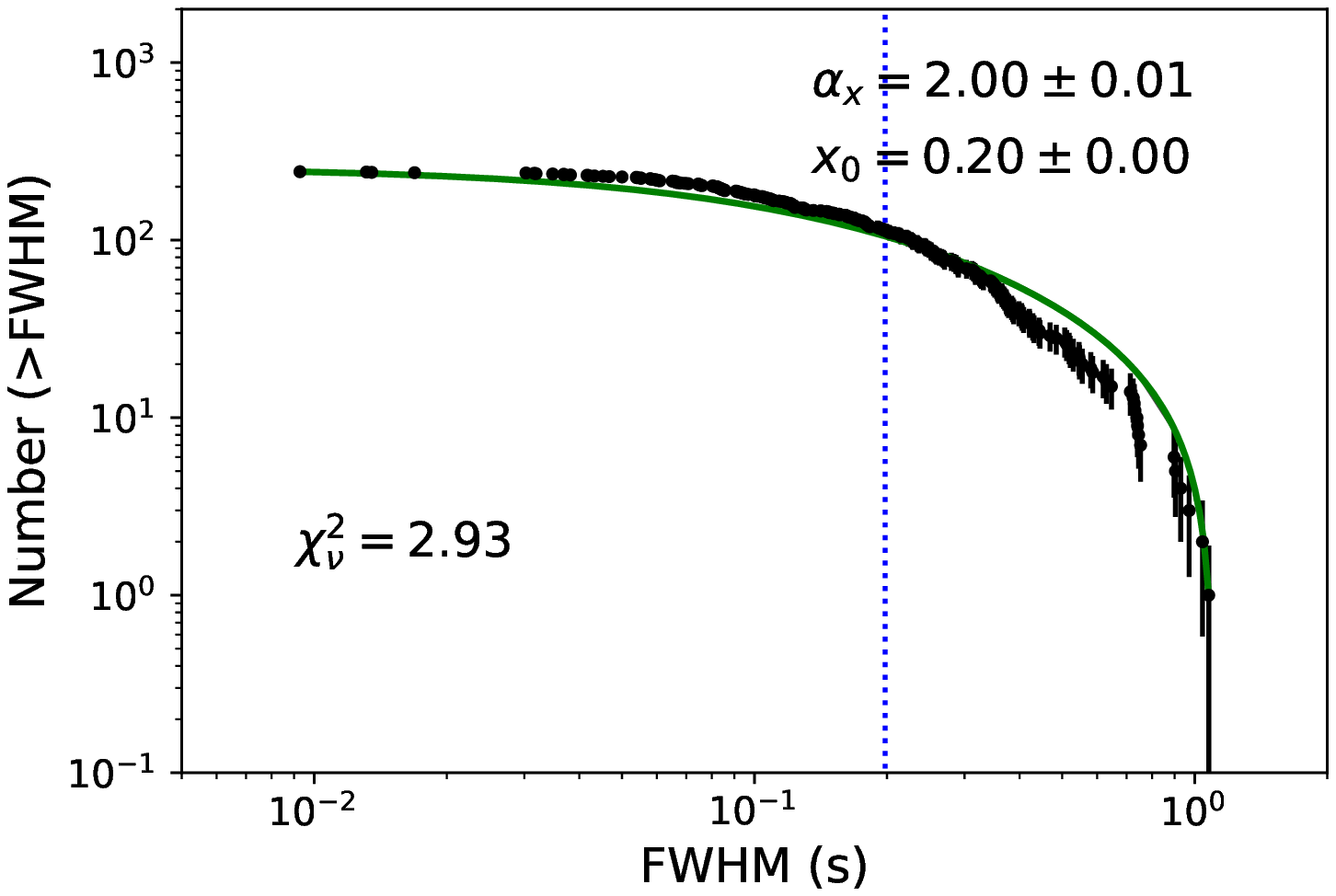}{0.5\linewidth}{(b)}
          }
 \gridline{
 \fig{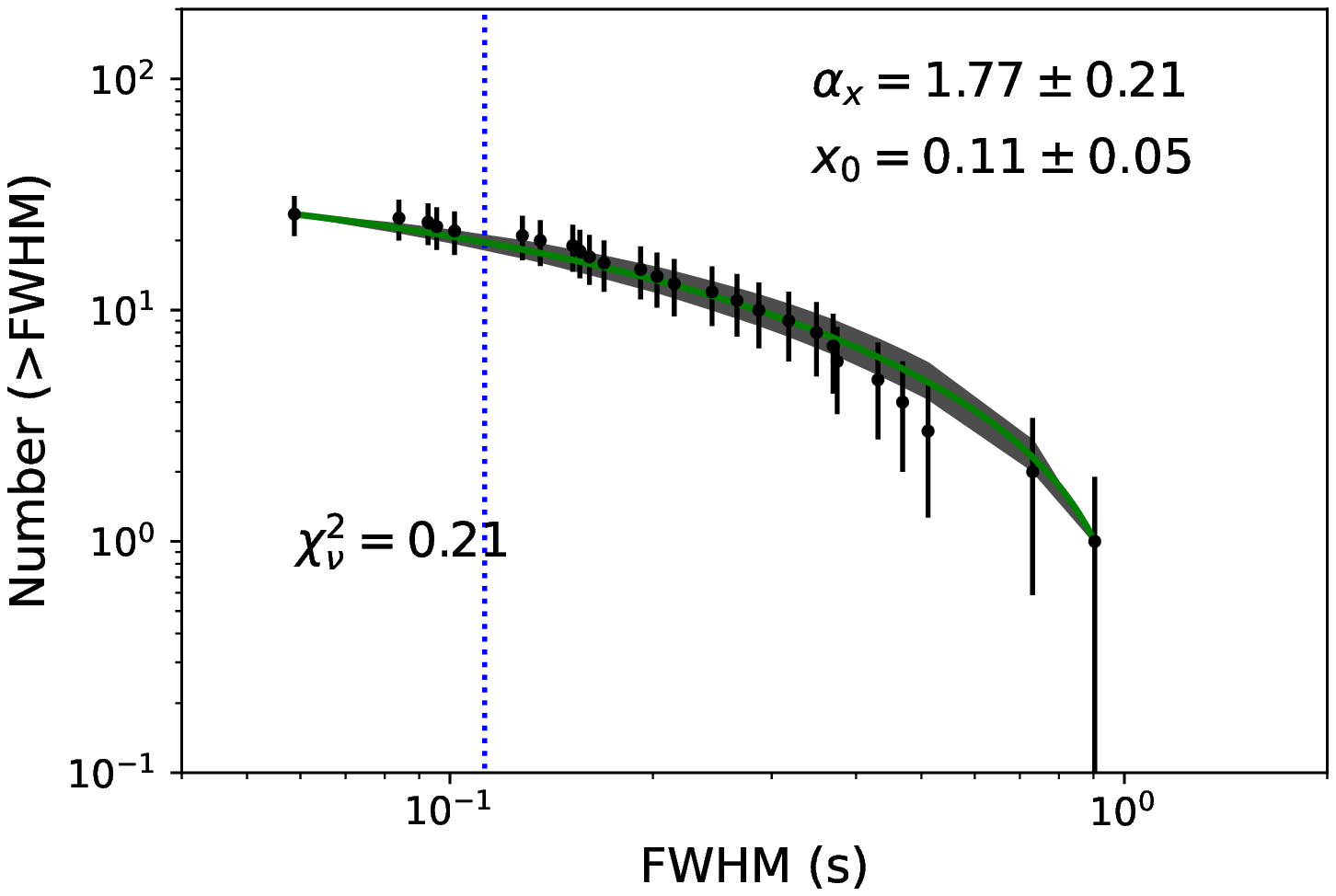}{0.5\textwidth}{(c)}
 \fig{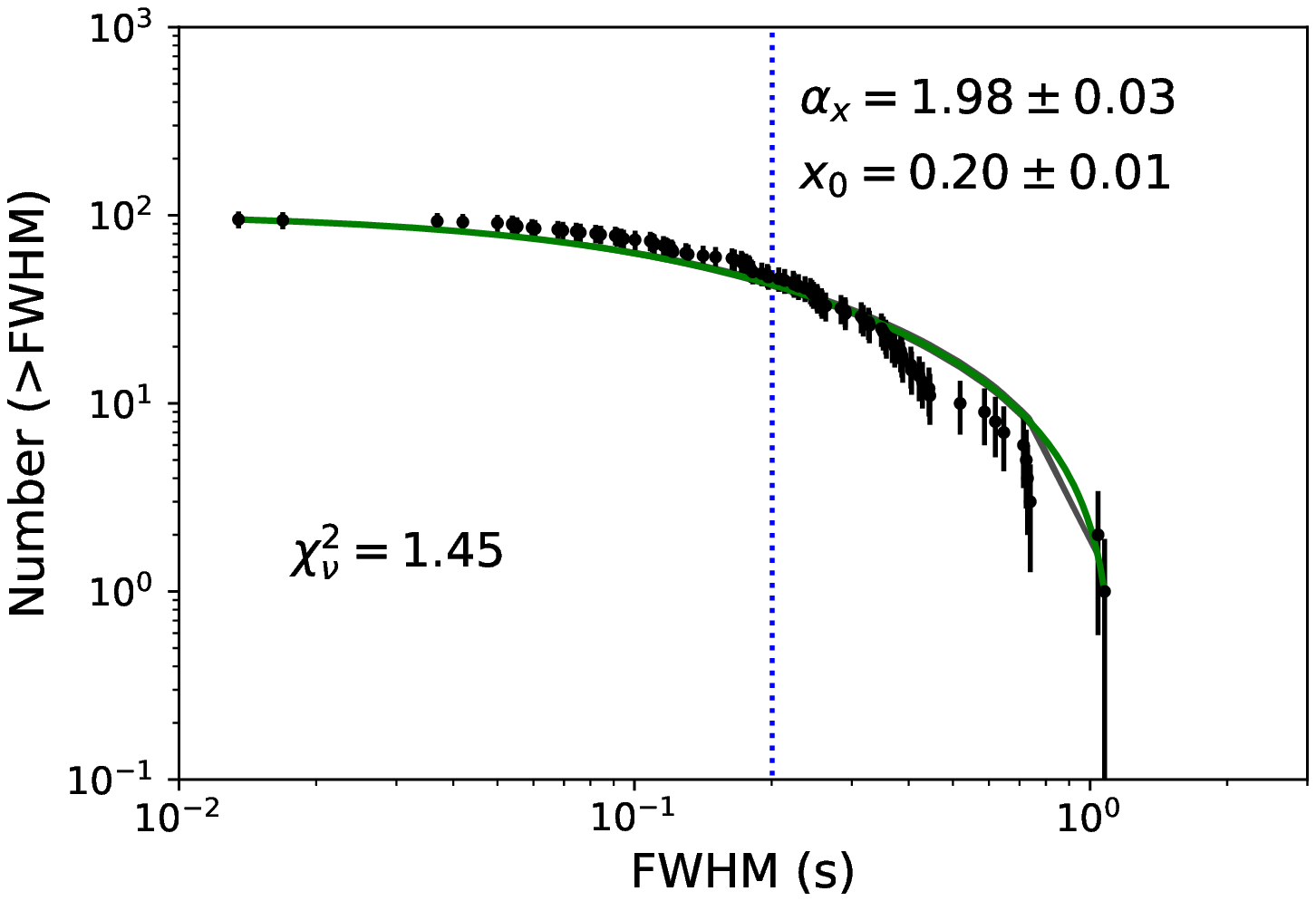}{0.5\textwidth}{(d)}
          }
           \gridline{
 \fig{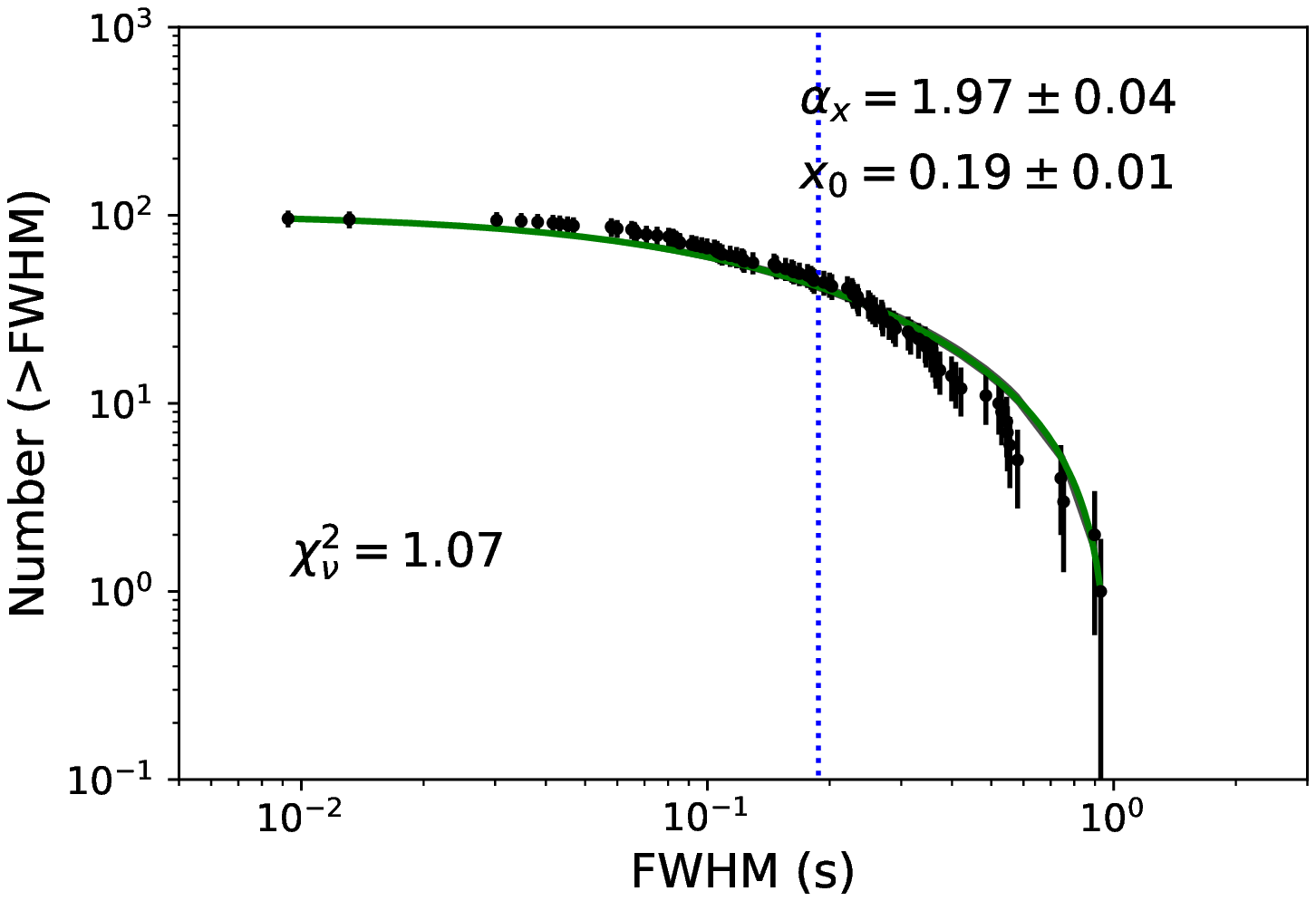}{0.5\textwidth}{(e)}
 \fig{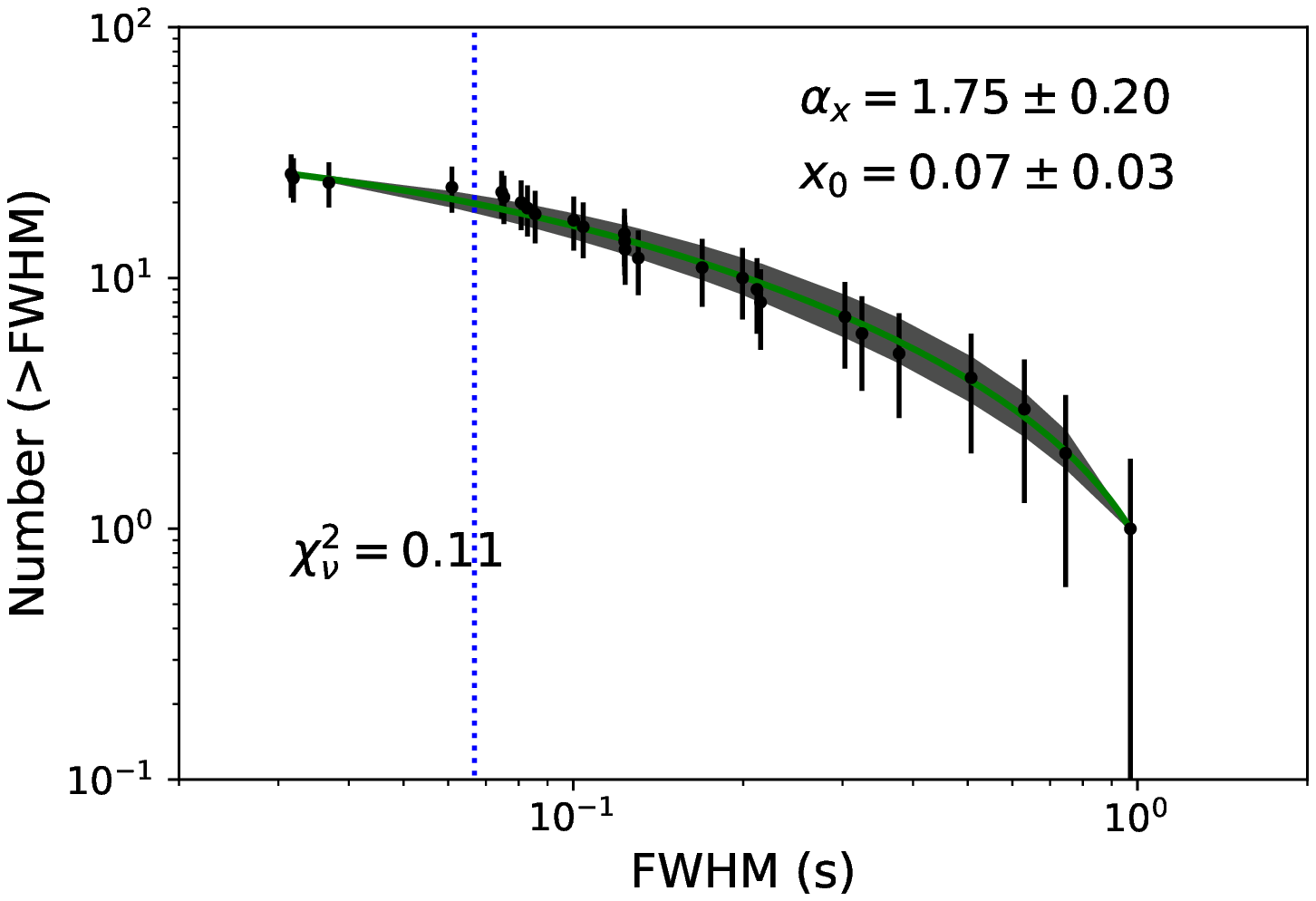}{0.5\textwidth}{(f)}
          }
\caption{The distributions of FWHM for BATSE GRBs. The symbols are the same as those in Figure 1. \label{FWHMdff}}
\end{figure*}

\begin{figure*}
\centering
\gridline{
\fig{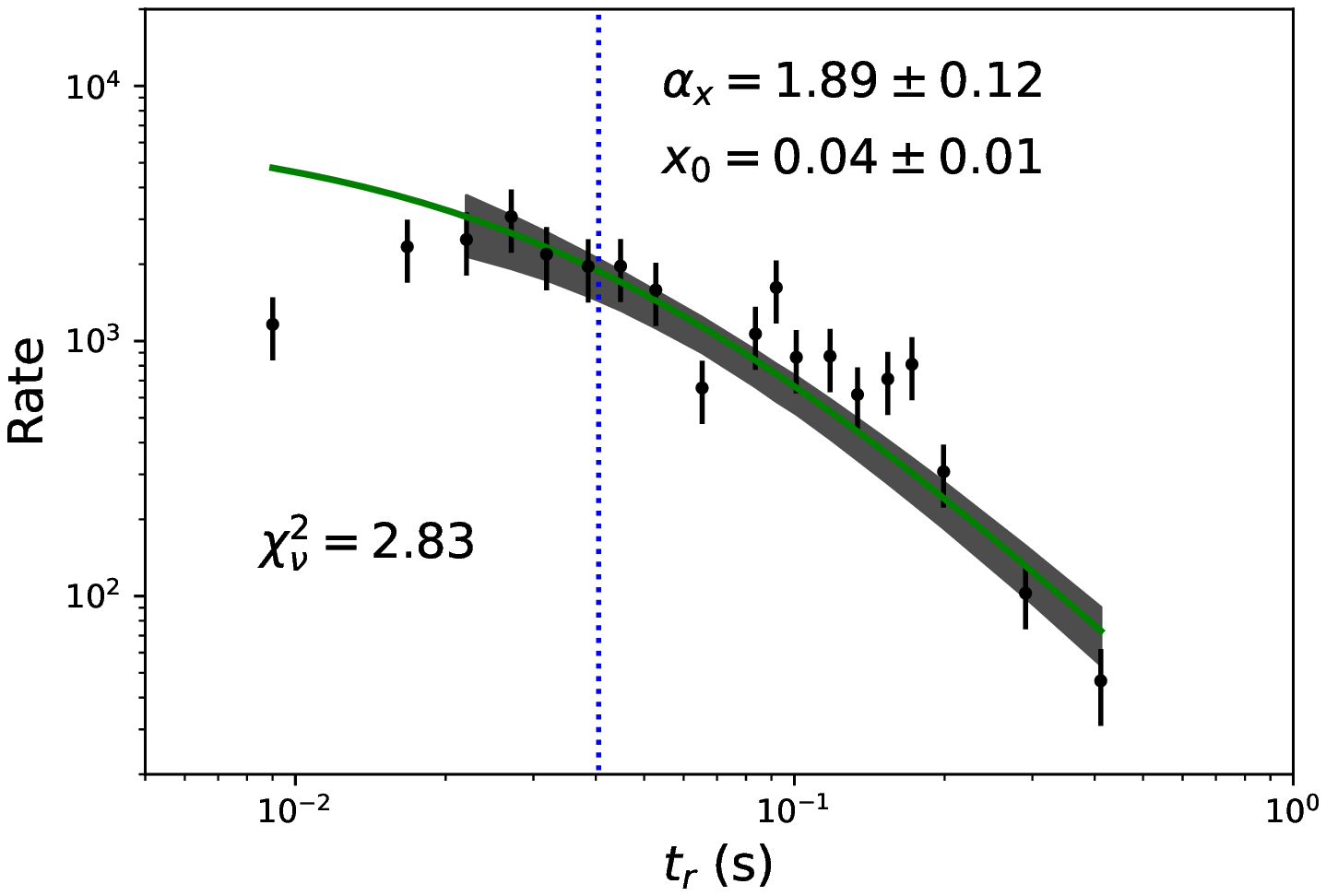}{0.5\textwidth}{(a)}
\fig{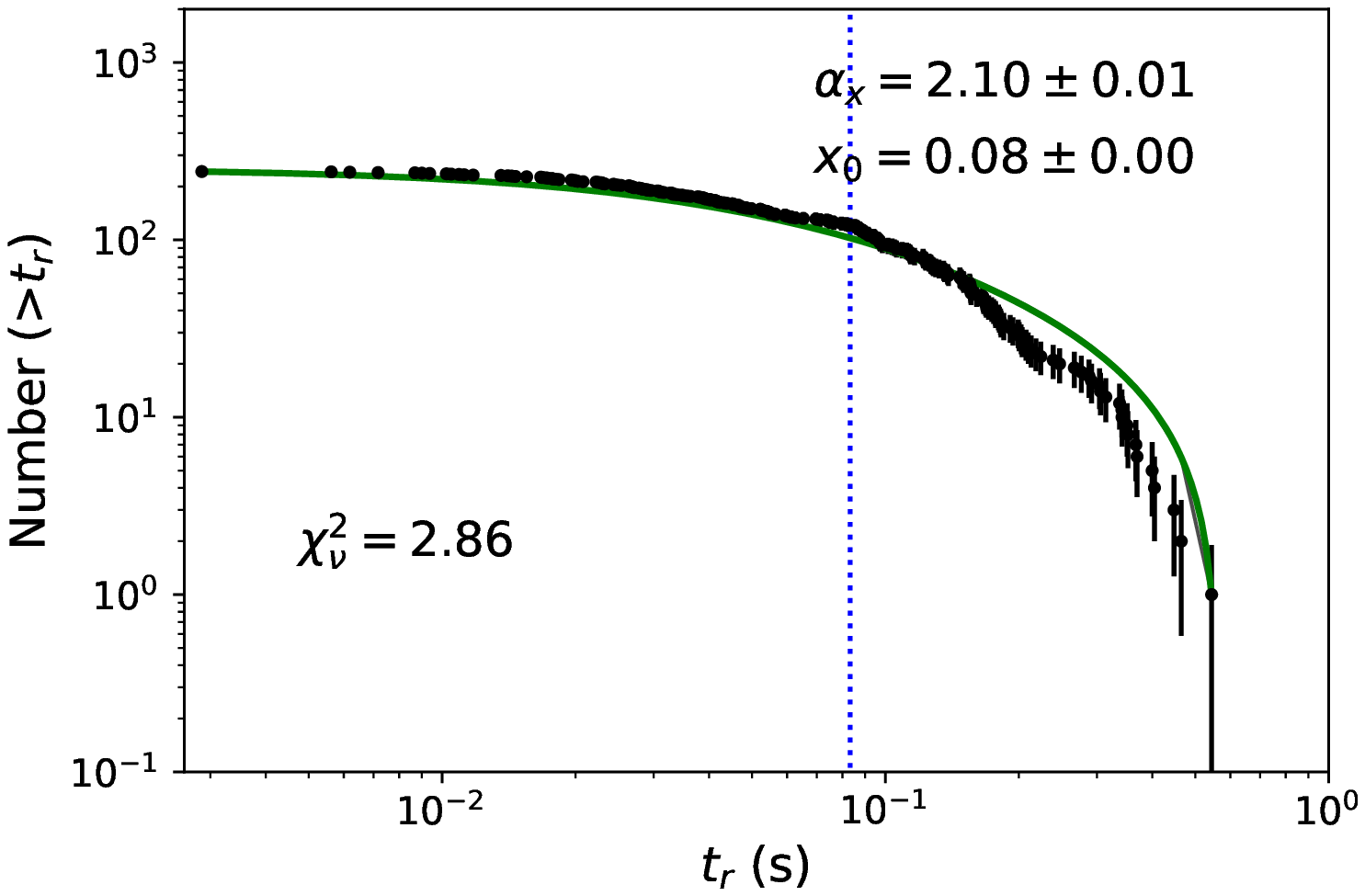}{0.5\linewidth}{(b)}
          }
 \gridline{
 \fig{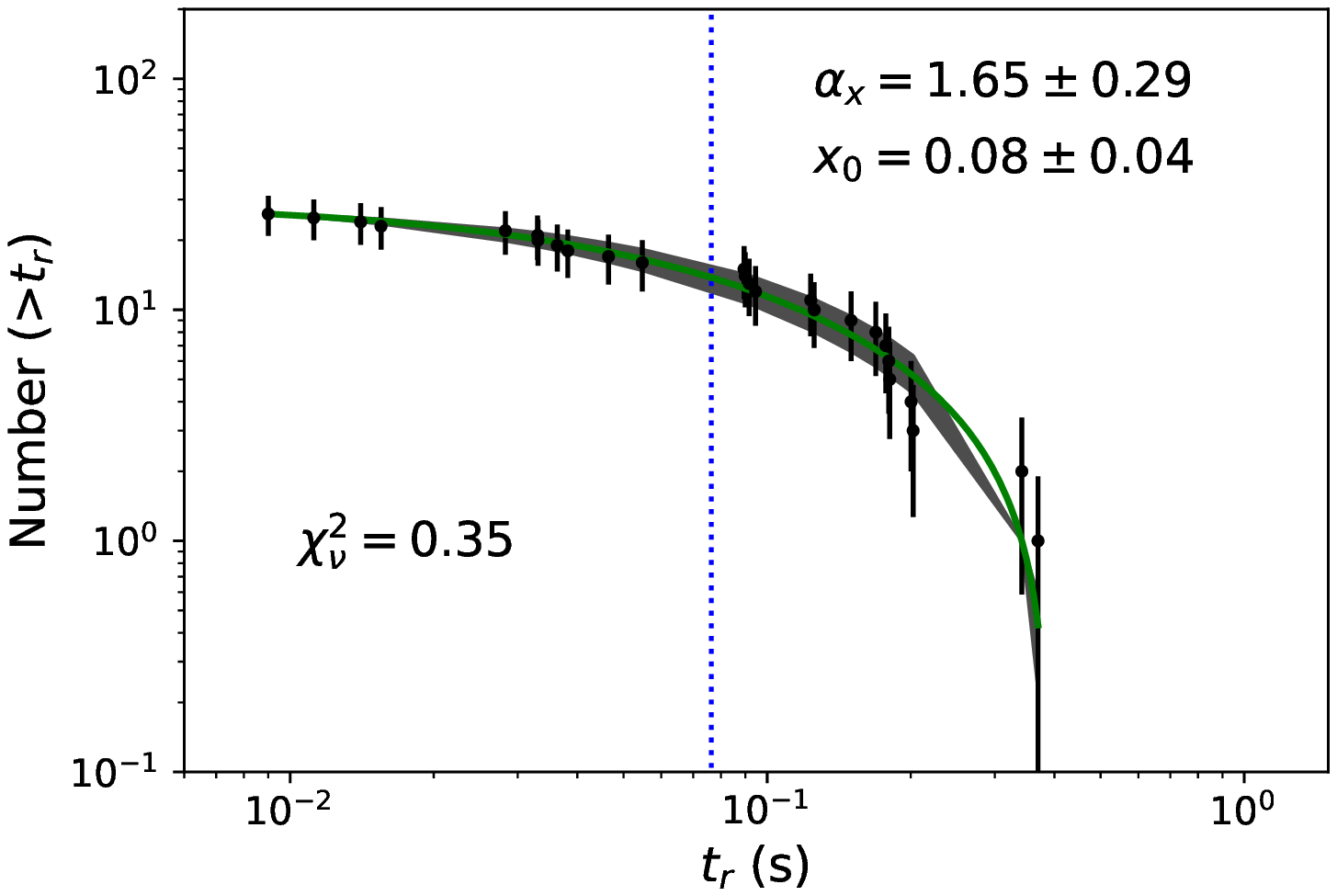}{0.5\textwidth}{(c)}
 \fig{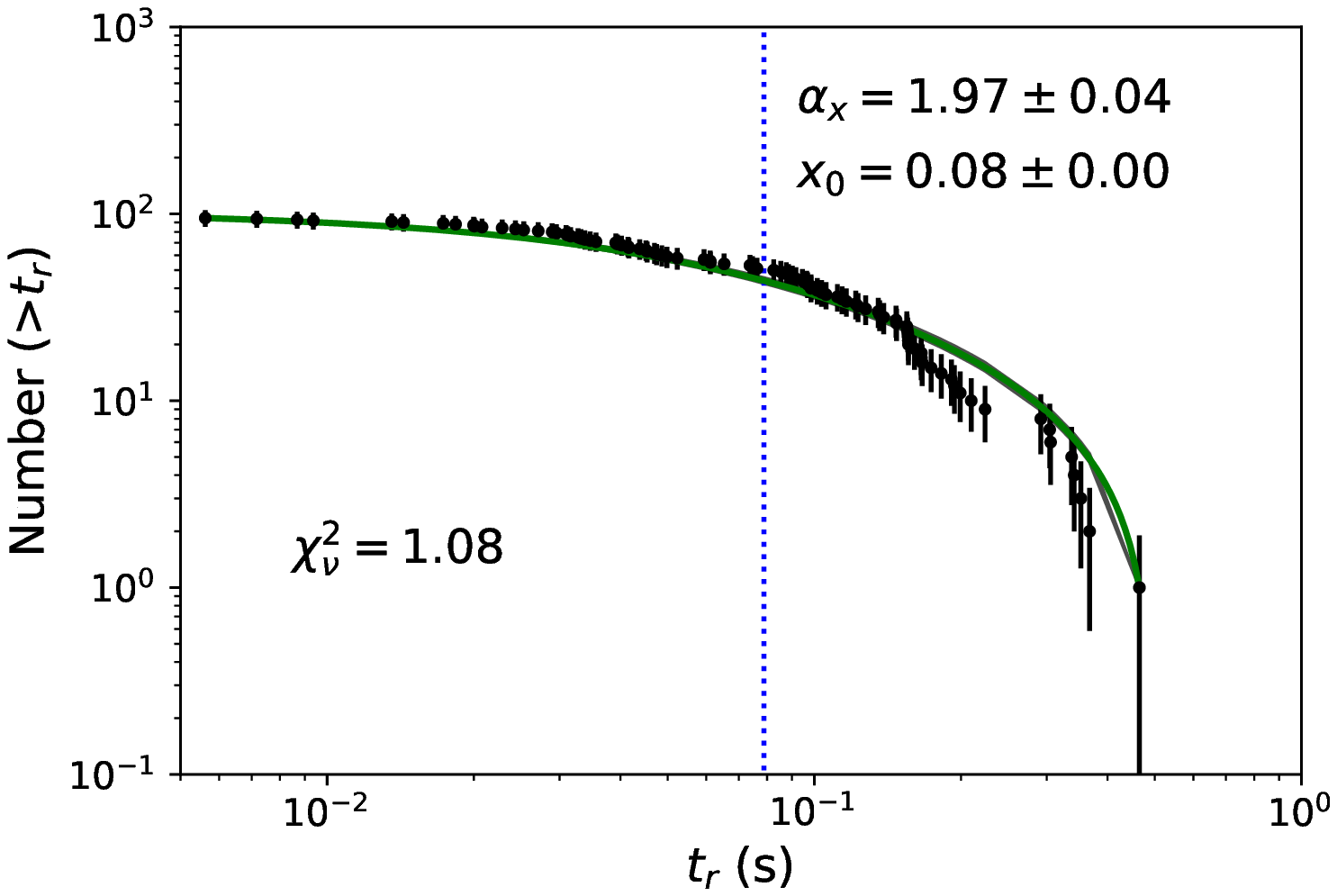}{0.5\textwidth}{(d)}
          }
           \gridline{
 \fig{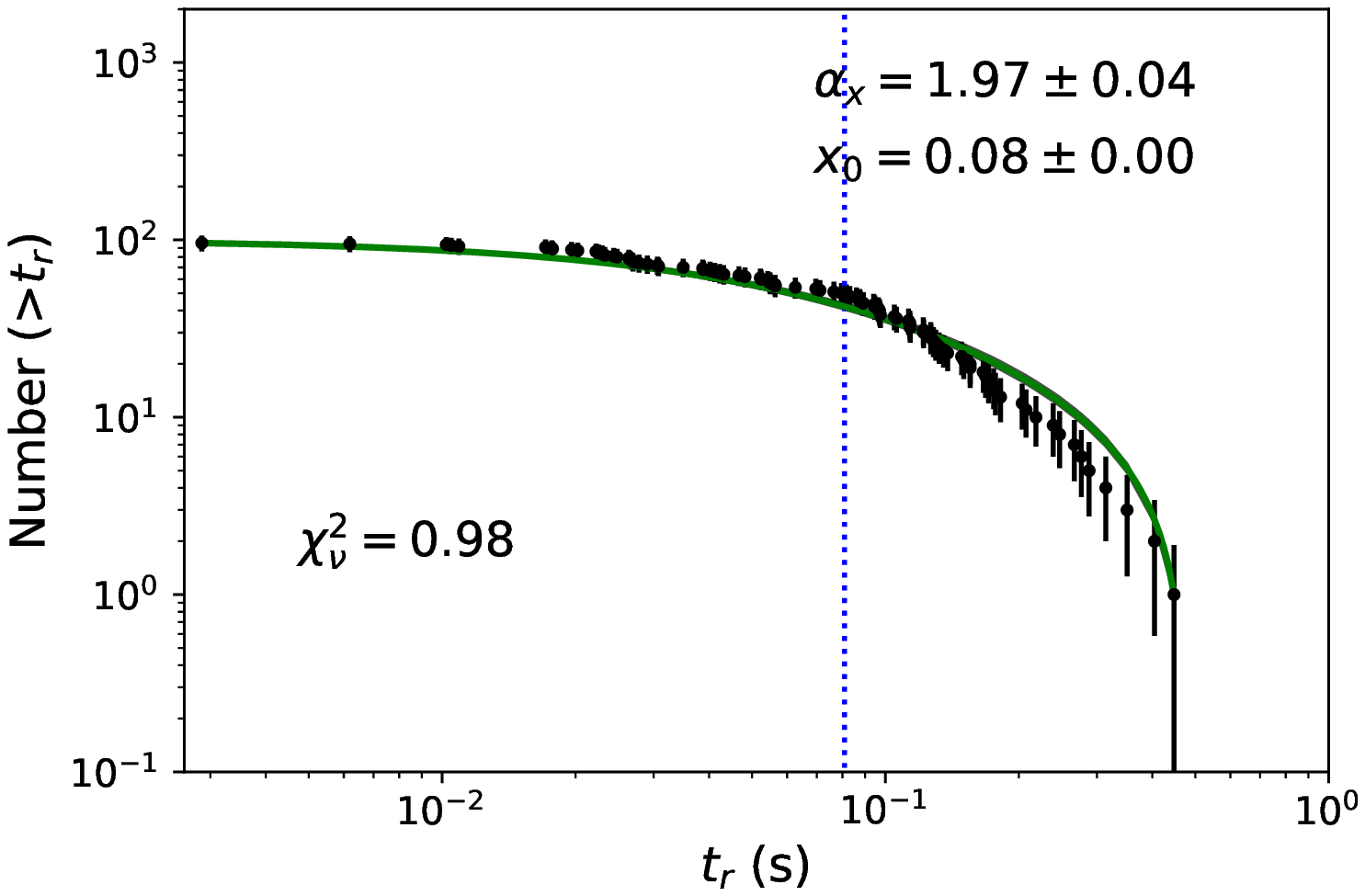}{0.5\textwidth}{(e)}
 \fig{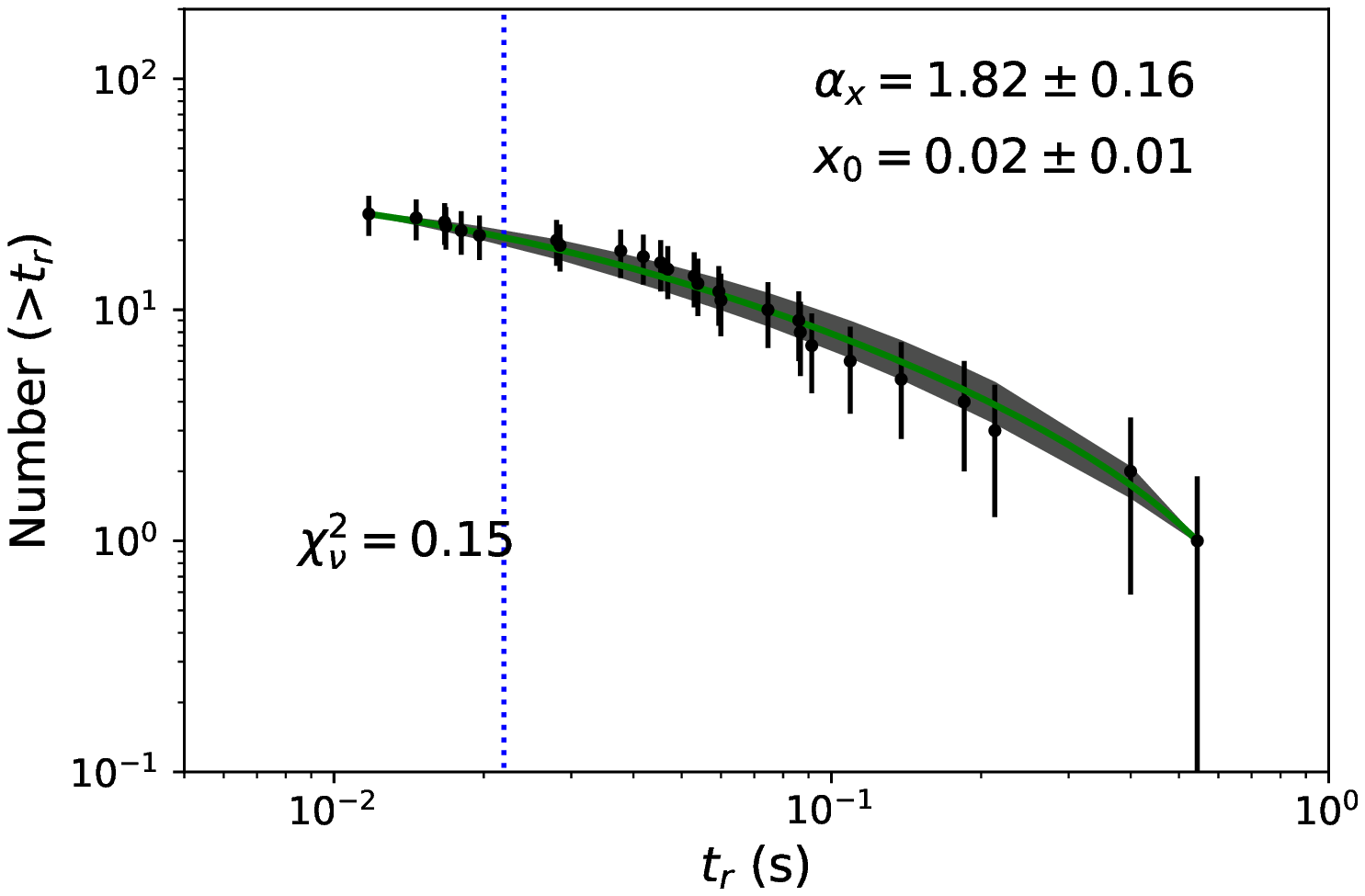}{0.5\textwidth}{(f)}
          }
\caption{The distributions of $t_{\rm r}$ for BATSE GRBs. The symbols are the same as those in Figure 1. \label{trdff}}
\end{figure*}

\begin{figure*}
\centering
\gridline{
\fig{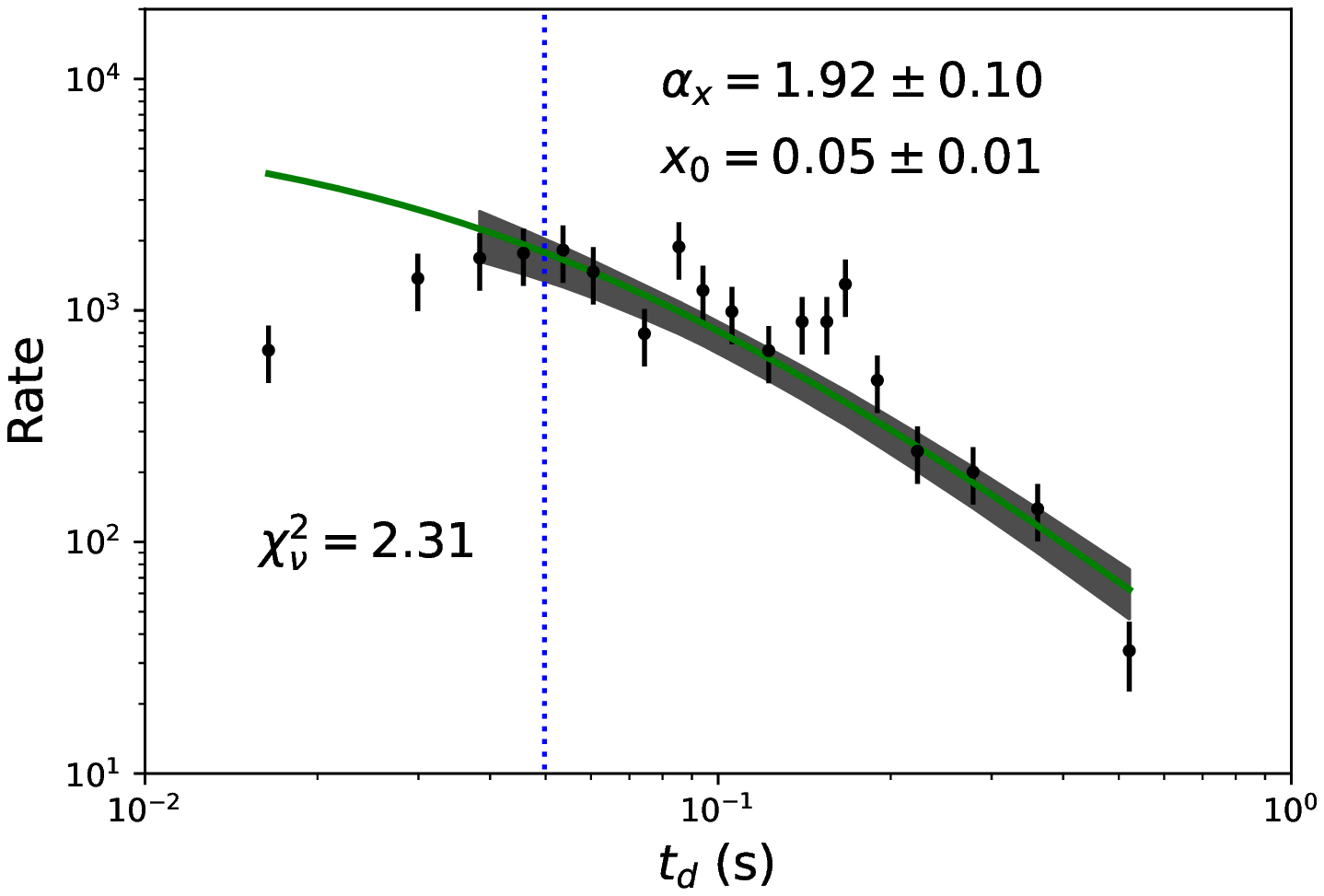}{0.5\textwidth}{(a)}
\fig{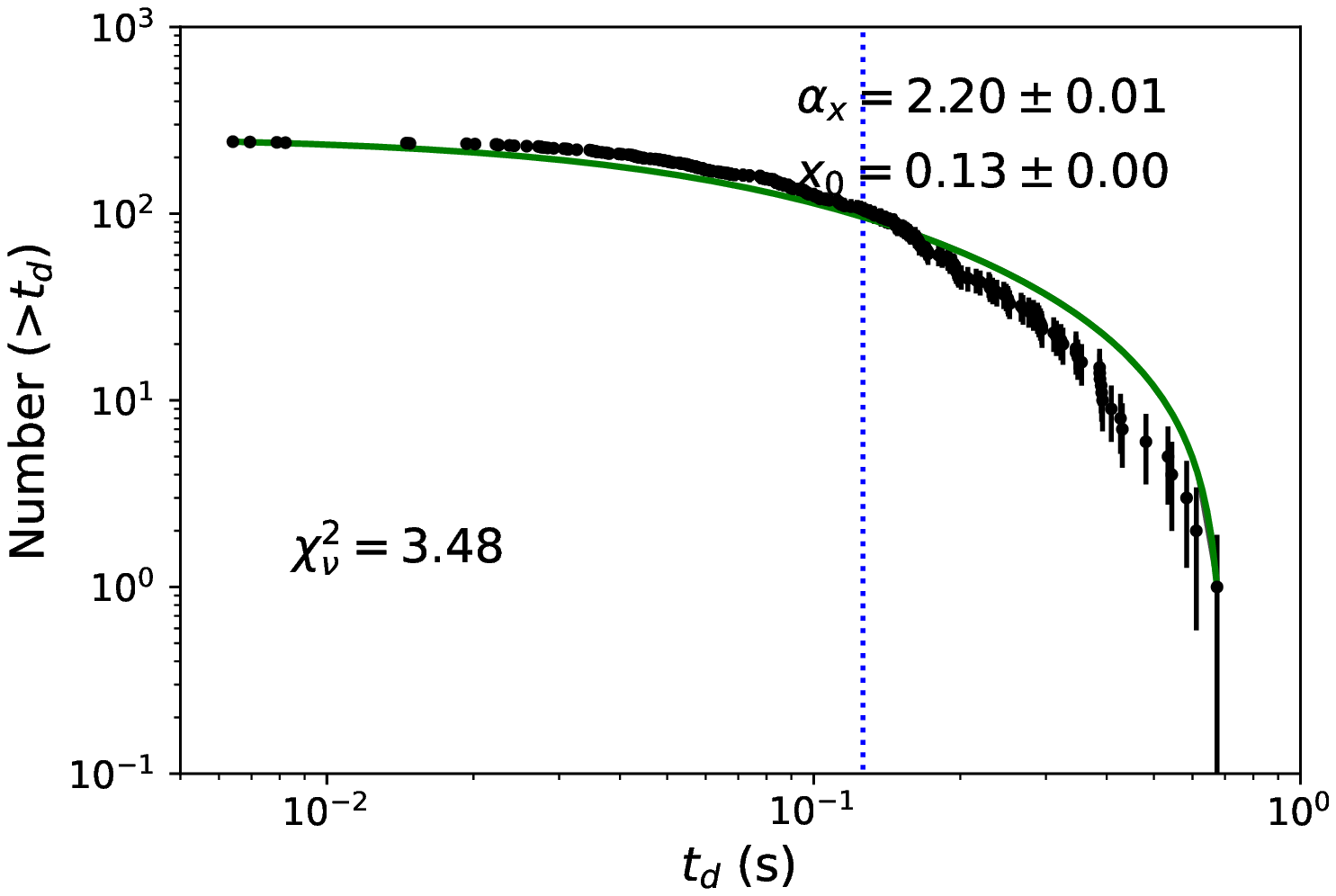}{0.5\linewidth}{(b)}
          }
 \gridline{
 \fig{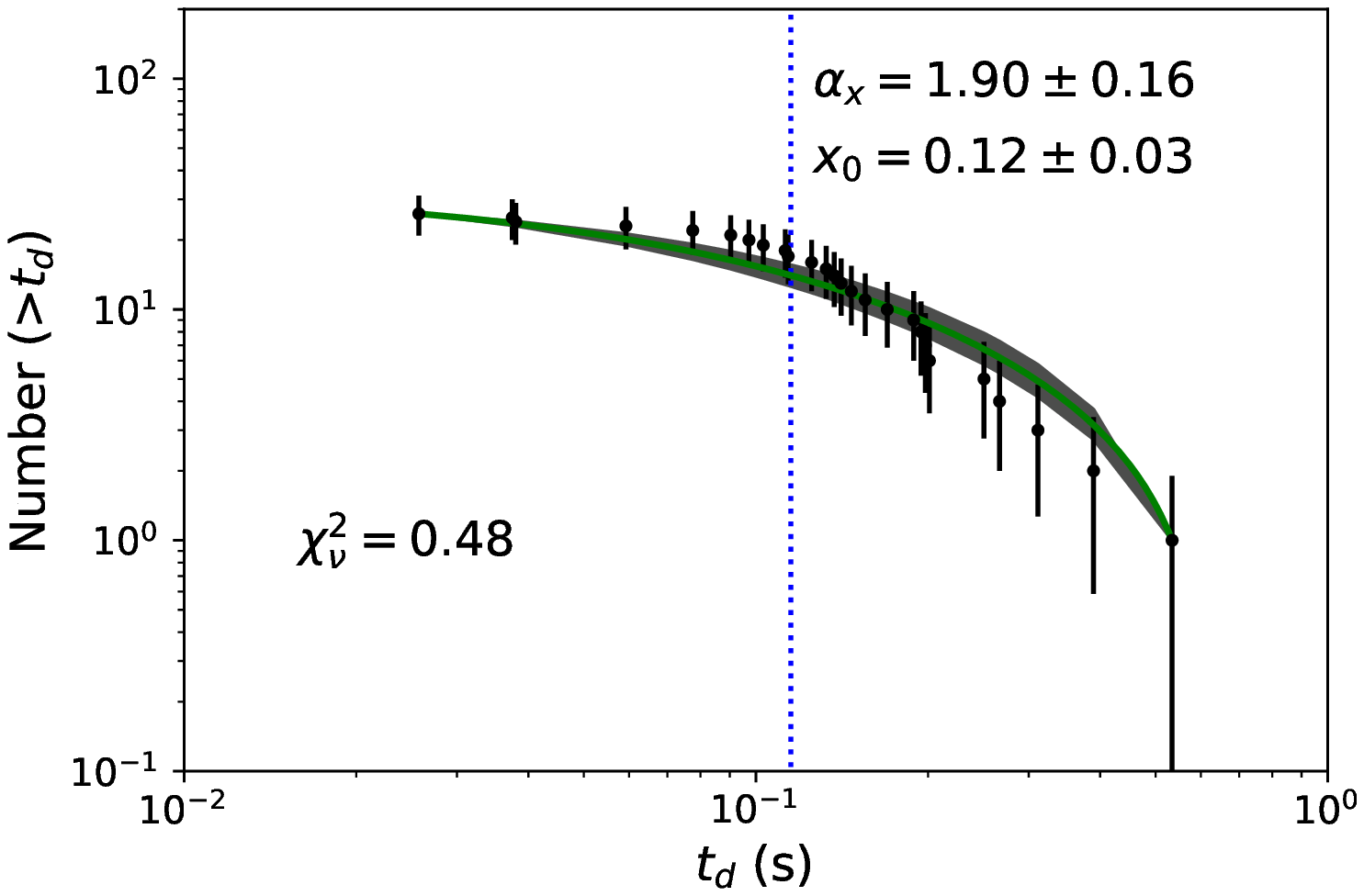}{0.5\textwidth}{(c)}
 \fig{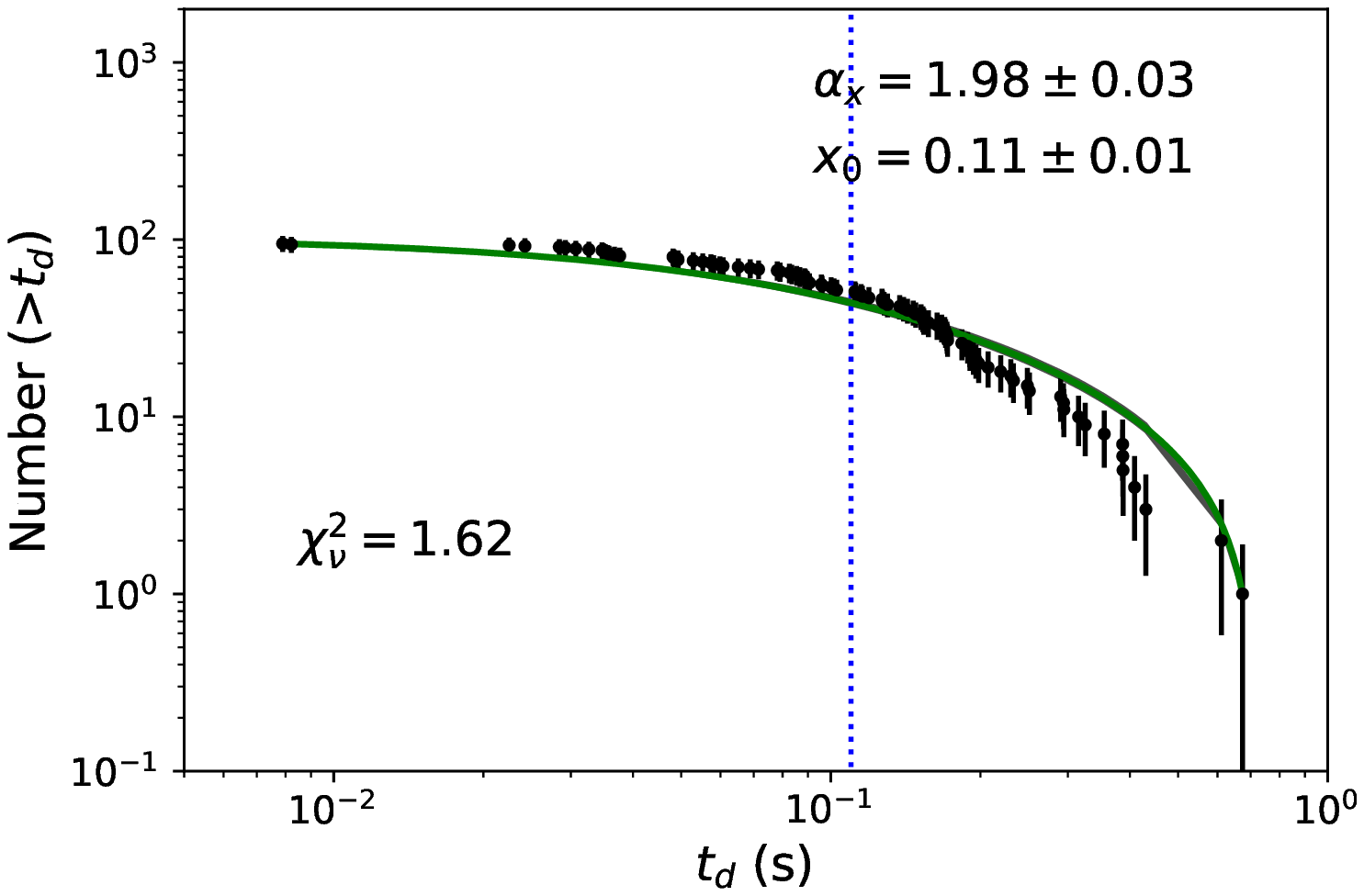}{0.5\textwidth}{(d)}
          }
           \gridline{
 \fig{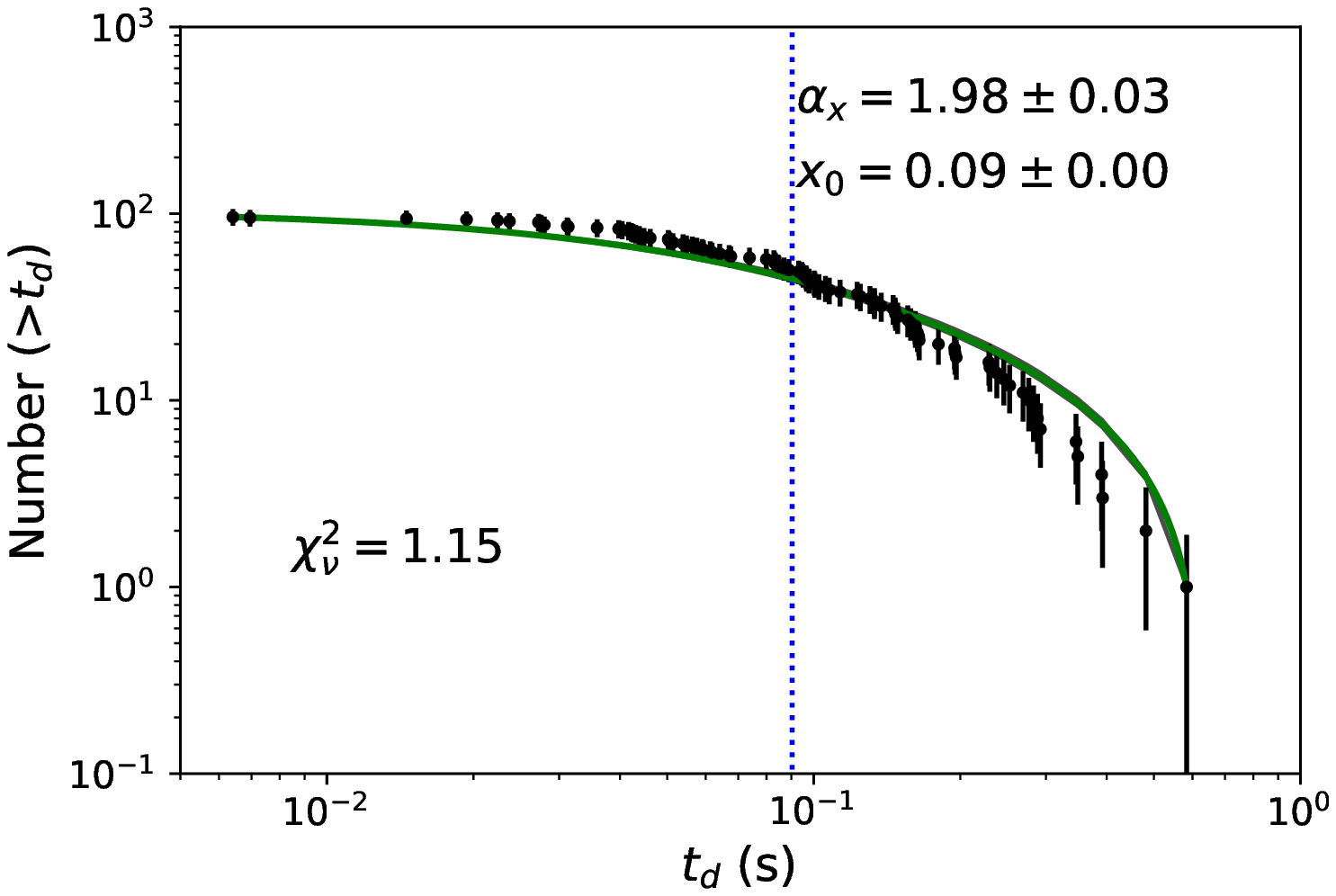}{0.5\textwidth}{(e)}
 \fig{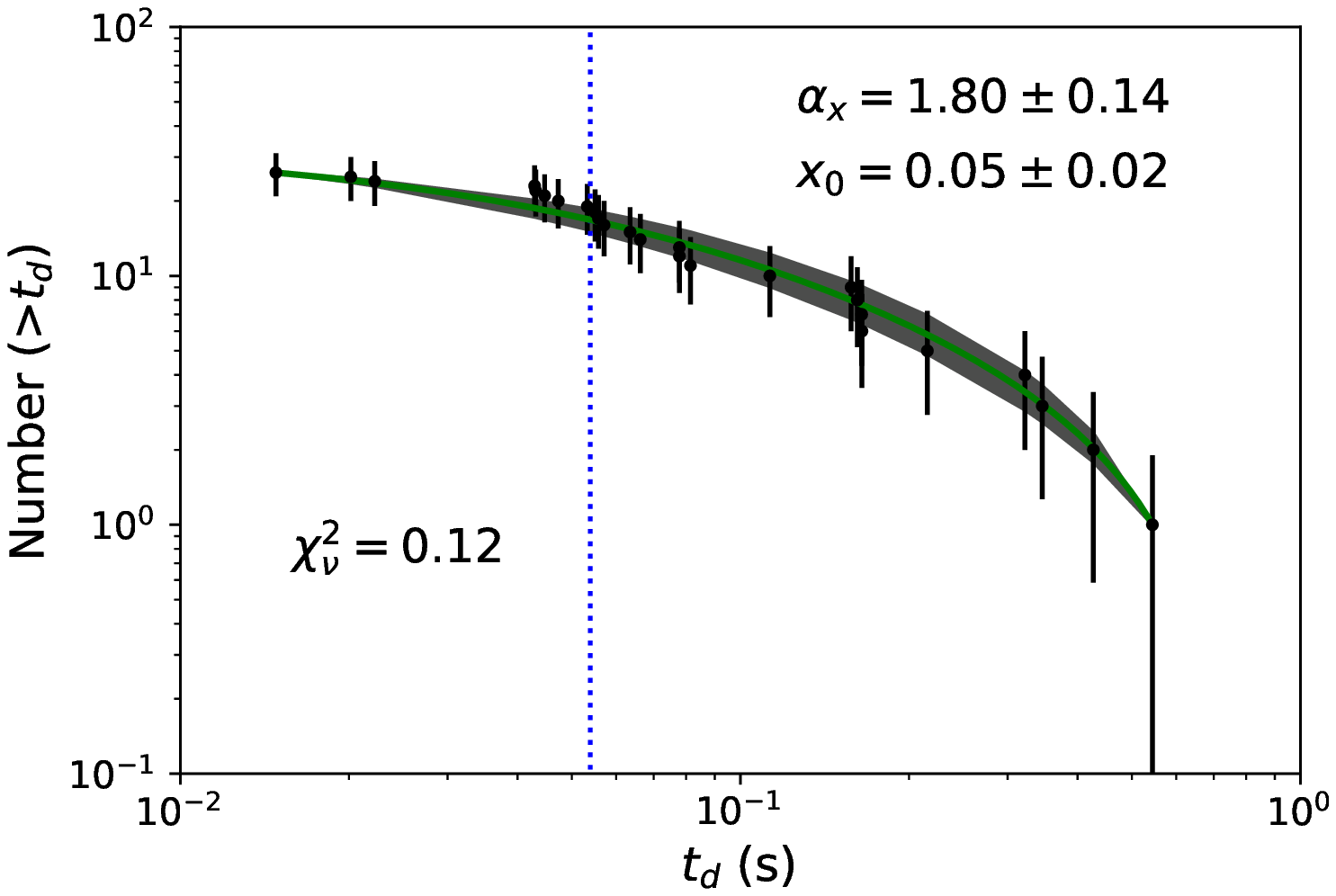}{0.5\textwidth}{(f)}
          }
\caption{The distributions of $t_{\rm d}$ for BATSE GRBs. The symbols are the same as those in Figure 1. \label{tddff}}
\end{figure*}

\begin{figure*}
\centering
\gridline{
\fig{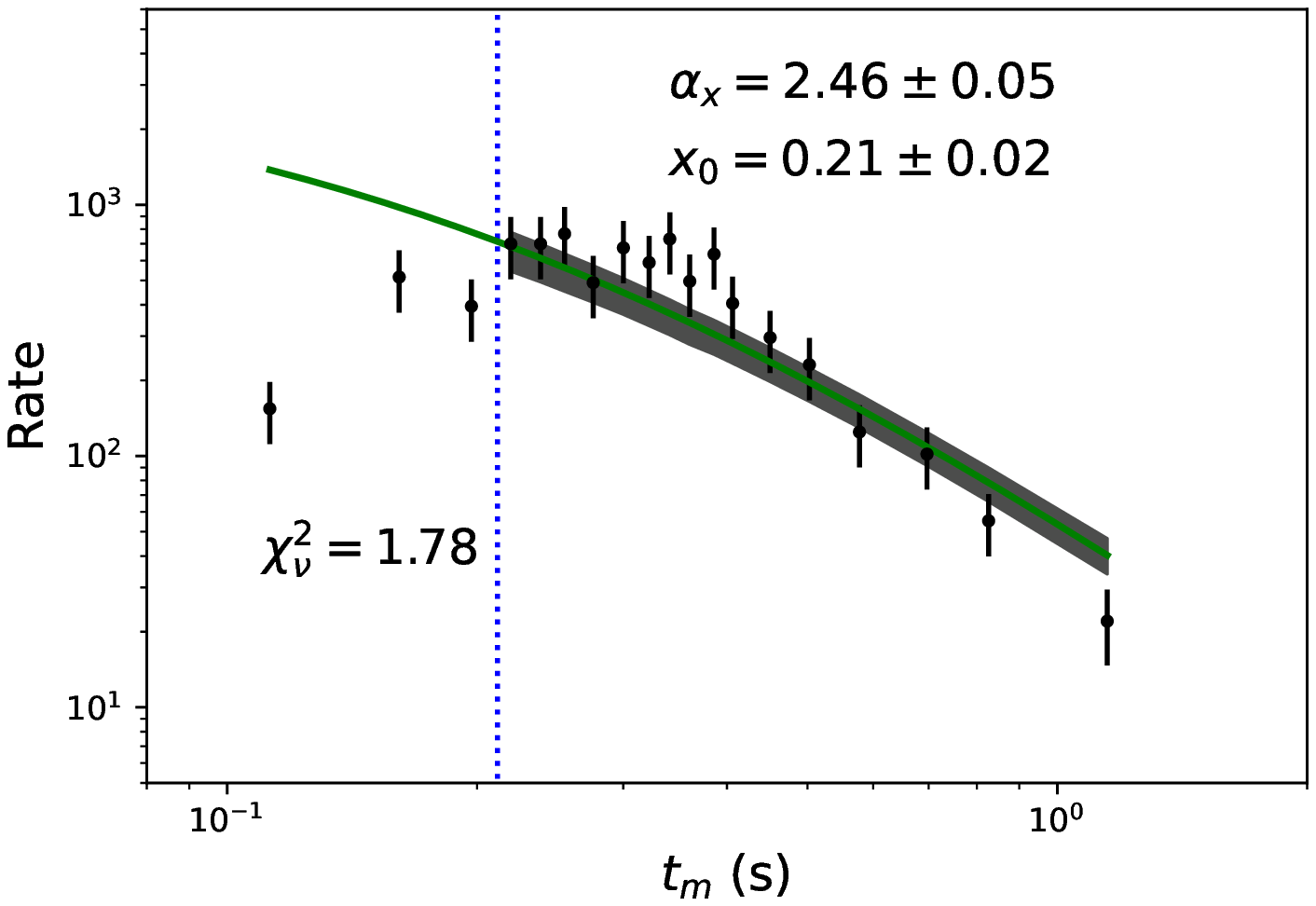}{0.5\textwidth}{(a)}
\fig{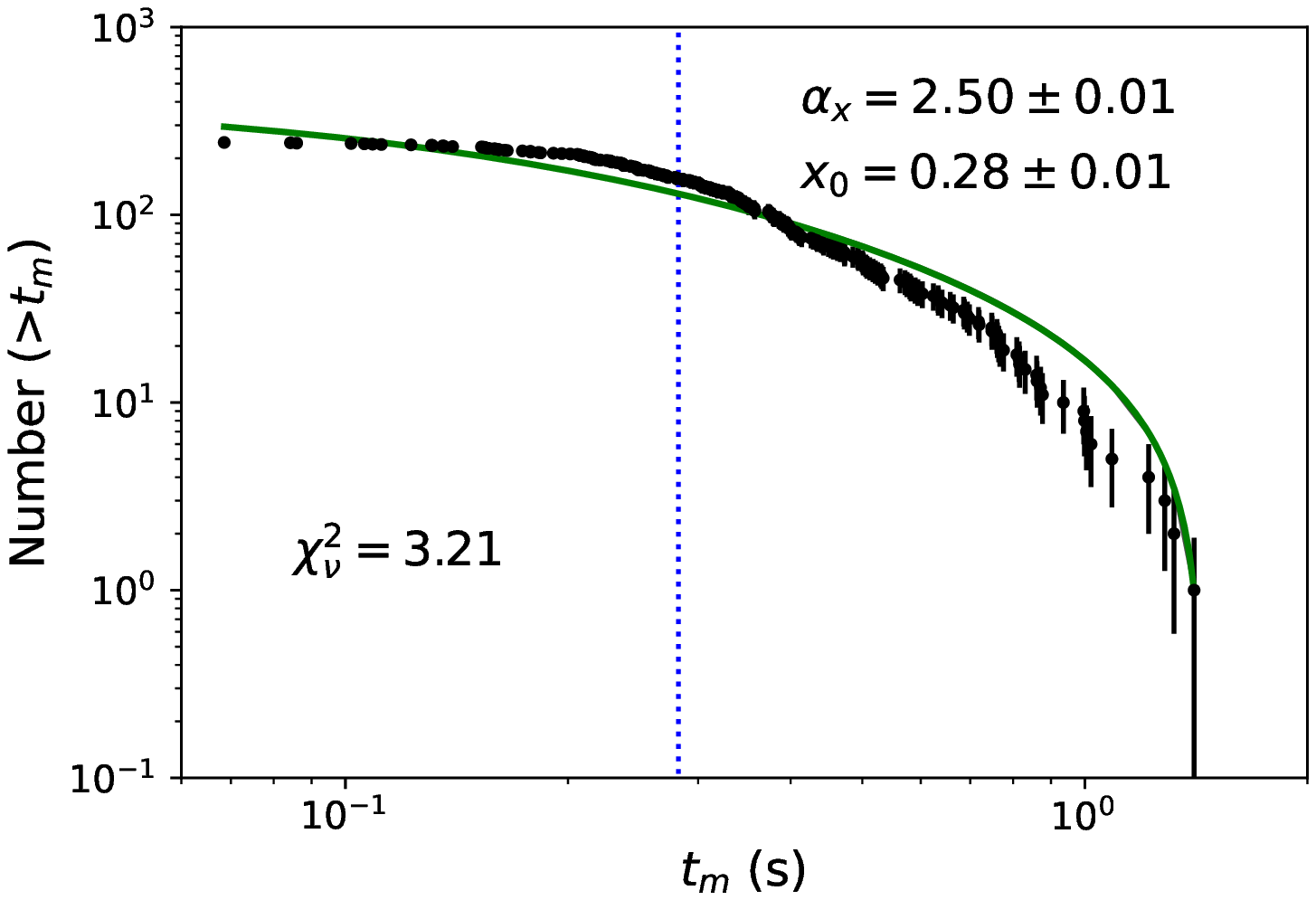}{0.5\linewidth}{(b)}
          }
 \gridline{
 \fig{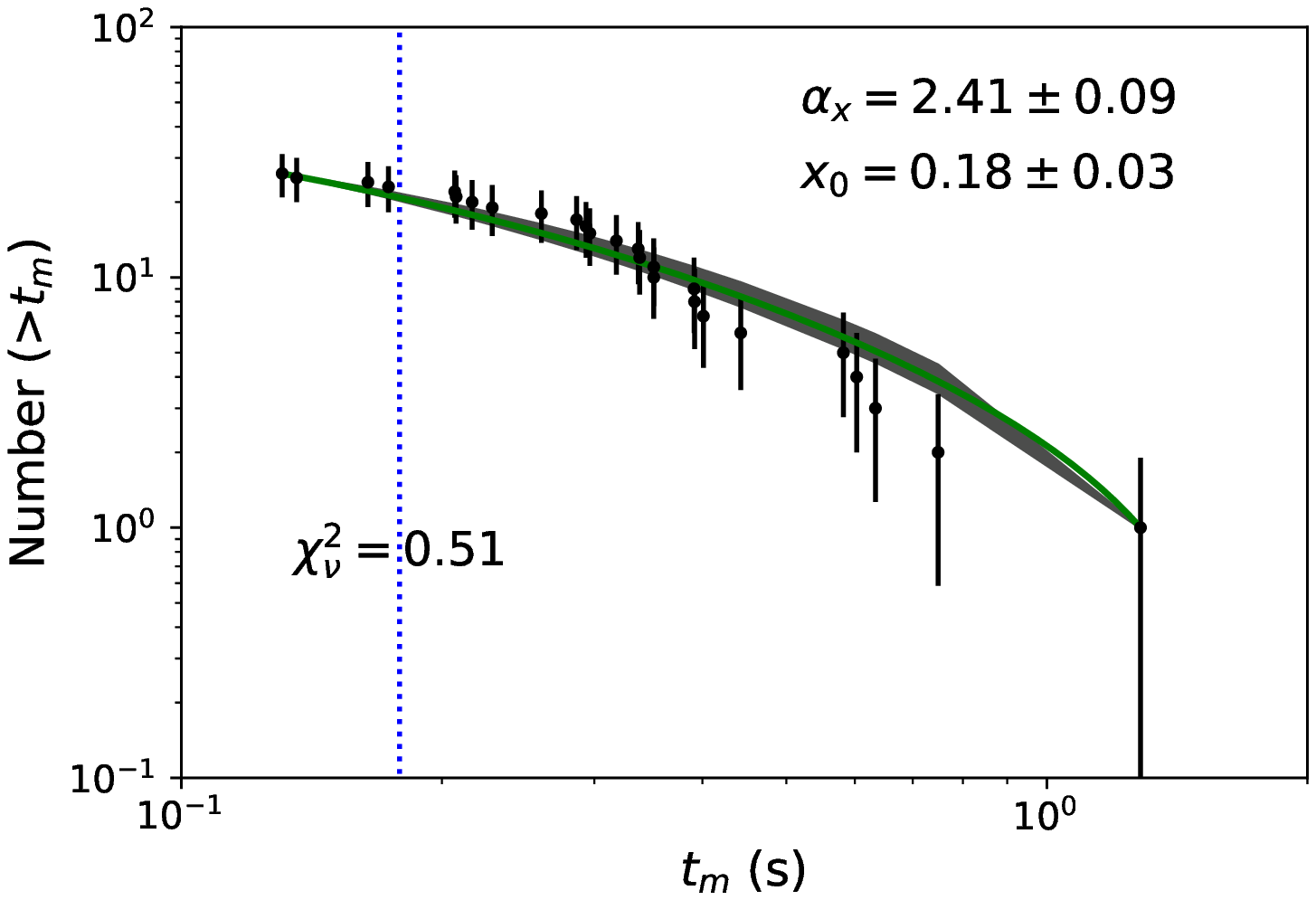}{0.5\textwidth}{(c)}
 \fig{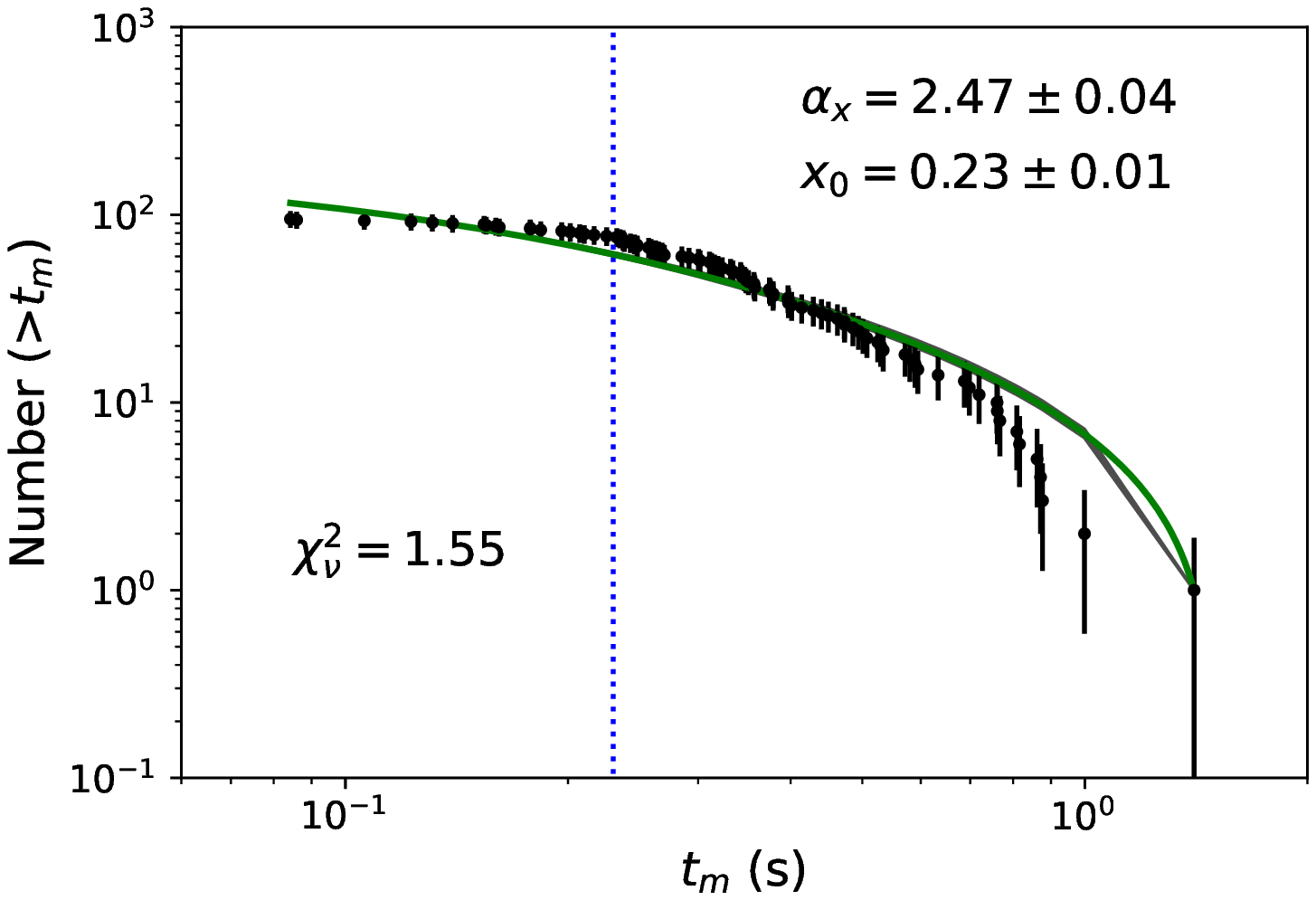}{0.5\textwidth}{(d)}
          }
           \gridline{
 \fig{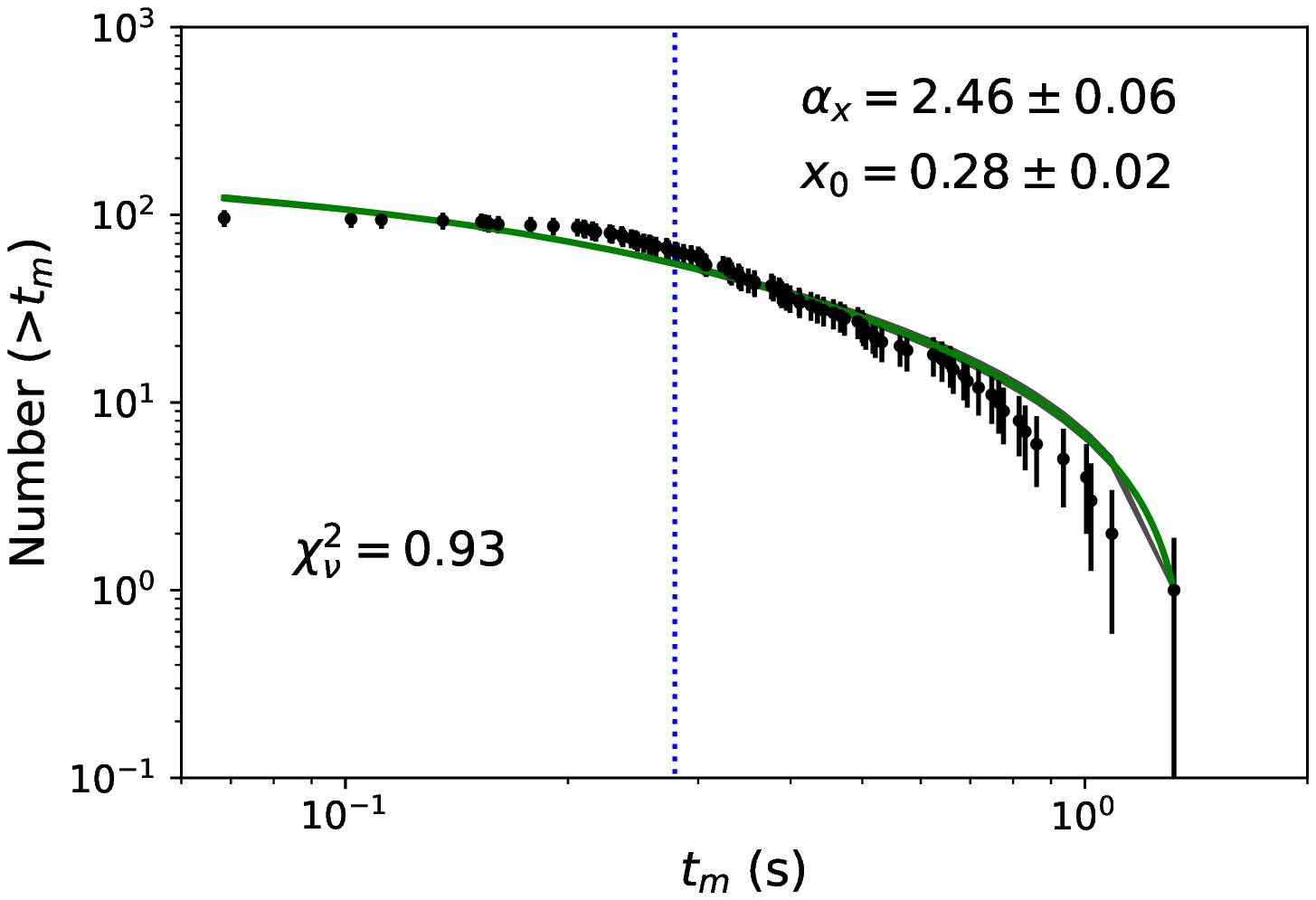}{0.5\textwidth}{(e)}
 \fig{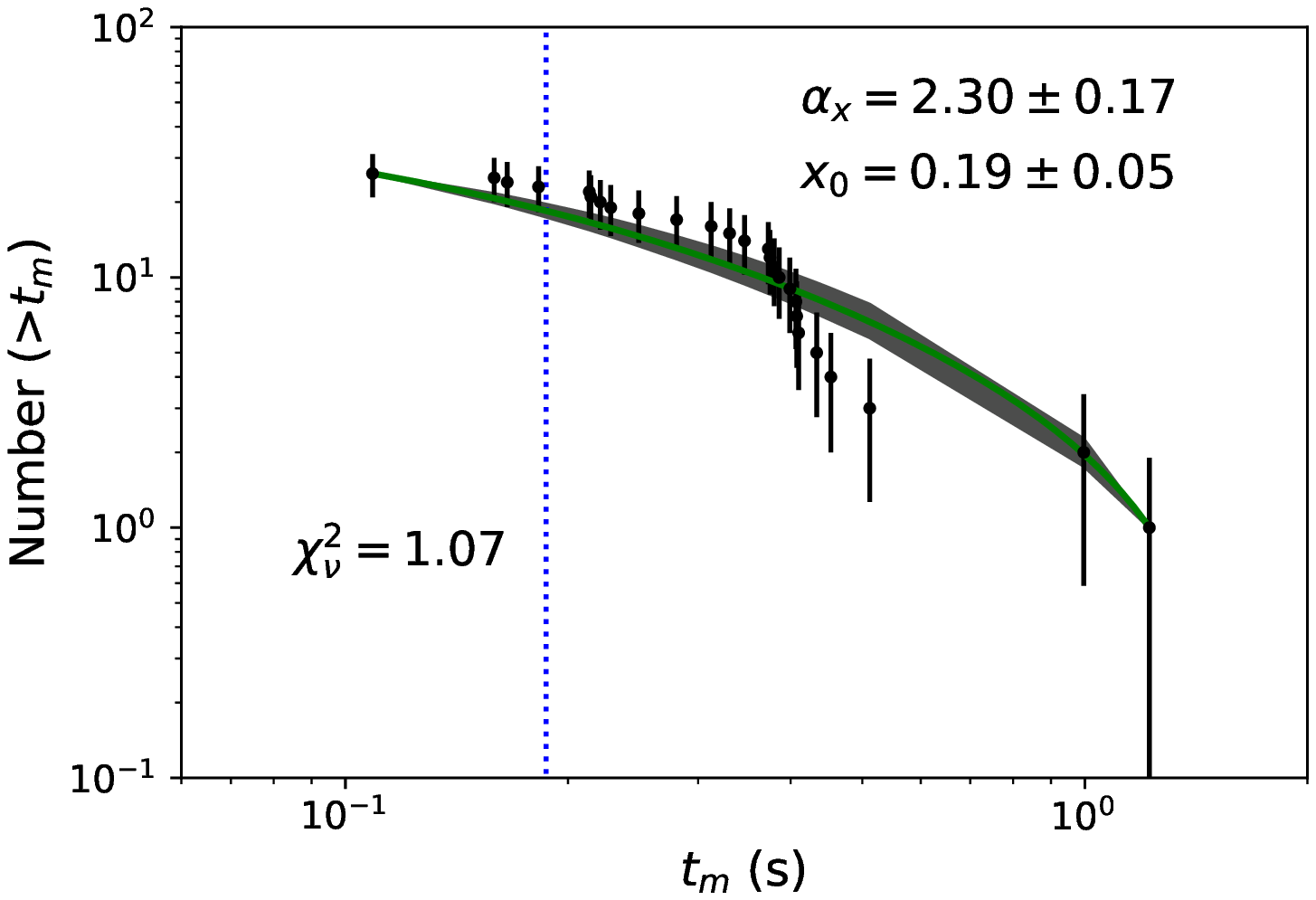}{0.5\textwidth}{(f)}
          }
\caption{The distributions of $t_{\rm m}$ for BATSE GRBs. The symbols are the same as those in Figure 1. \label{tmdff}}
\end{figure*}

\begin{figure*}
\centering
\gridline{
\fig{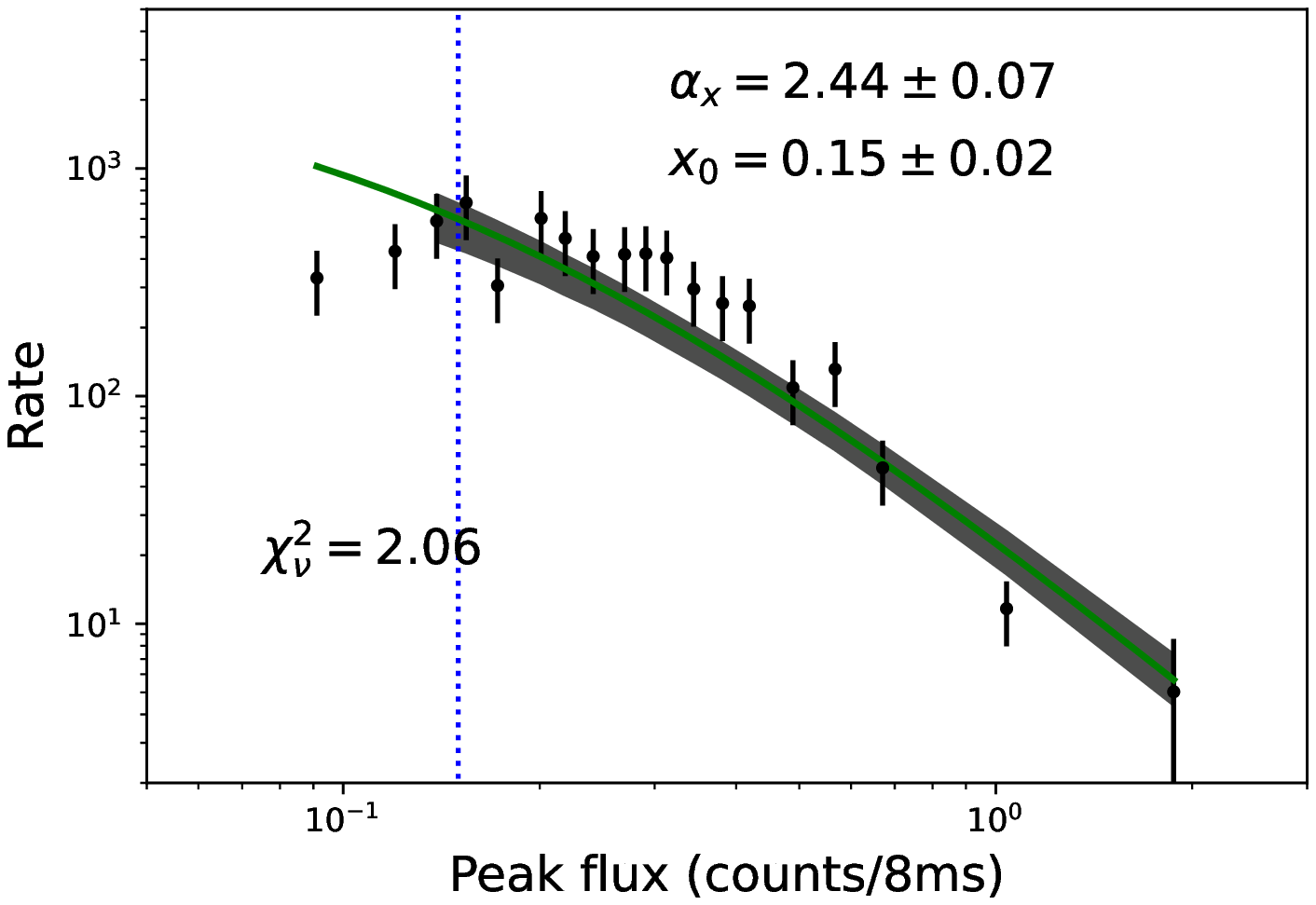}{0.5\textwidth}{(a)}
\fig{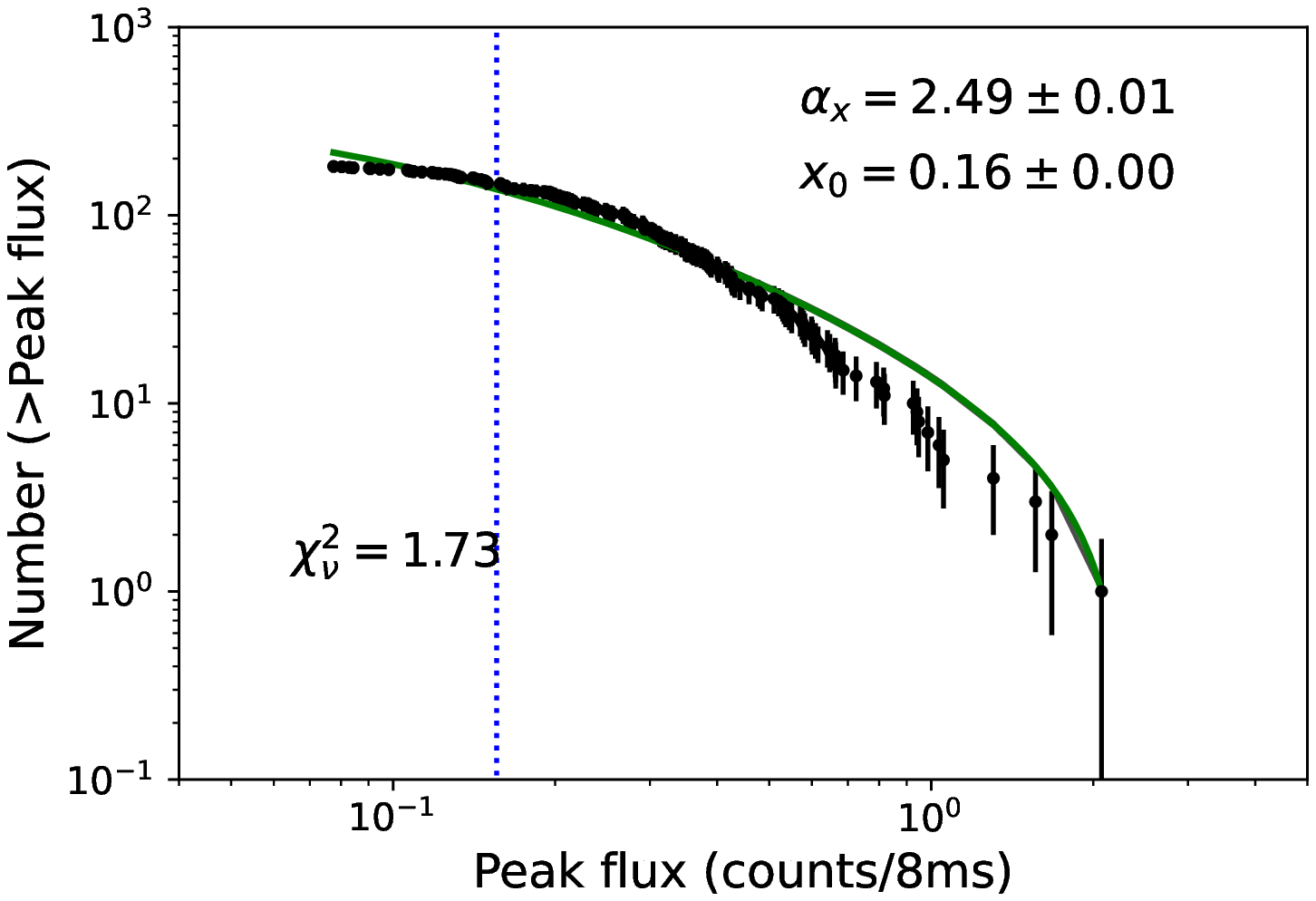}{0.5\linewidth}{(b)}
          }
 \gridline{
 \fig{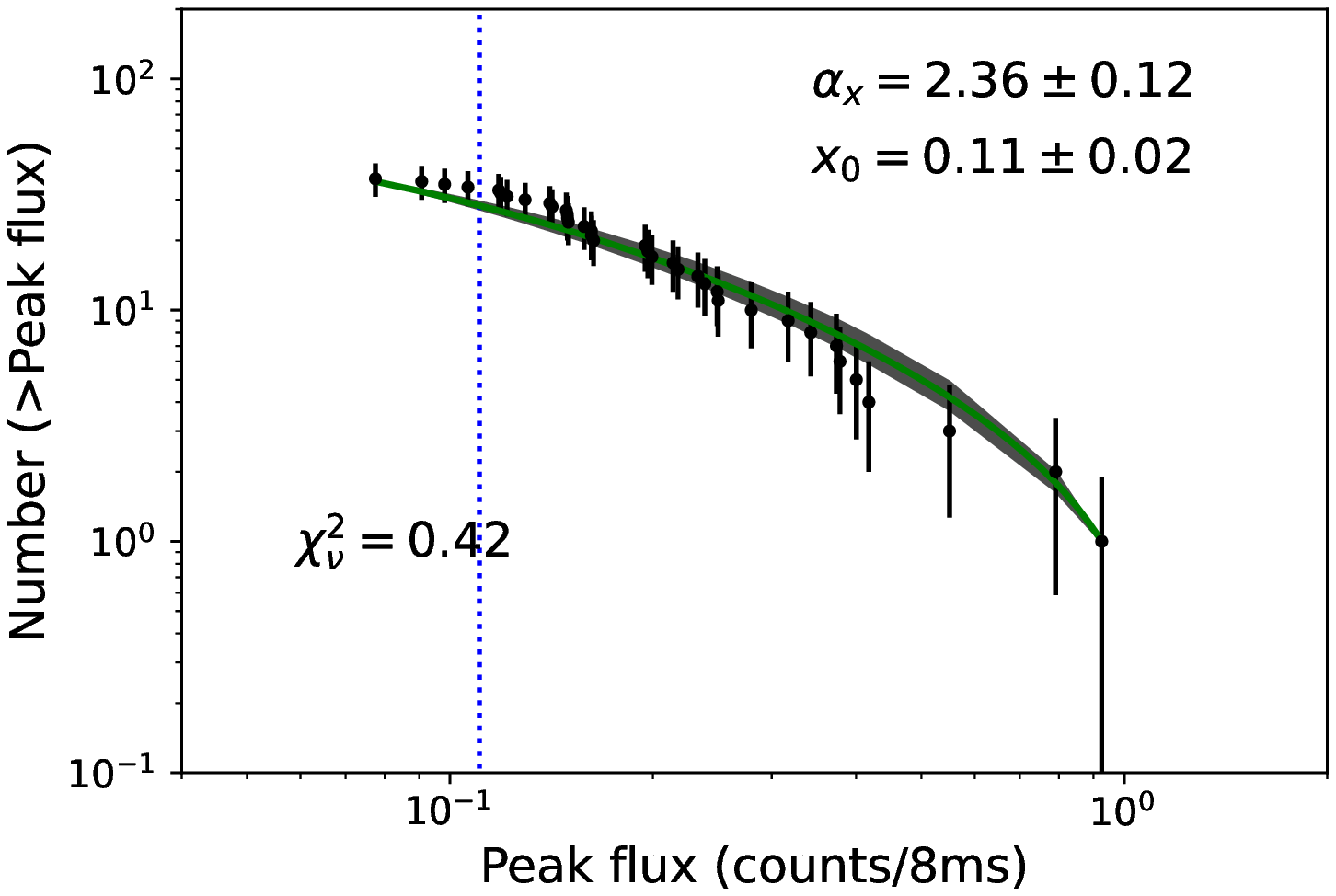}{0.5\textwidth}{(c)}
 \fig{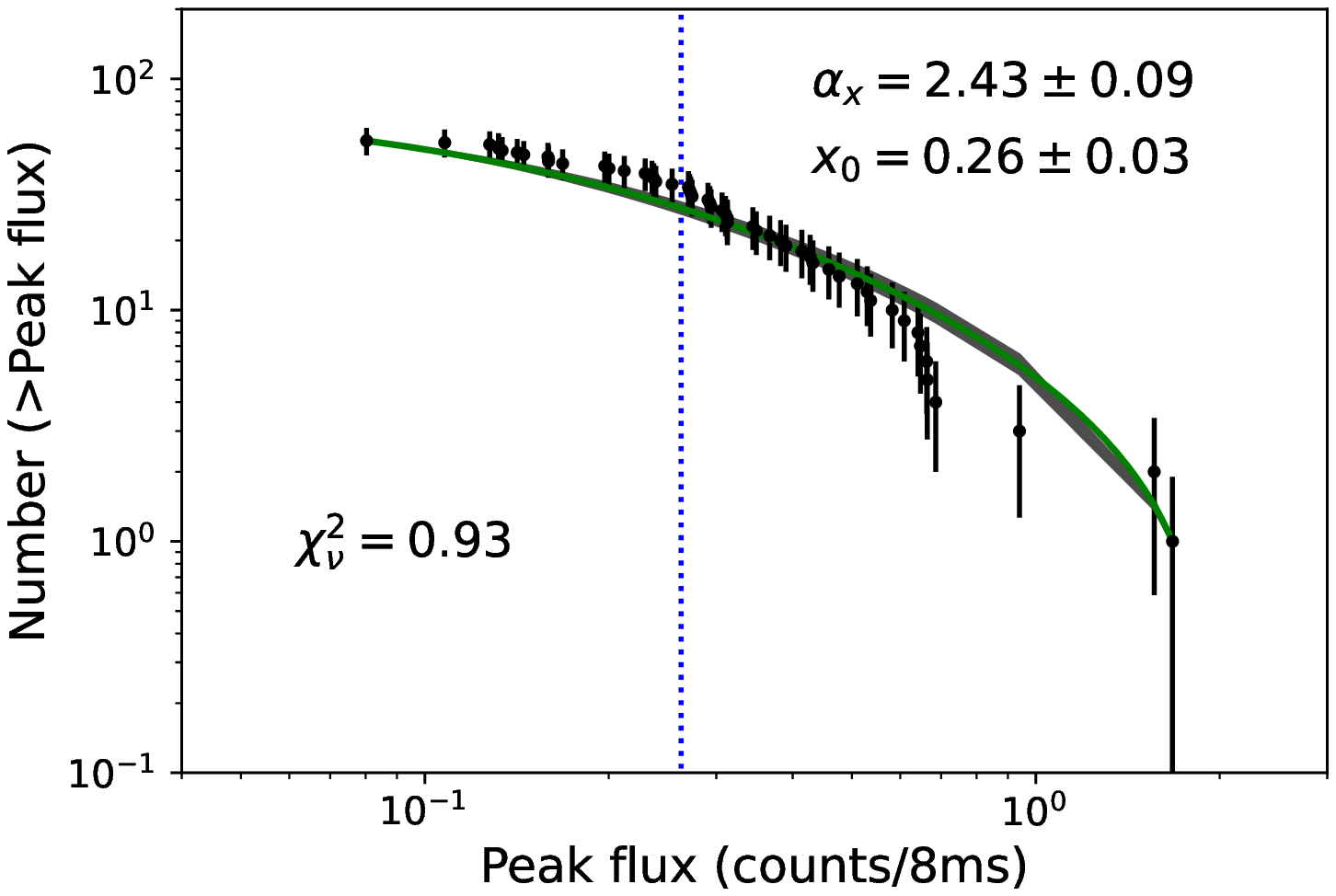}{0.5\textwidth}{(d)}
          }
           \gridline{
 \fig{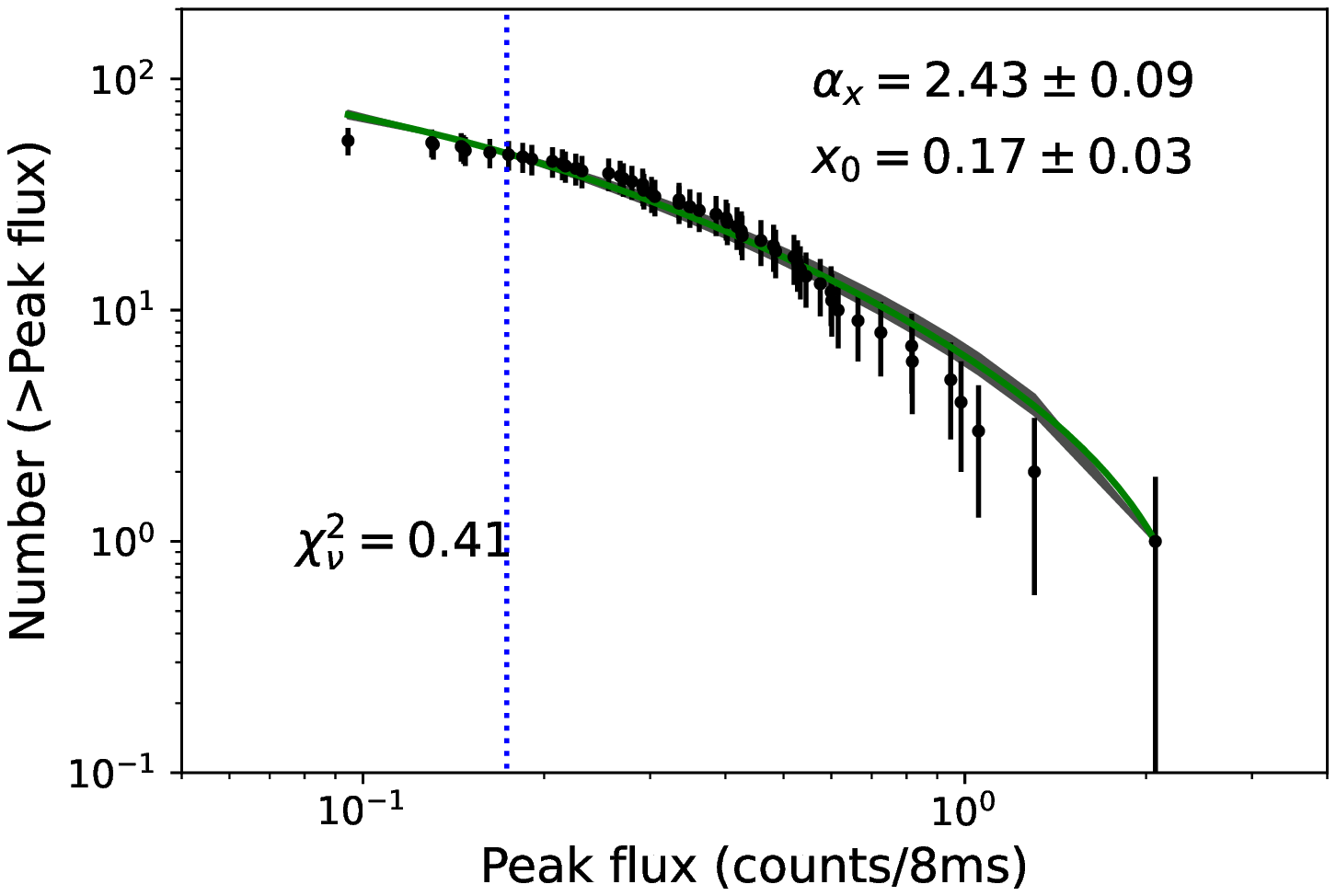}{0.5\textwidth}{(e)}
 \fig{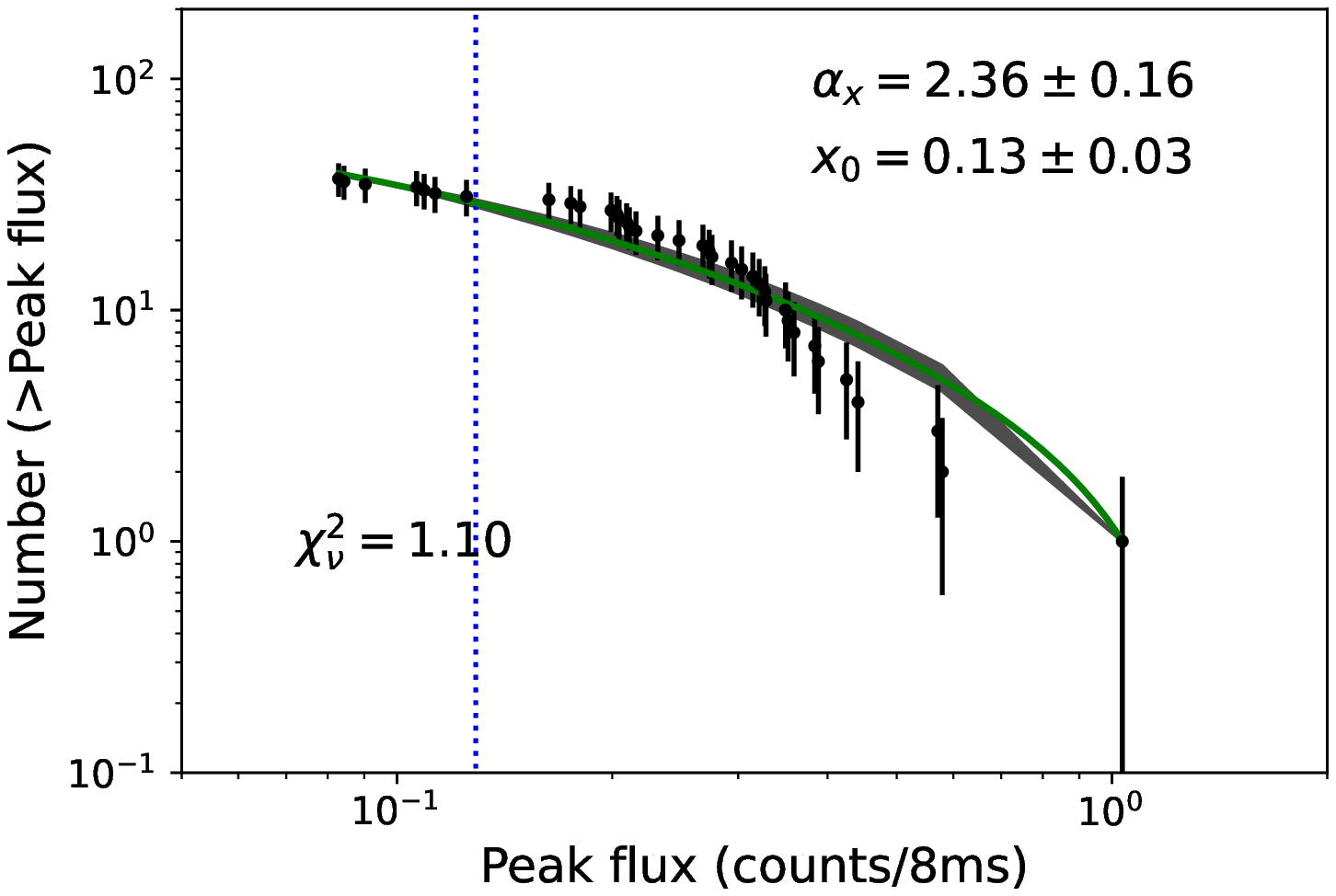}{0.5\textwidth}{(f)}
          }
\caption{The distributions of $f_{\rm m}$ for Swift GRBs. (a) The differential distribution of $f_{\rm m}$ for the total Swift sample. (b) The cumulative distribution of $f_{\rm m}$ for the total Swift sample. (c) The cumulative distribution of $f_{\rm m}$ in Ch1. (d) The cumulative distribution of $f_{\rm m}$ in Ch2. (d) The cumulative distribution of $f_{\rm m}$ in Ch3. (e) The cumulative distribution of $f_{\rm m}$ in Ch4. The gray region represents the 95\% confidence level, the green solid line is the best fit, and the blue dotted line is marked as the threshold $x_{\rm 0}$. \label{fmdff2}}
\end{figure*}

\begin{figure*}
\centering
\gridline{
\fig{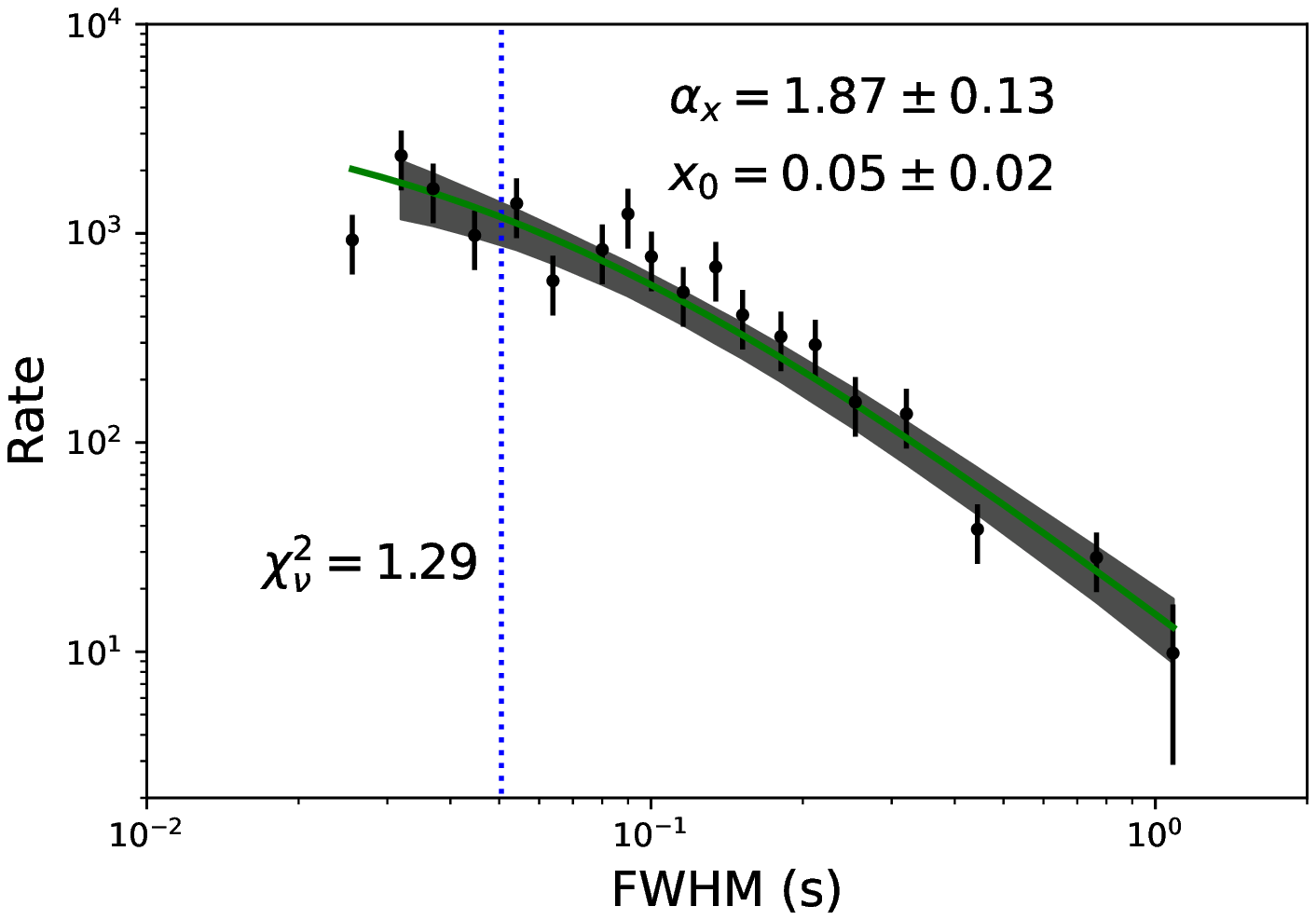}{0.5\textwidth}{(a)}
\fig{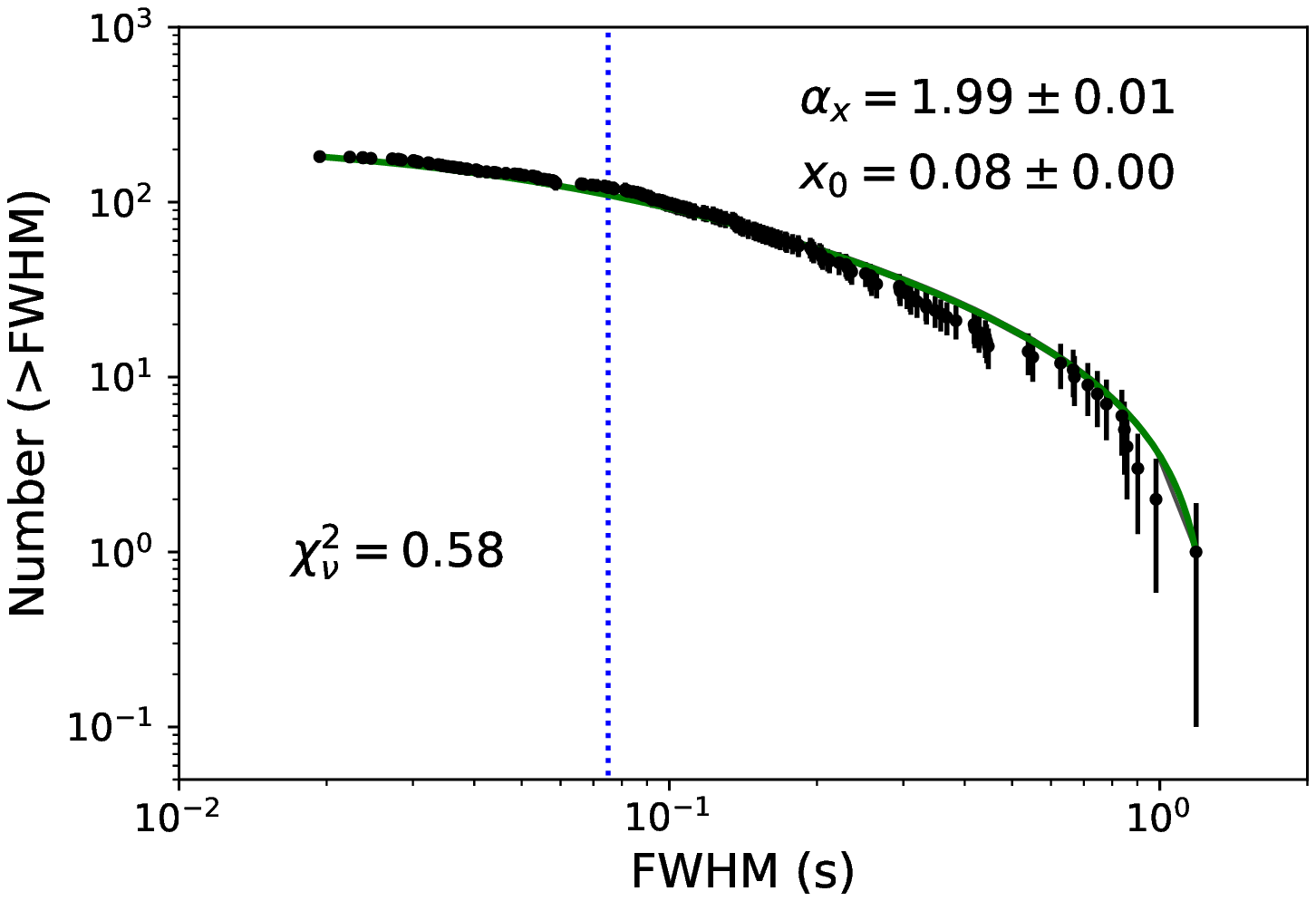}{0.5\linewidth}{(b)}
          }
 \gridline{
 \fig{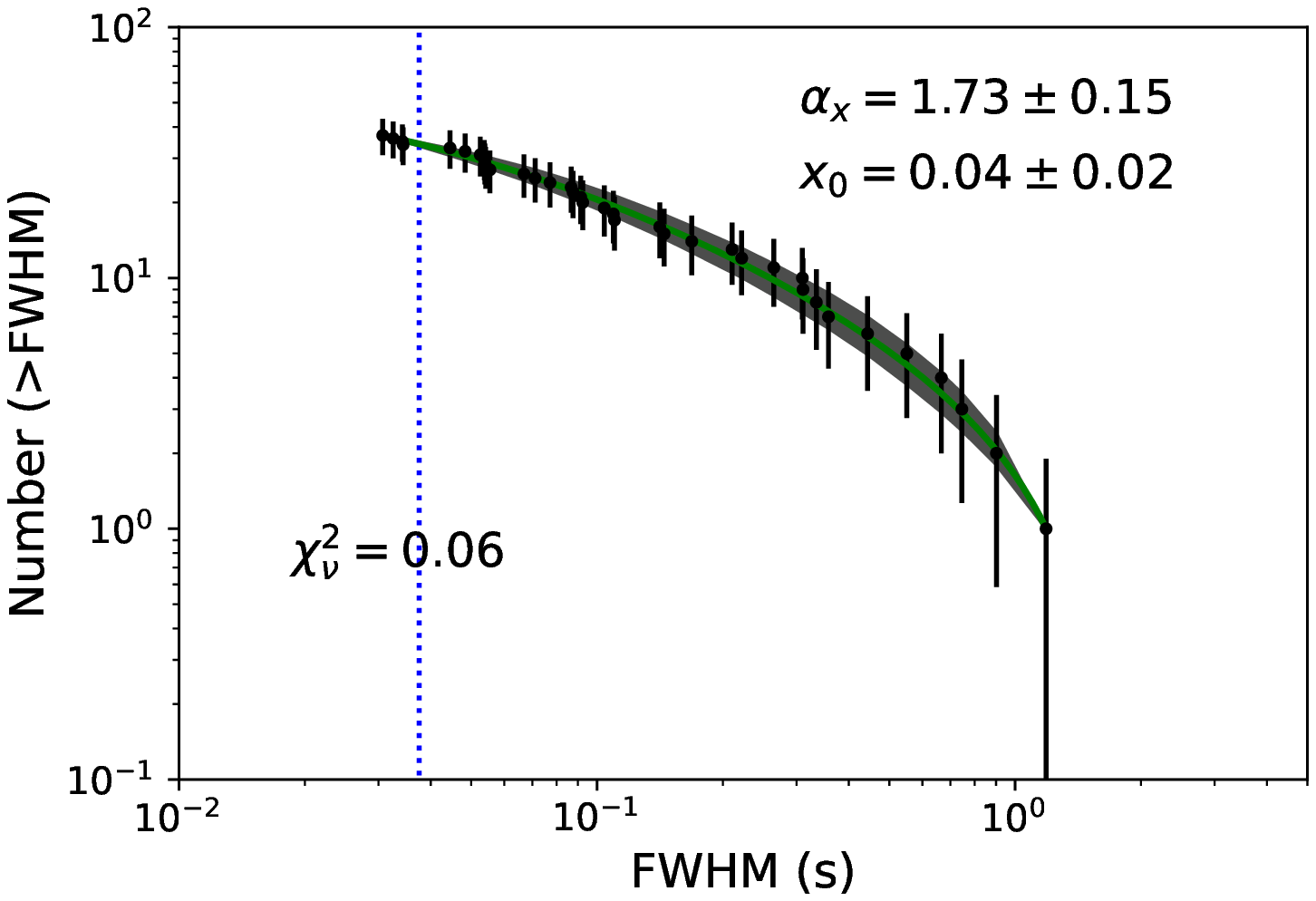}{0.5\textwidth}{(c)}
 \fig{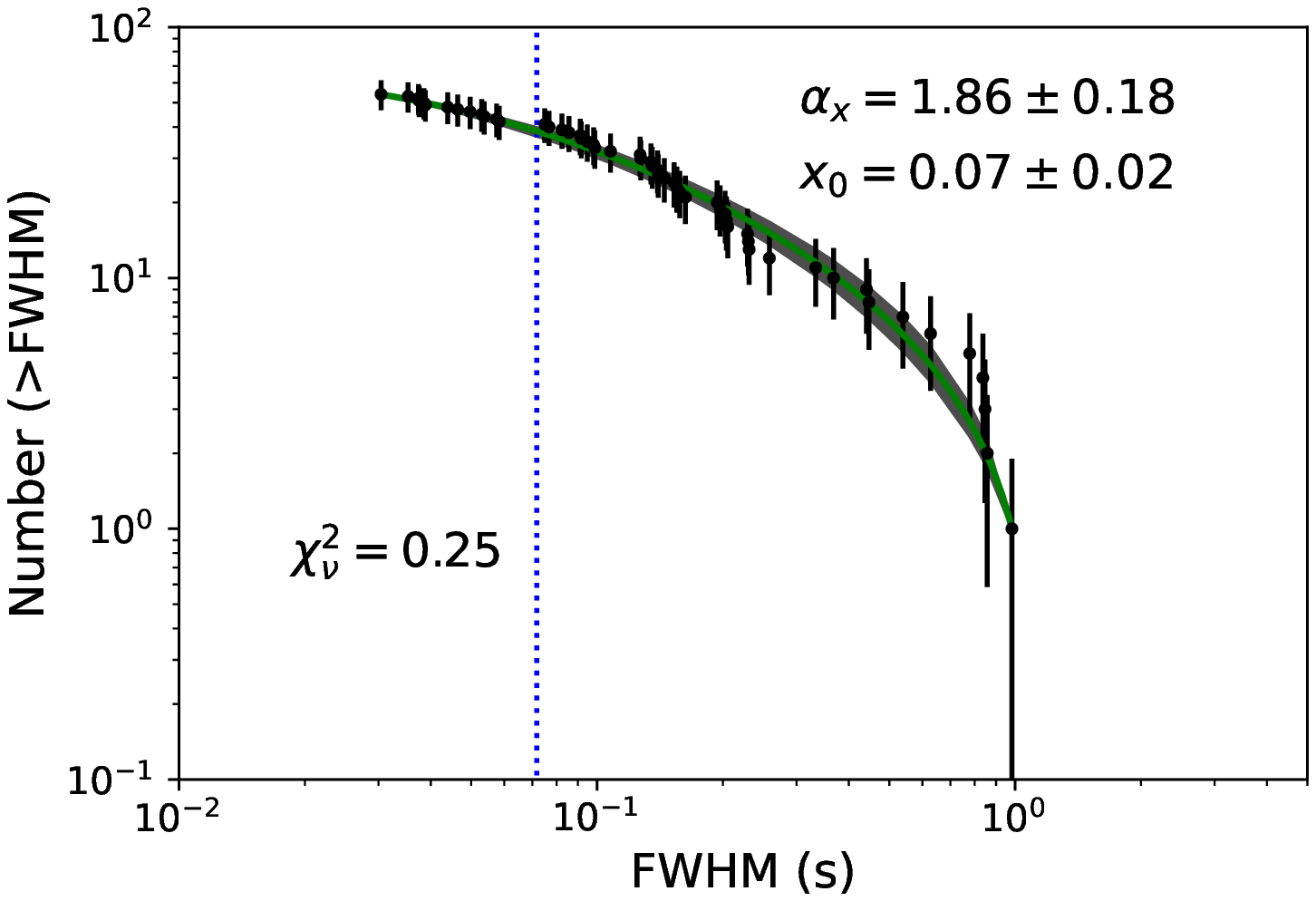}{0.5\textwidth}{(d)}
          }
           \gridline{
 \fig{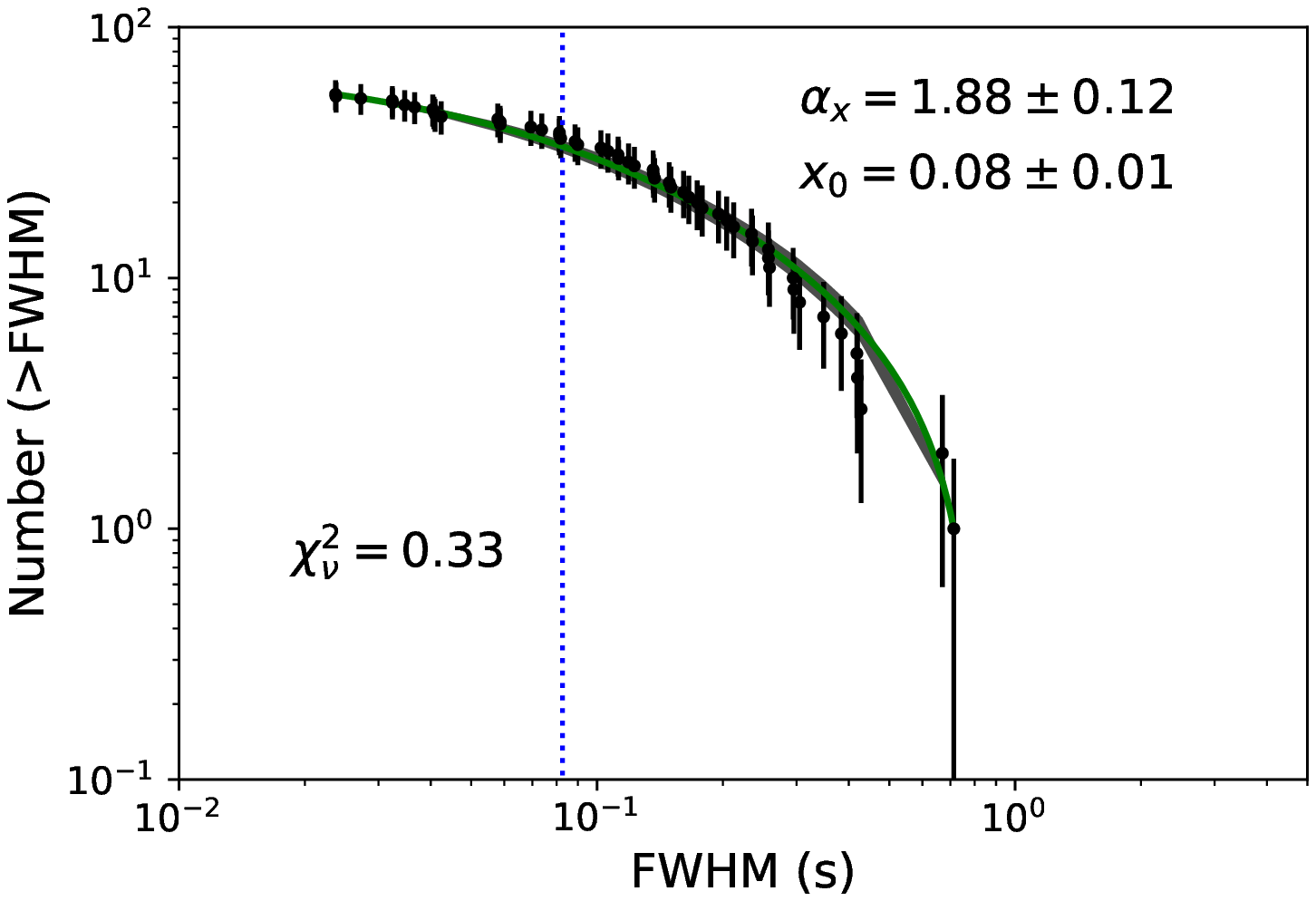}{0.5\textwidth}{(e)}
 \fig{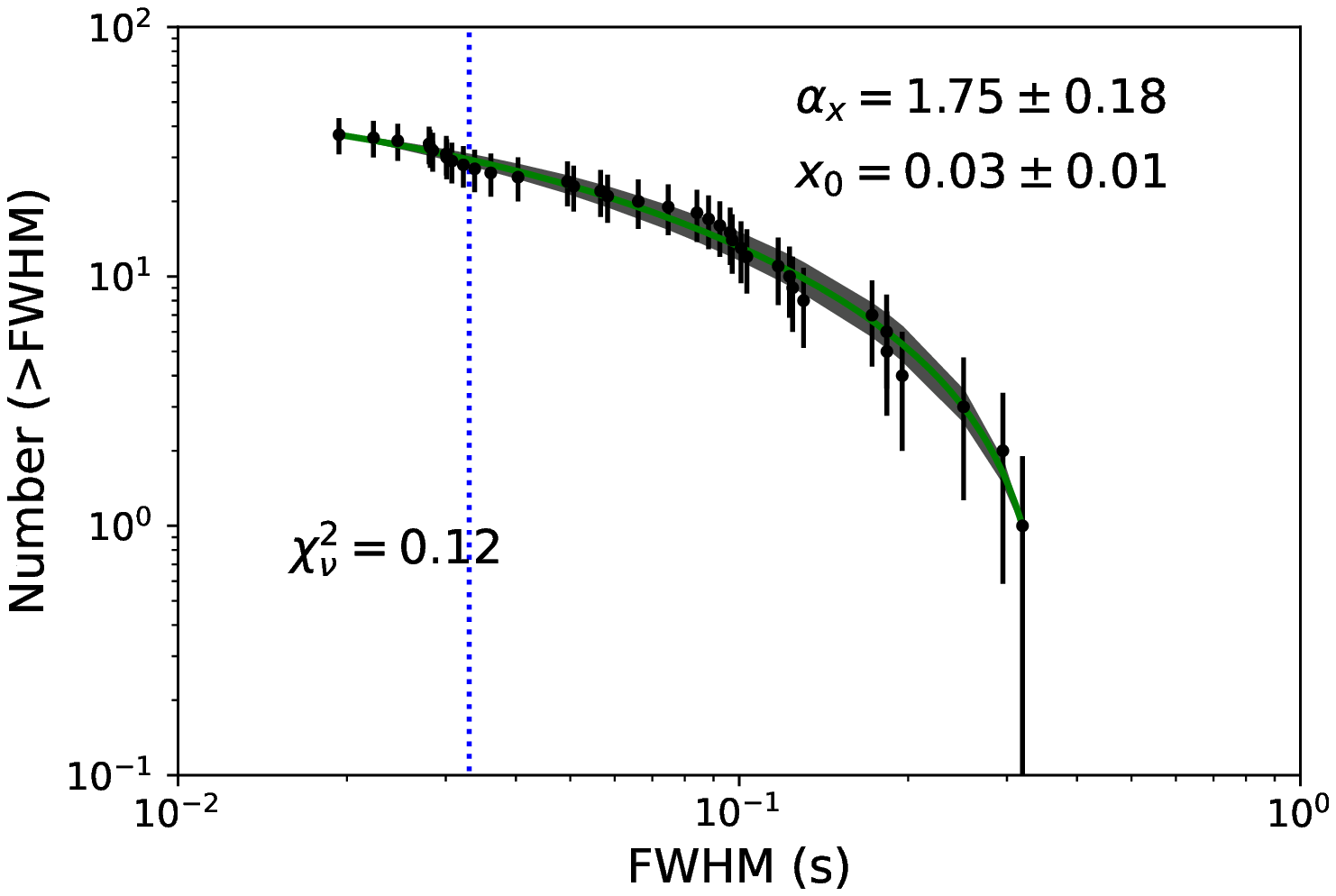}{0.5\textwidth}{(f)}
          }
\caption{The distributions of FWHM for Swift GRBs. The symbols are the same as those in Figure 6. \label{FWHMdff2}}
\end{figure*}

\begin{figure*}
\centering
\gridline{
\fig{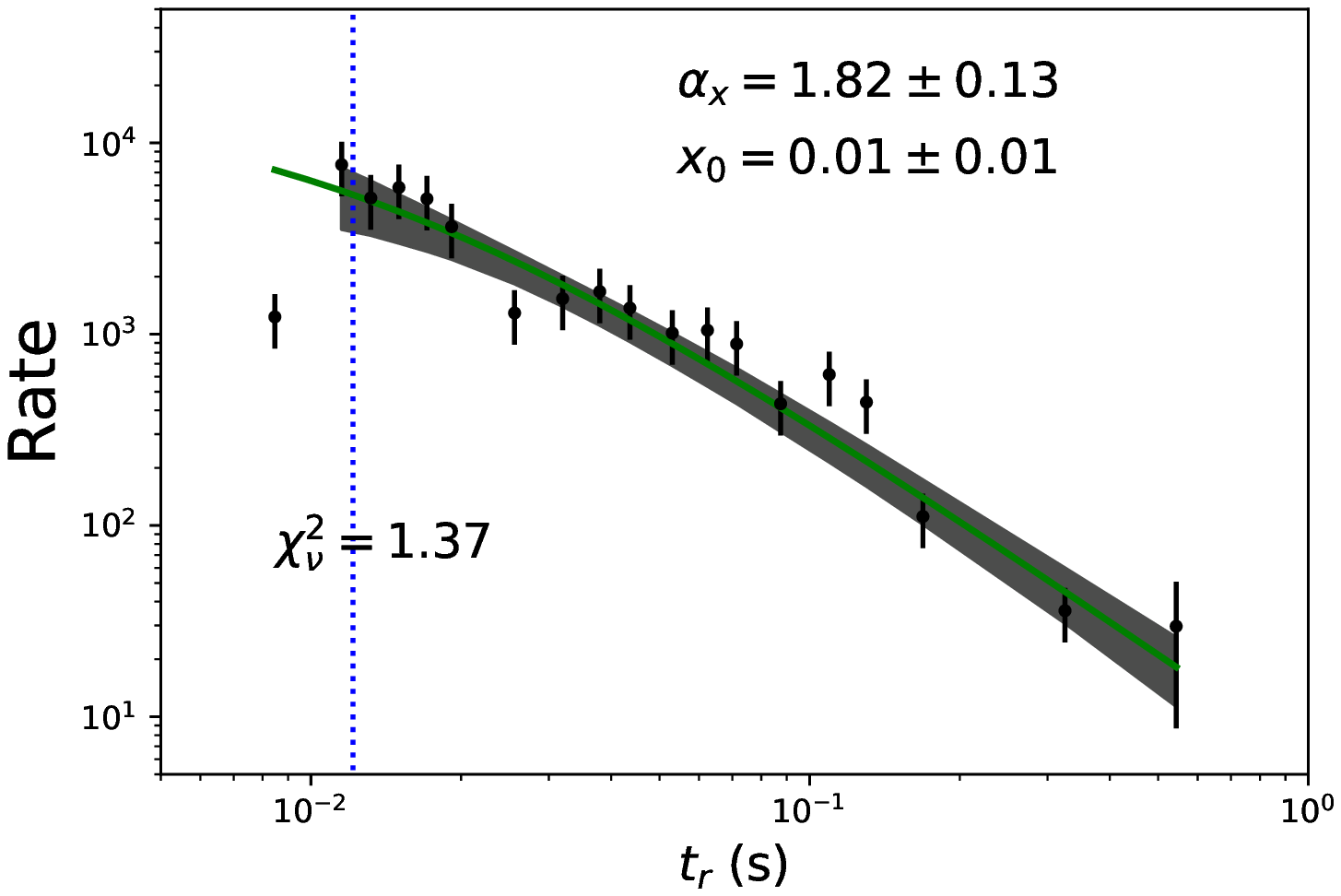}{0.5\textwidth}{(a)}
\fig{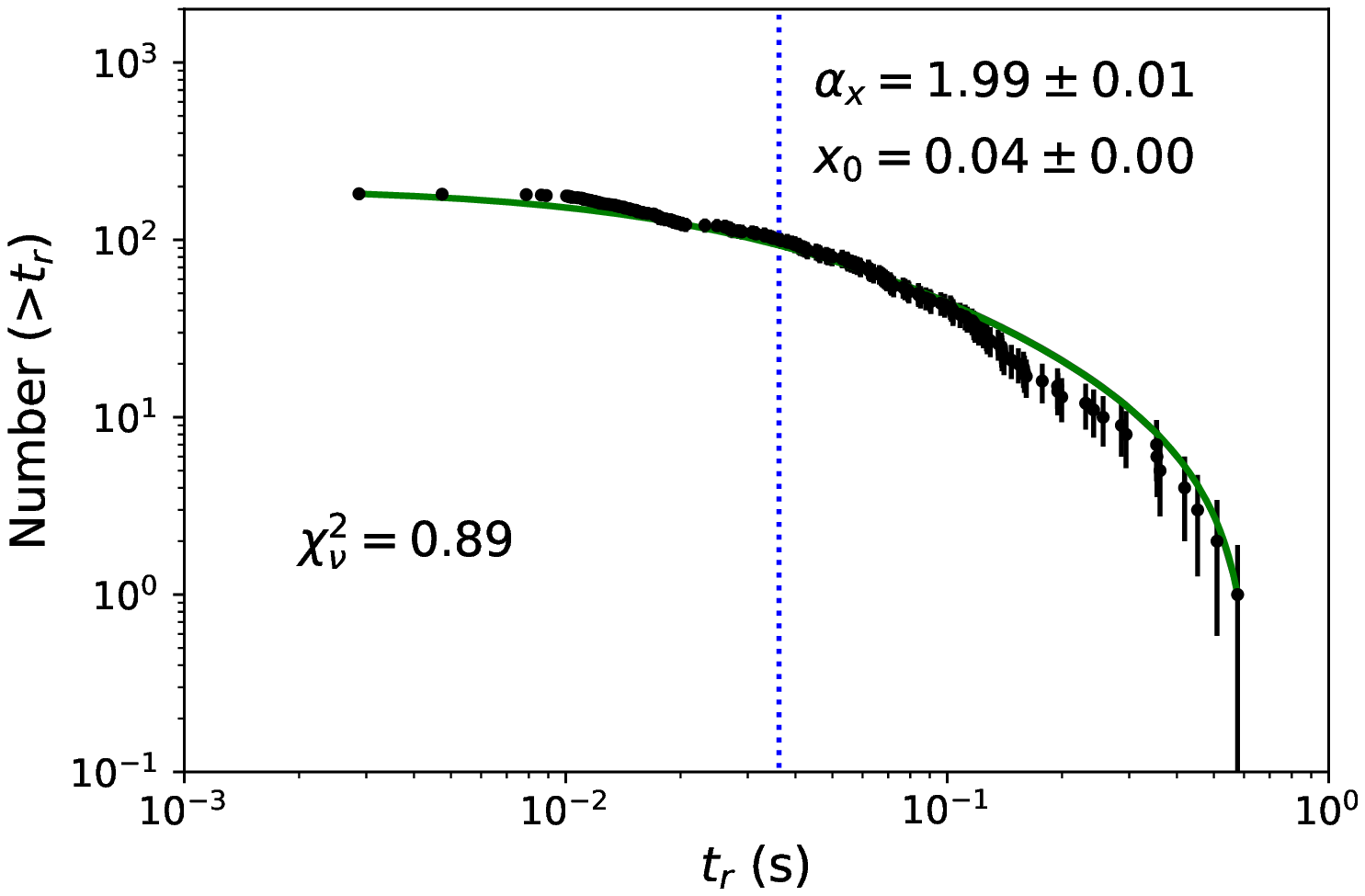}{0.5\linewidth}{(b)}
          }
 \gridline{
 \fig{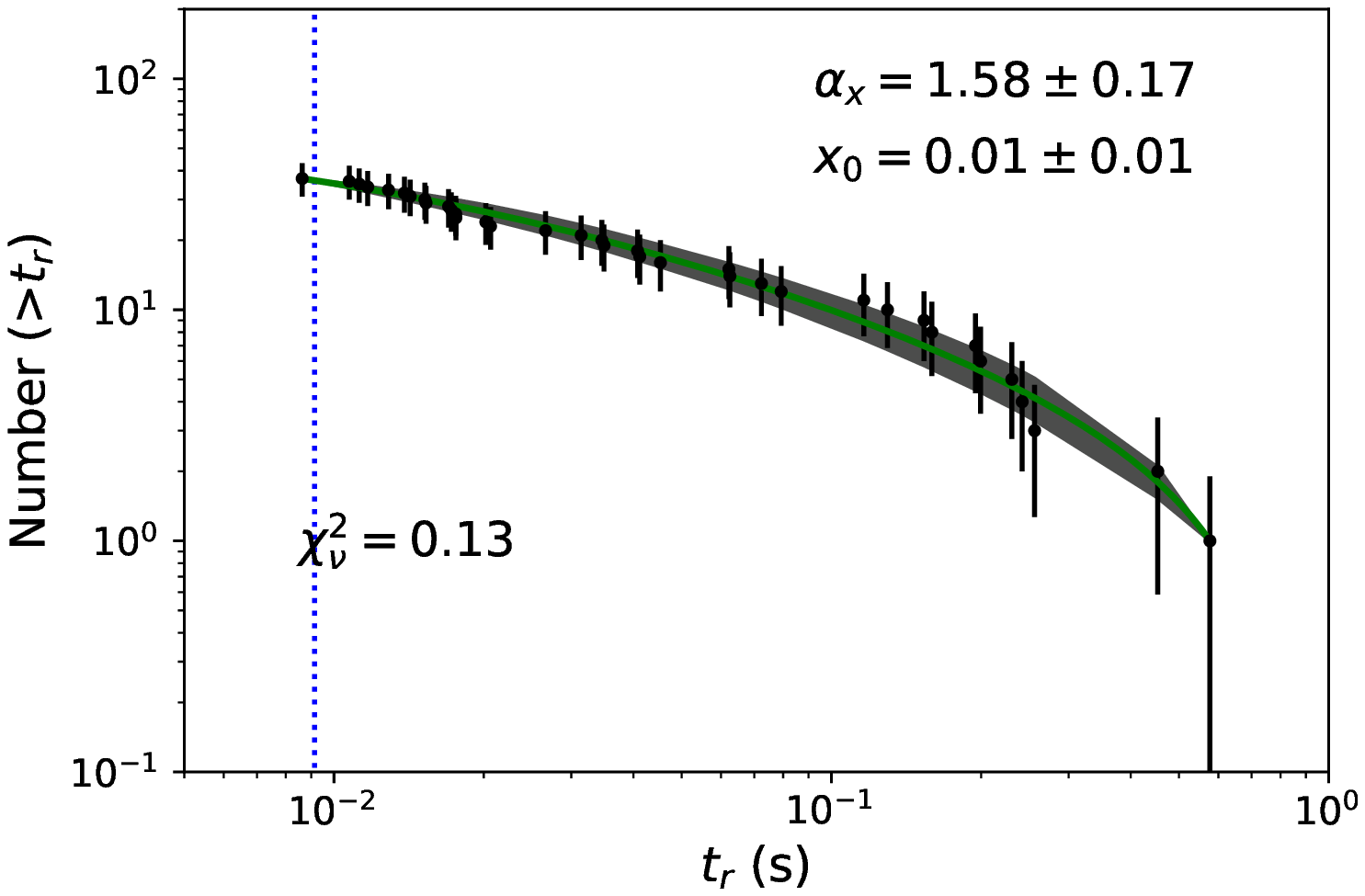}{0.5\textwidth}{(c)}
 \fig{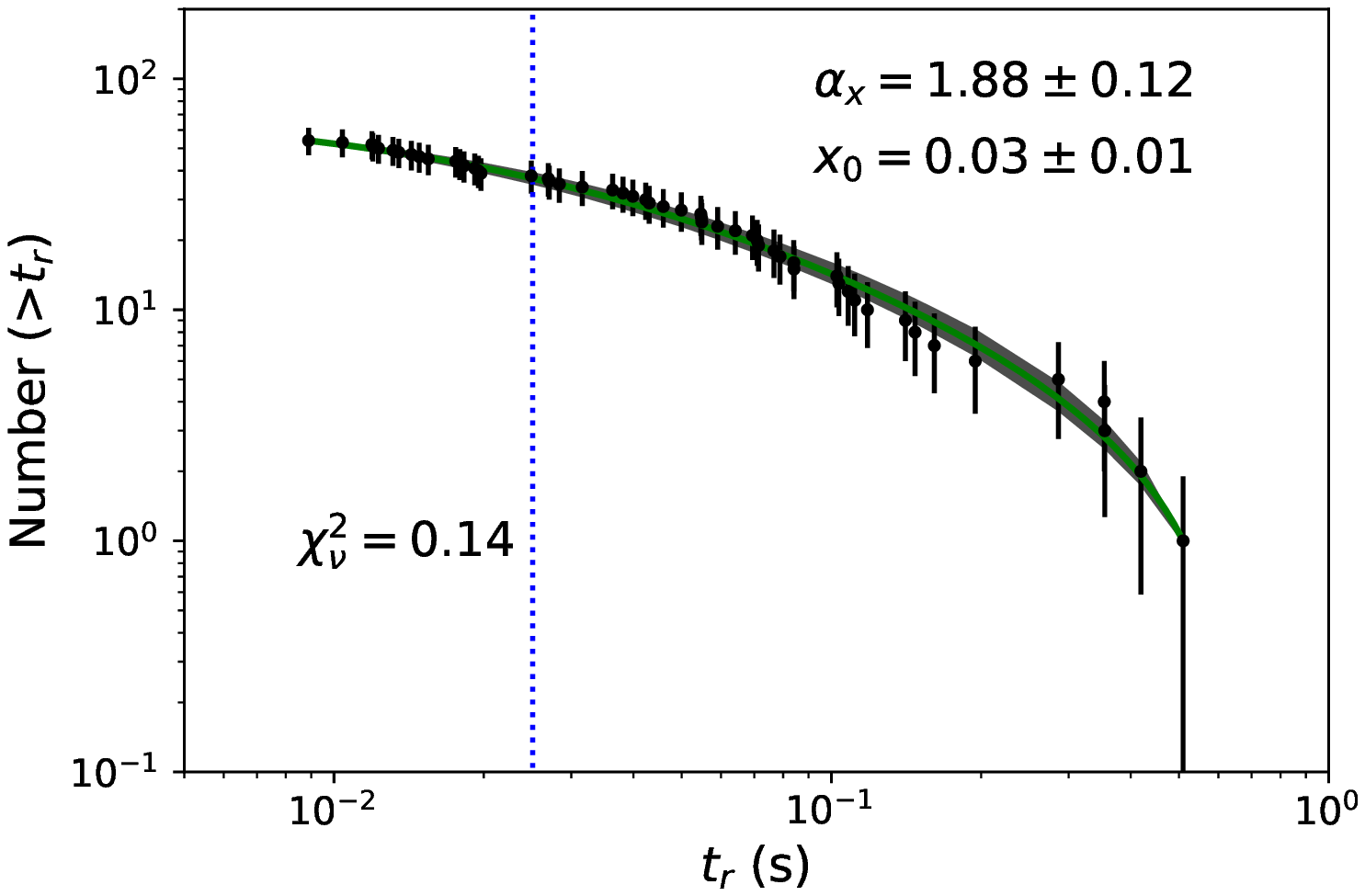}{0.5\textwidth}{(d)}
          }
           \gridline{
 \fig{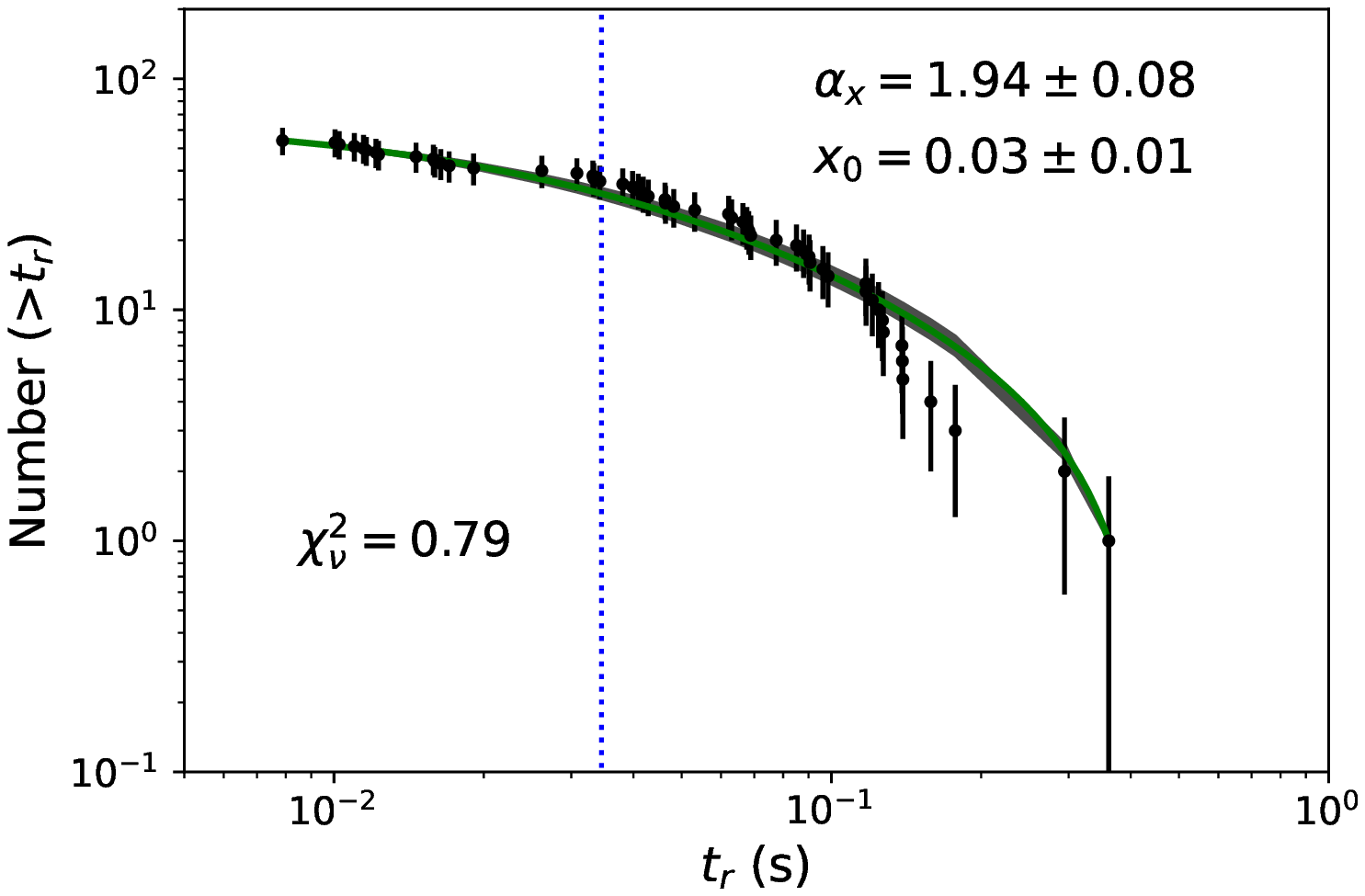}{0.5\textwidth}{(e)}
 \fig{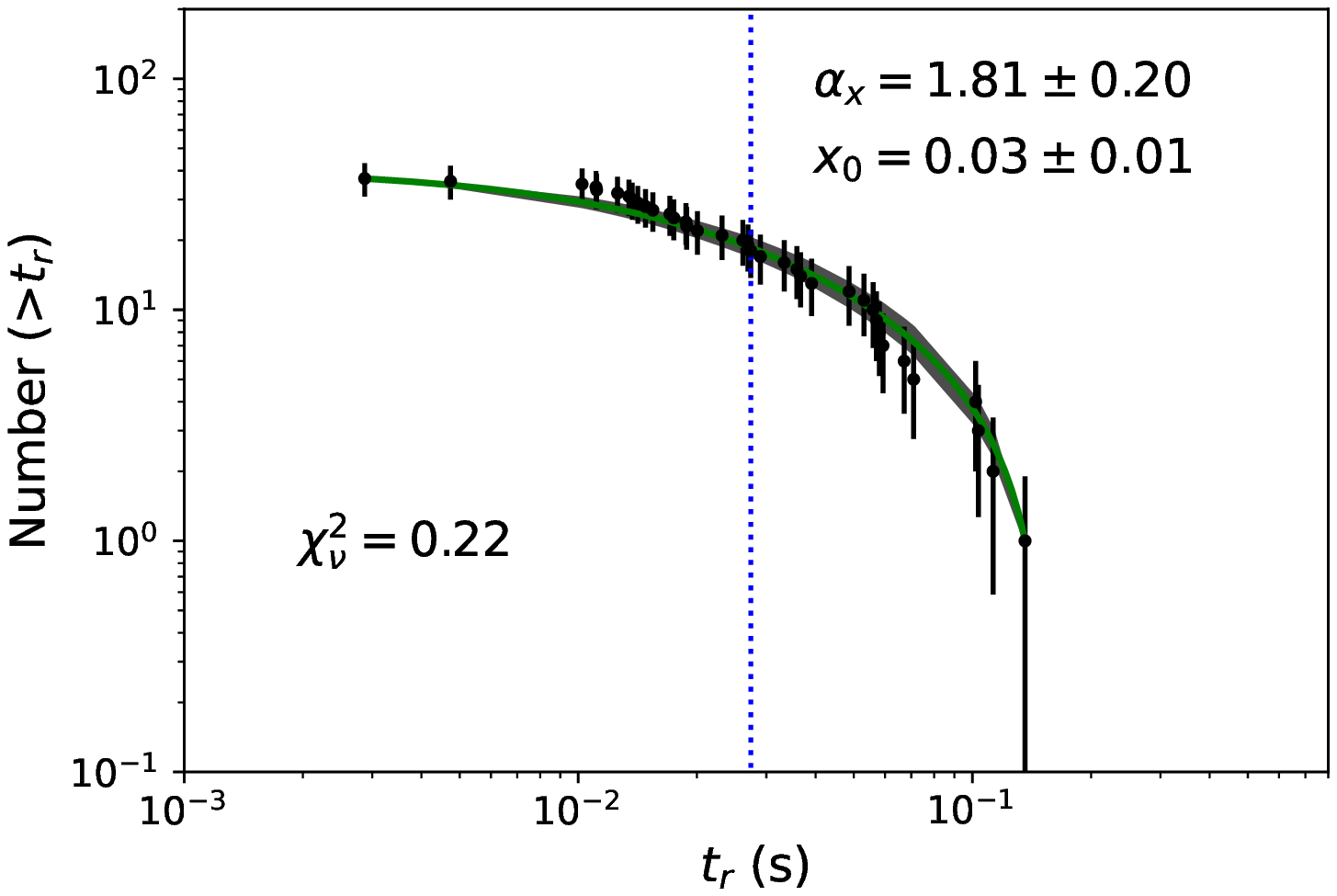}{0.5\textwidth}{(f)}
          }
\caption{The distributions of $t_{\rm r}$ for Swift GRBs. The symbols are the same as those in Figure 6. \label{trdff2}}
\end{figure*}

\begin{figure*}
\centering
\gridline{
\fig{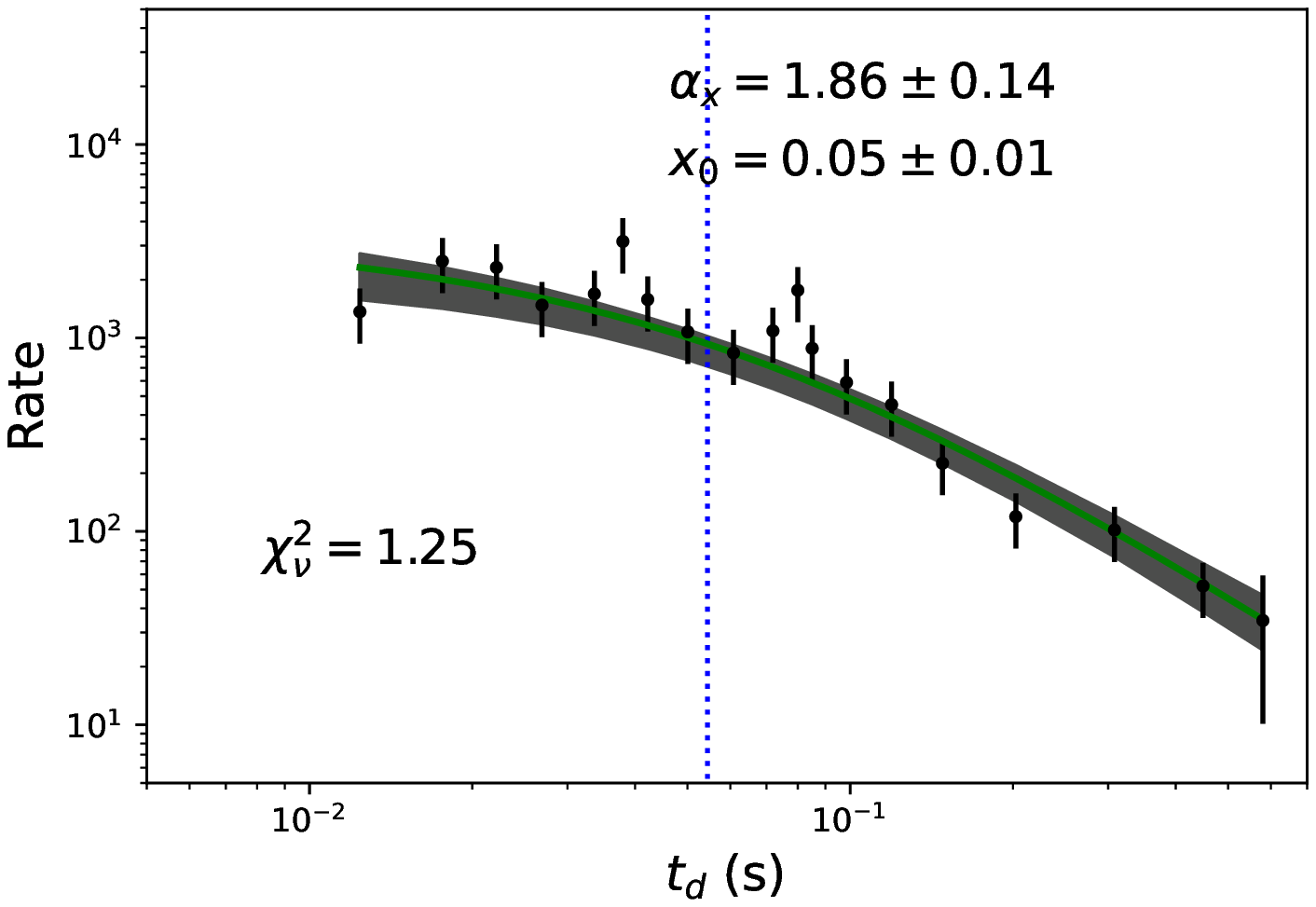}{0.5\textwidth}{(a)}
\fig{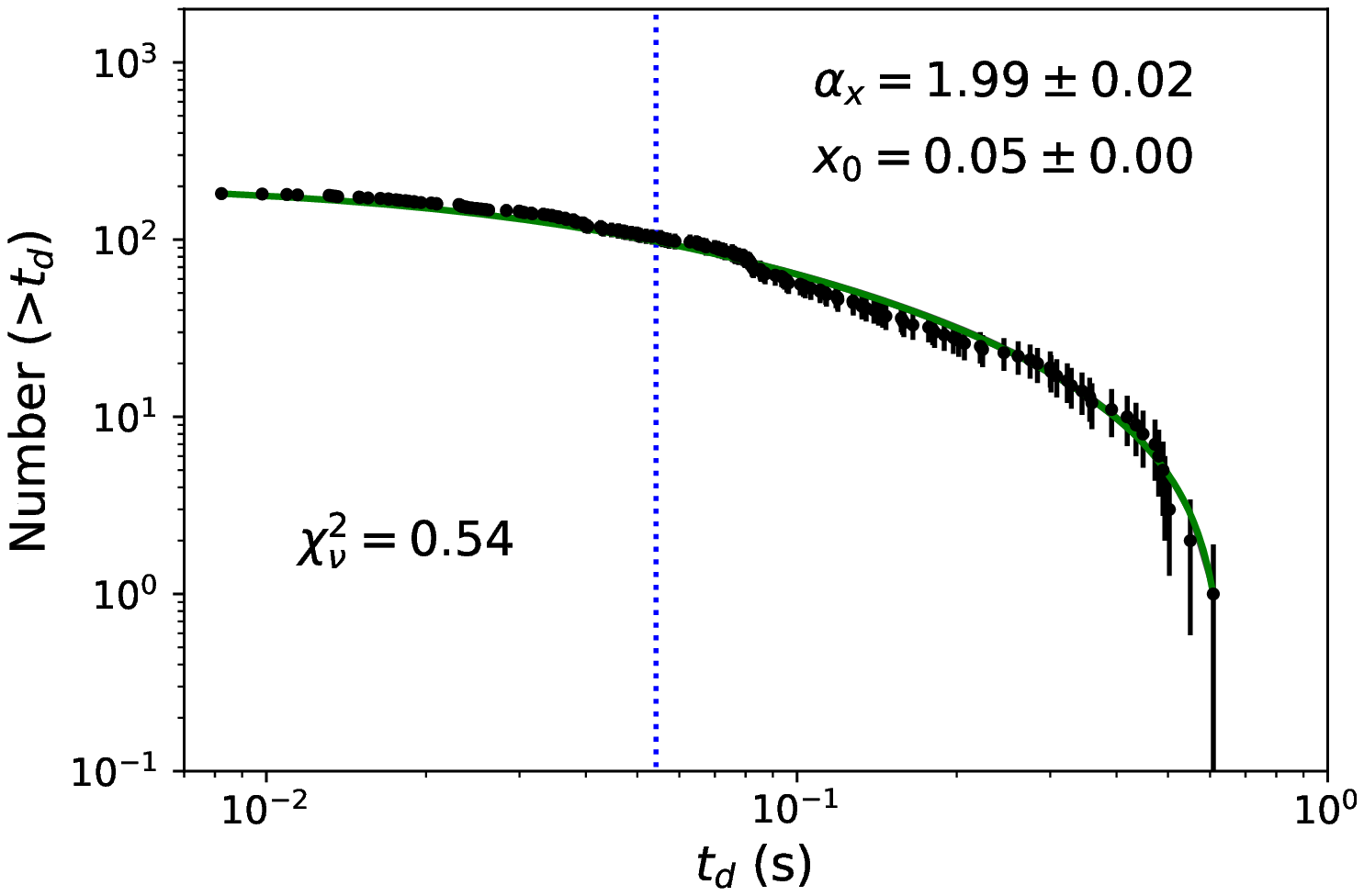}{0.5\linewidth}{(b)}
          }
 \gridline{
 \fig{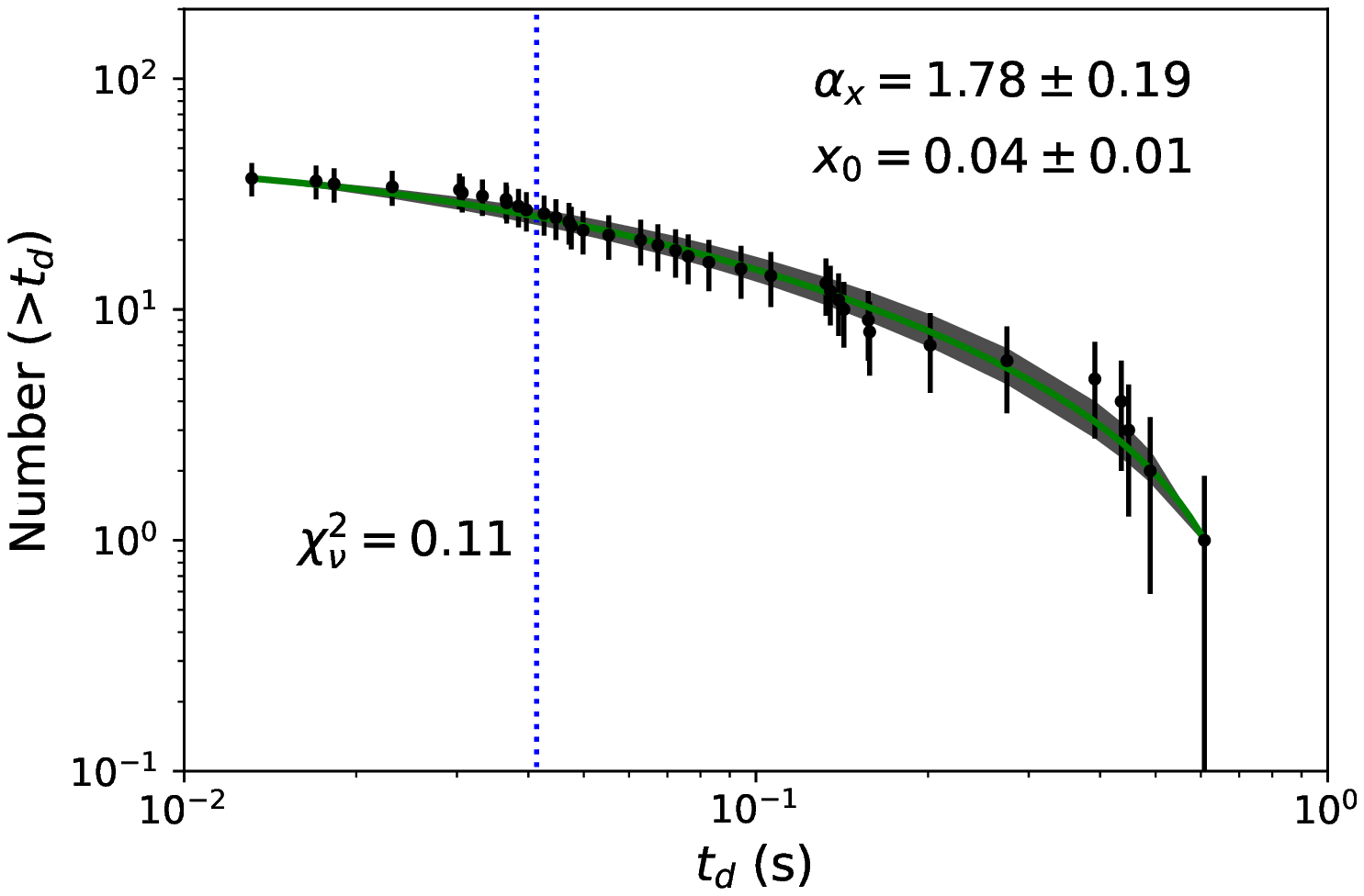}{0.5\textwidth}{(c)}
 \fig{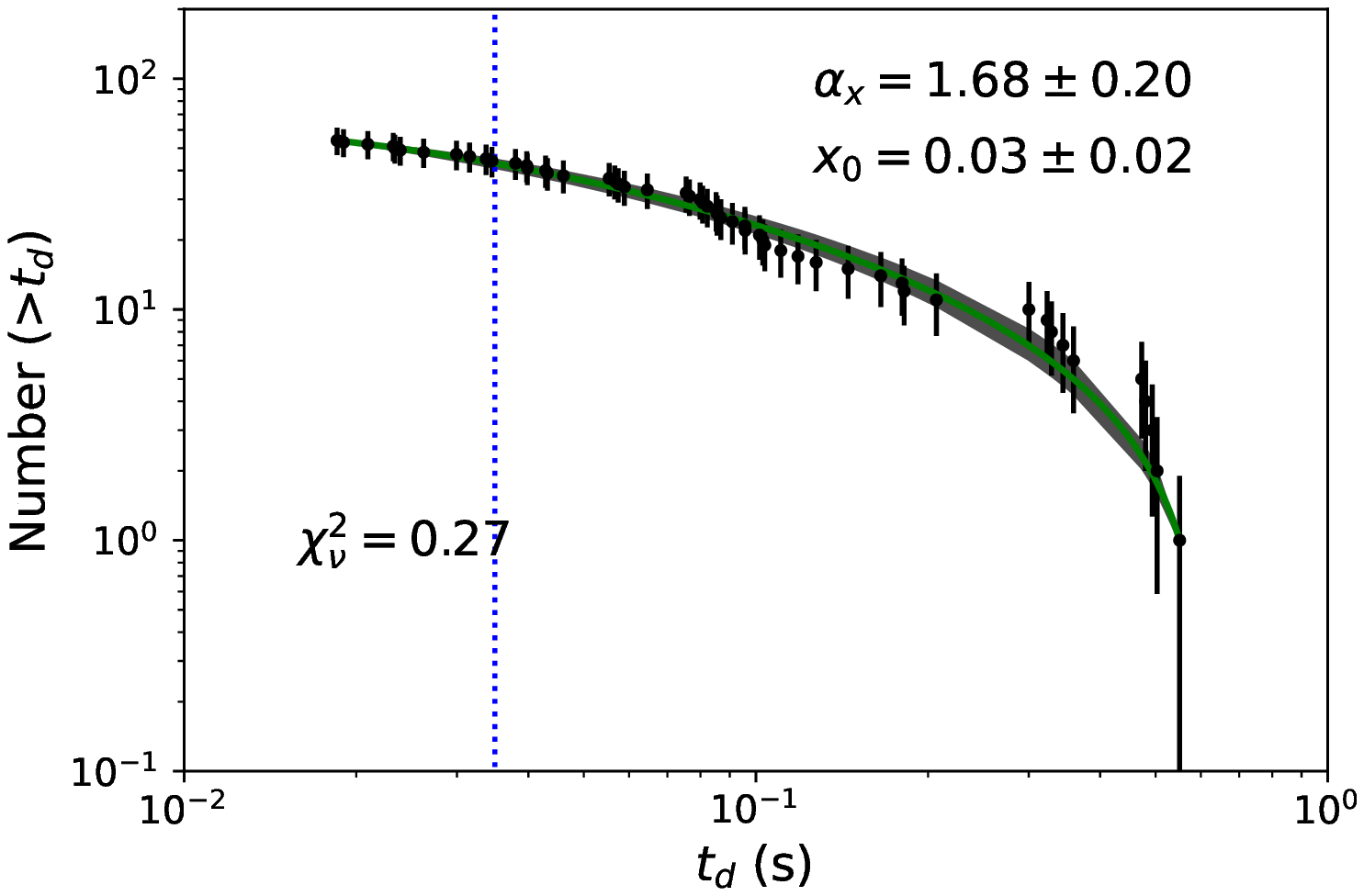}{0.5\textwidth}{(d)}
          }
 \gridline{
 \fig{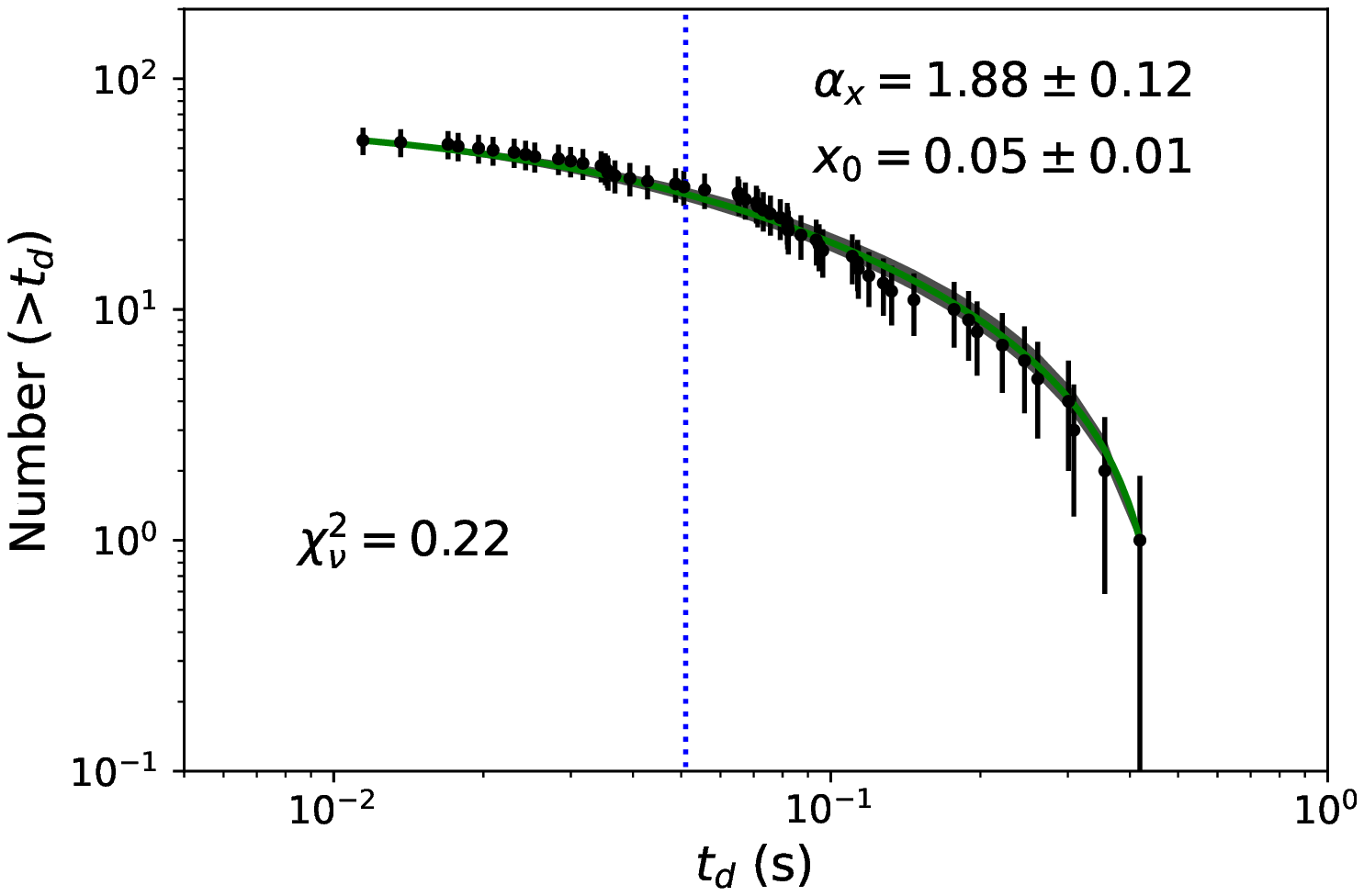}{0.5\textwidth}{(e)}
 \fig{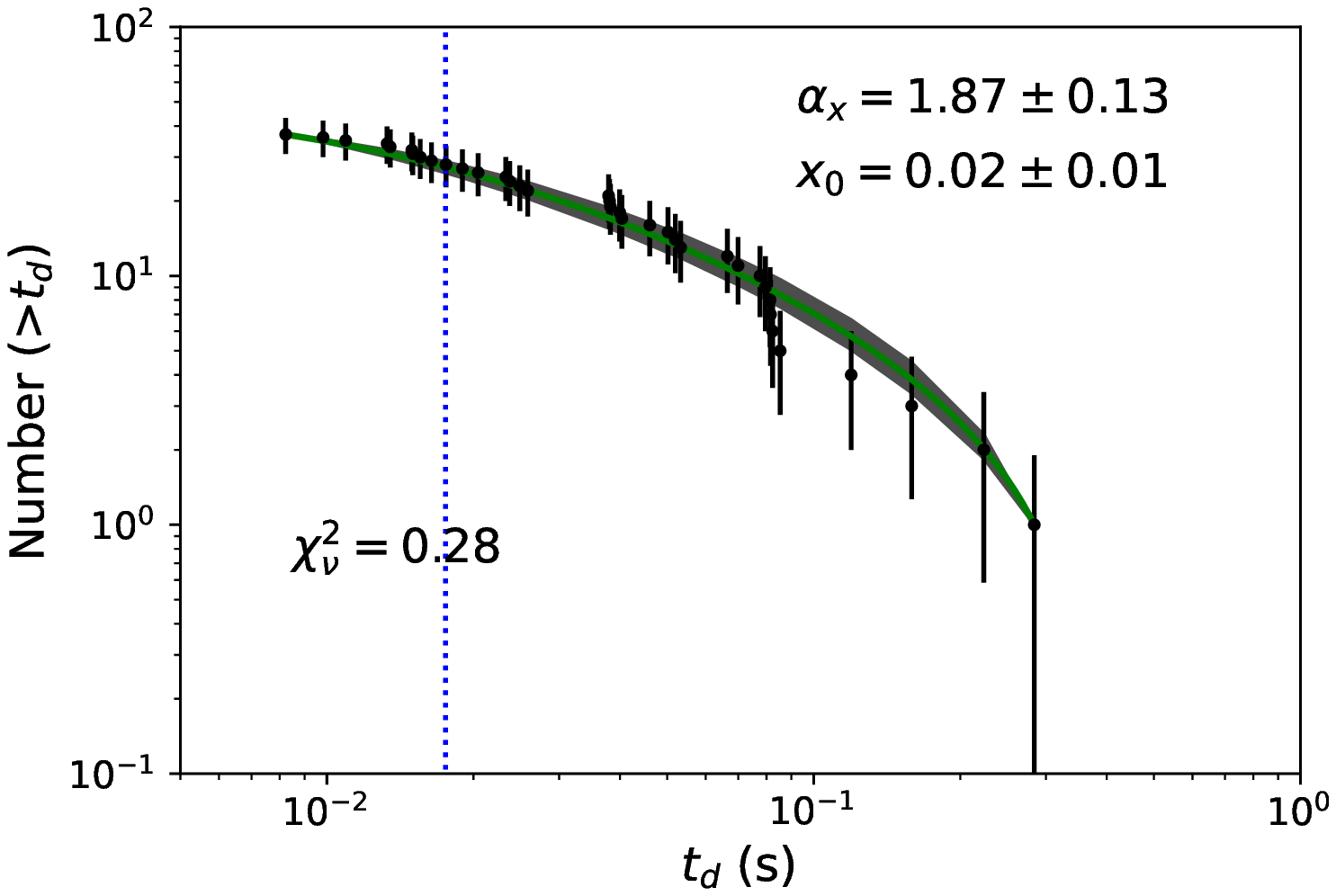}{0.5\textwidth}{(f)}
          }
\caption{The distributions of $t_{\rm d}$ for Swift GRBs. The symbols are the same as those in Figure 6. \label{tddff2}}
\end{figure*}

\begin{figure*}
\centering
\gridline{
\fig{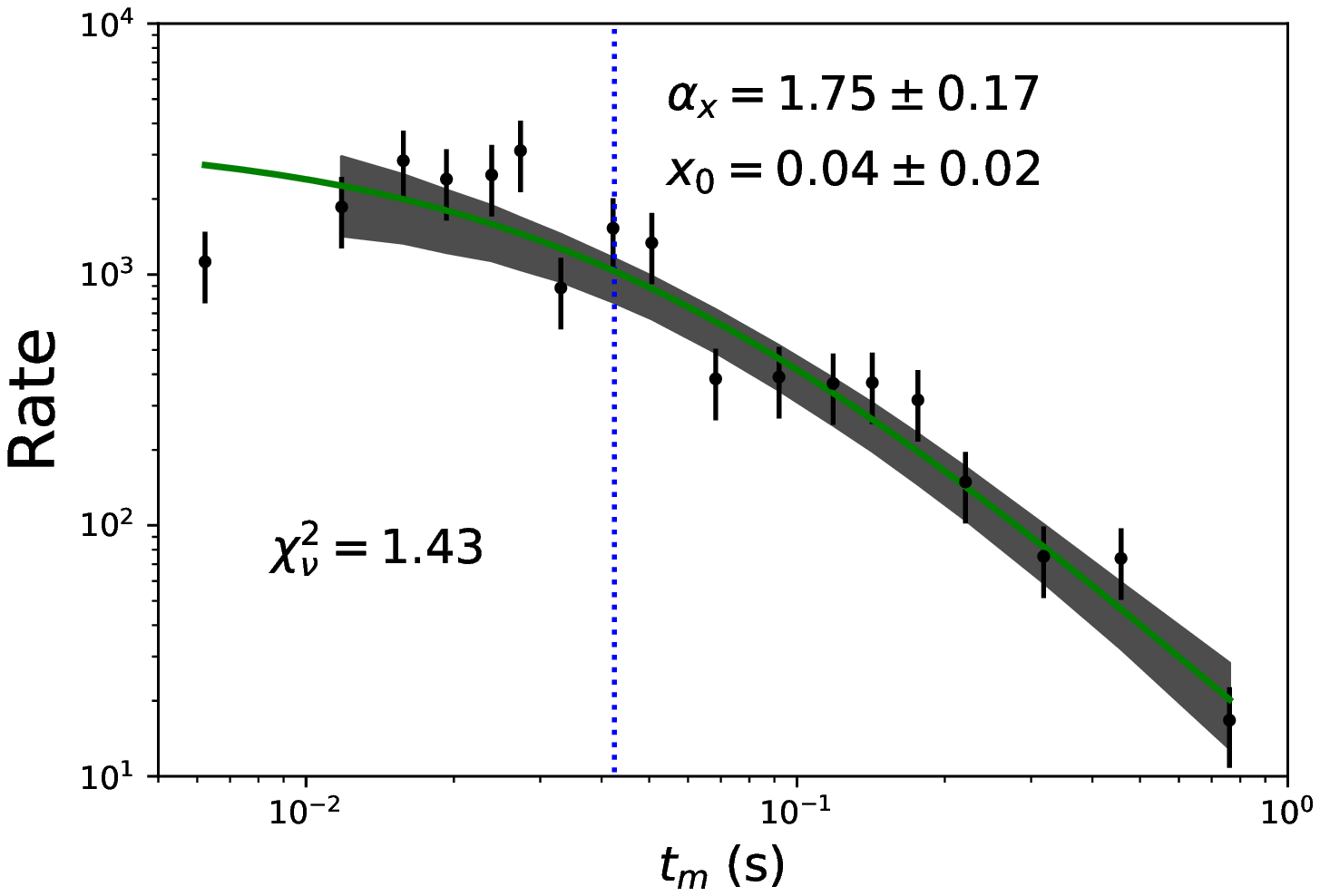}{0.5\textwidth}{(a)}
\fig{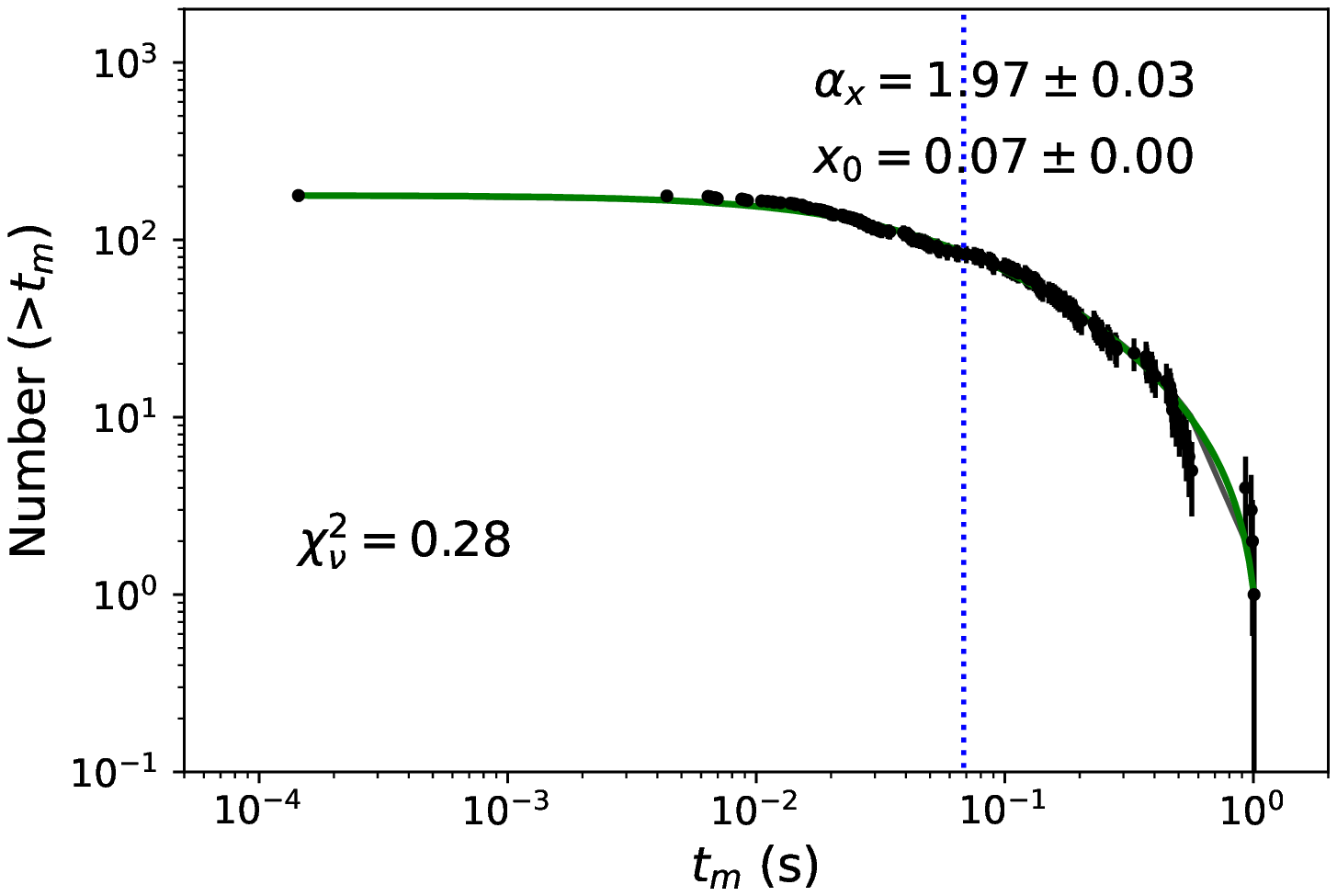}{0.5\linewidth}{(b)}
          }
 \gridline{
 \fig{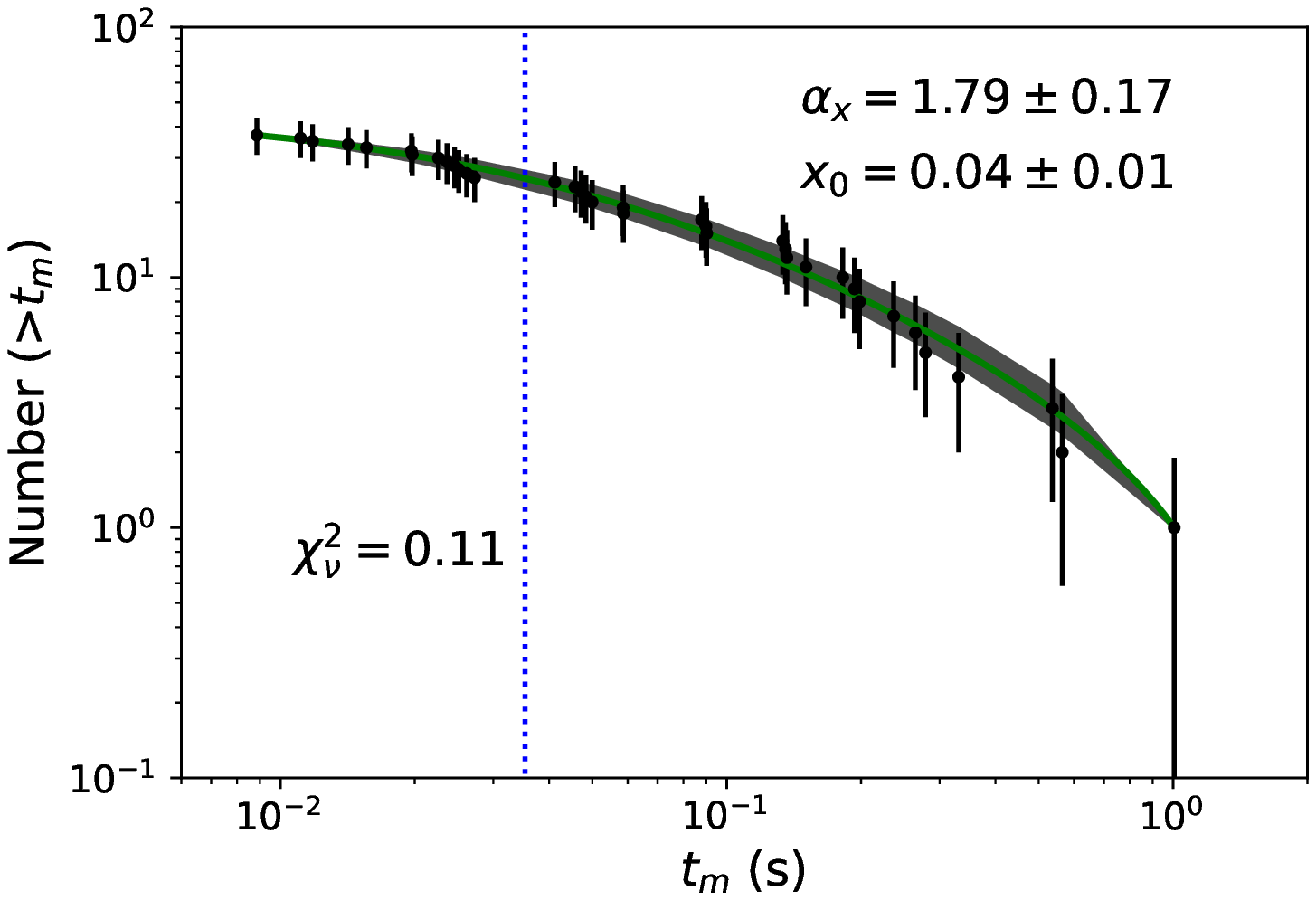}{0.5\textwidth}{(c)}
 \fig{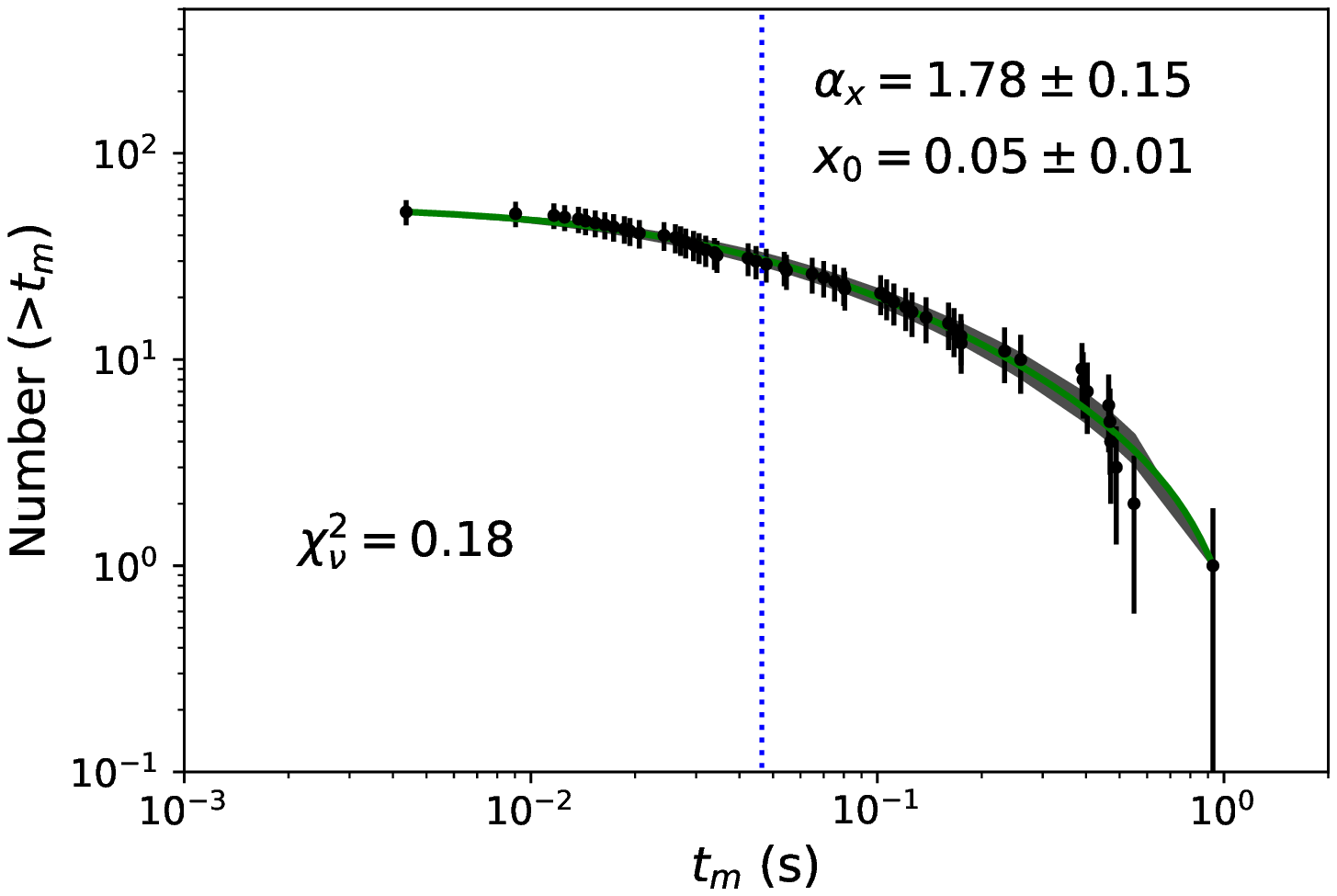}{0.5\textwidth}{(d)}
          }
 \gridline{
 \fig{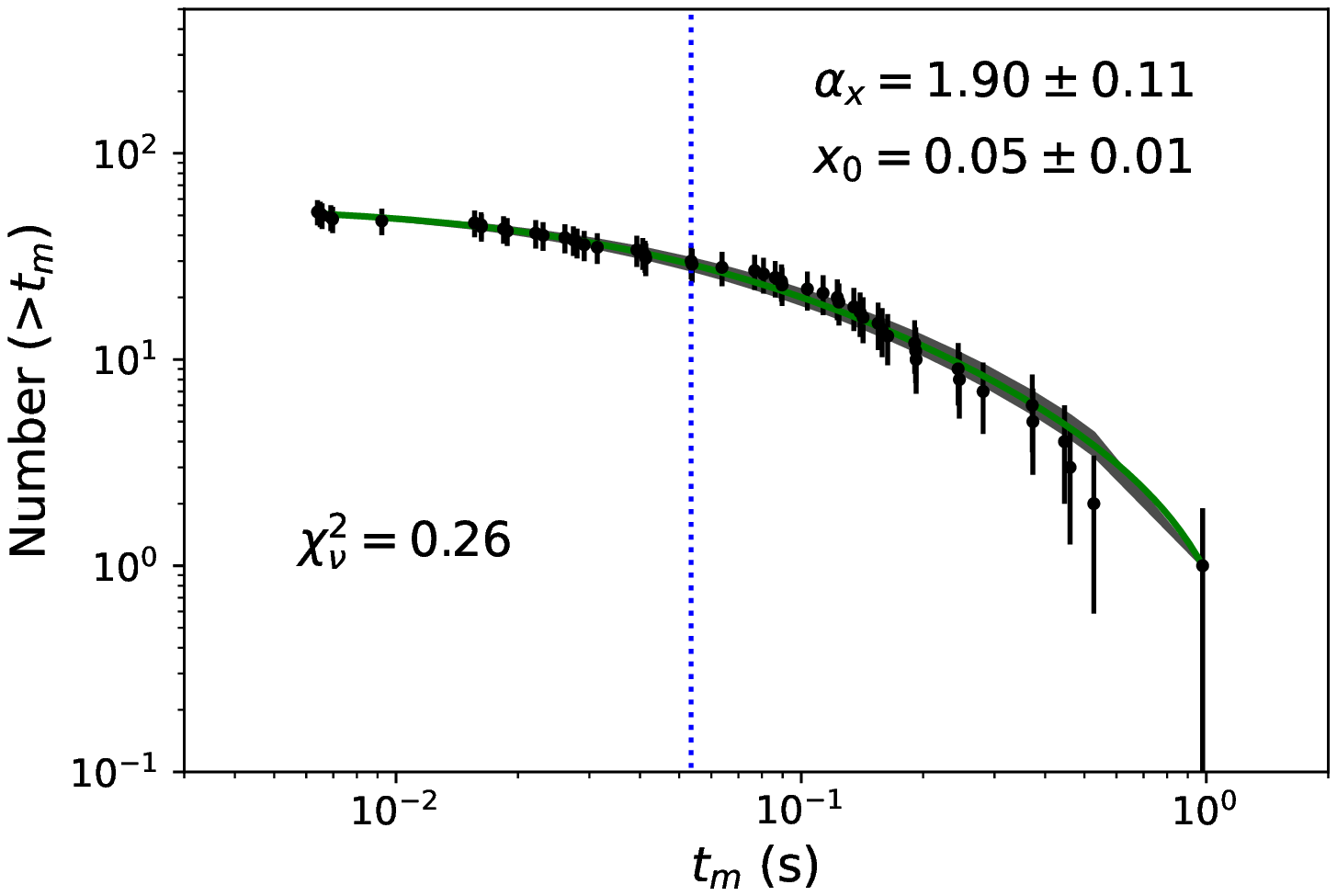}{0.5\textwidth}{(e)}
 \fig{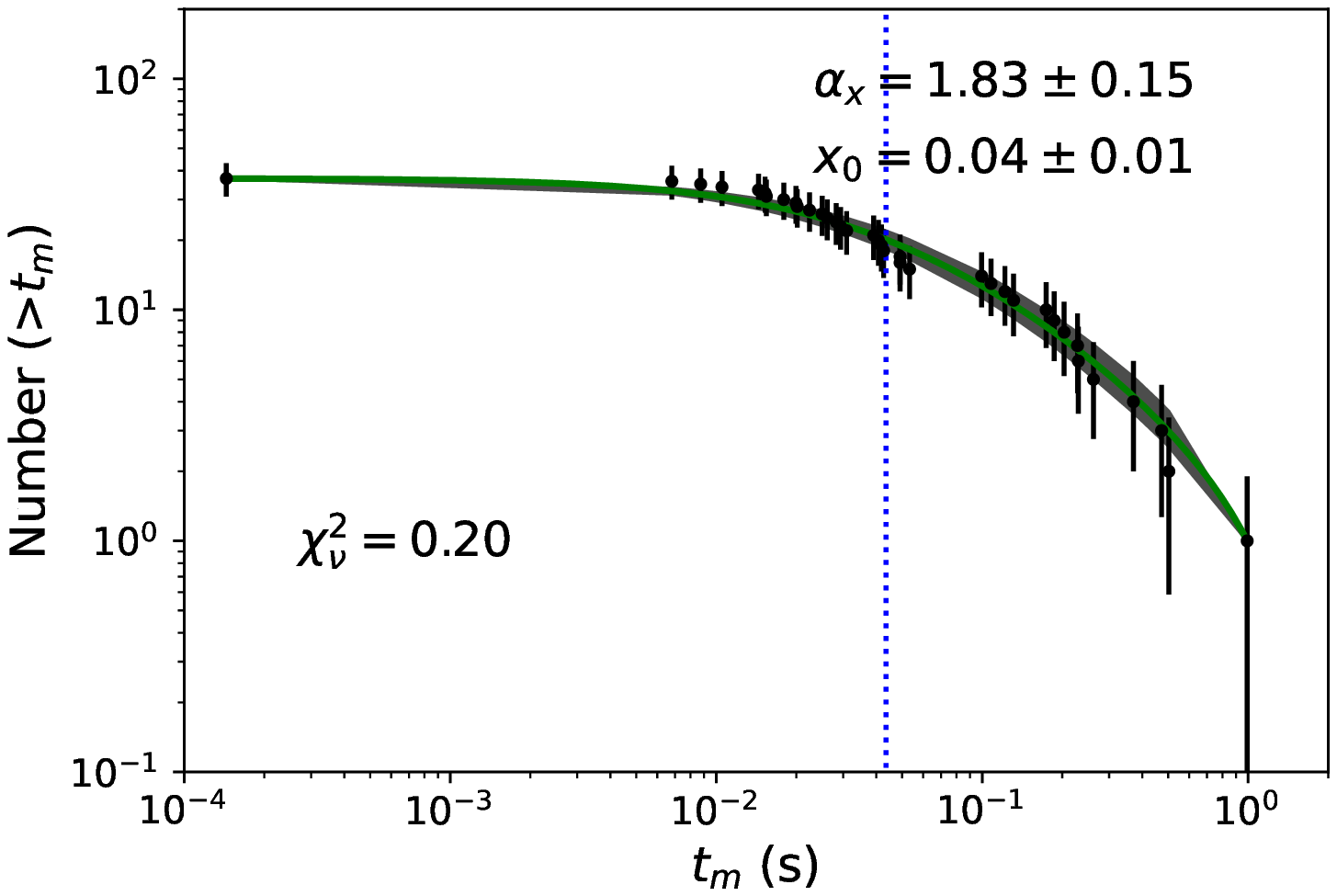}{0.5\textwidth}{(f)}
          }
\caption{The distributions of $t_{\rm m}$ for Swift GRBs. The symbols are the same as those in Figure 6. \label{tmdff2}}
\end{figure*}

The differential and cumulative distributions are fitted with Equations~\ref{eq1} and ~\ref{eq2}. The fitting results are shown in Figures~\ref{fmdff} - ~\ref{tmdff} for BATSE bursts and Figures~\ref{fmdff2} - ~\ref{tmdff2} for Swift bursts, respectively. In each Figure, the differential and cumulative distributions for the total sample are shown in panels (a) and (b). Note that we adopt a rank-order
plot to get the differential distributions. The detailed method refers to Section 7.1.3 in \cite{Aschwanden2011}. In addition, in order to analyse the possible evolution from low to high energy channel, the distributions of different energy channels for both satellite
bursts are fitted. The fitting results are shown in panels (c) - (f). It's important to note that the cumulative
distribution rather than differential distribution is used due to the fact that the sample number of individual energy channel
is not sufficient to bin the data. The detailed fitting results are summarized in Tables~\ref{tab1} and ~\ref{tab2}.

From Figures~\ref{fmdff} -~\ref{tmdff} (a), we find that all the parameters of BATSE GRBs possess the similar power-law differential distributions. In Figure~\ref{fmdff} (a), the best-fitting power-law index of peak flux for the total BATSE sample is 2.48 $\pm$ 0.02. The result is larger than the value of 2.09 $\pm$ 0.18 measured by \cite{Lv2020} for a sample of BATSE bursts whose lightcurves have more than two pulses. However, it can be found from Figure~\ref{FWHMdff} (a) that the best-fitting power-law index of FWHM for the total BATSE sample is 1.89 $\pm$ 0.12, which is
quite close to the value of 1.82 $^{+0.14}_{-0.15}$ given by \cite{Lv2020}.
In addition, the similar power-law differential distributions of rise time, decay time, and peak time with indexes of 1.89 $\pm$ 0.12, 1.92 $\pm$ 0.10, and 2.46 $\pm$ 0.05 can be found in Figures~\ref{trdff} -~\ref{tmdff} (a), respectively.

For cumulative distributions, the best-fitting power-law index of peak flux for the total BATSE sample is 3.05 $\pm$ 0.02 in Figure~\ref{fmdff} (b), which is larger than the value of 1.99 $^{+0.16}_{-0.19}$ reported by \cite{Lv2020}. Particularly, it can be seen from Figure~\ref{FWHMdff} (b) that the best-fitting power-law index of FWHM for the total BATSE sample is 2.00 $\pm$ 0.01, which is also larger than the value of 1.75 $^{+0.11}_{-0.13}$ given by \cite{Lv2020}. The best-fitting power-law indexes of rise time, decay time, and peak time are 2.10 $\pm$ 0.01, 2.20 $\pm$ 0.01, and 2.50 $\pm$ 0.01 for the total BATSE sample (see Figures~\ref{trdff} - \ref{tmdff} (b) ).

Figure~\ref{fmdff} (c) - (f) show the cumulative distributions of peak flux from BATSE Ch1 to Ch4. Similar results can be seen from Figures~\ref{FWHMdff} - \ref{tmdff} (c) - (f).
The mean values of these power-law indexes of peak flux, FWHM, rise time, decay time, and peak time of four individual channels are 2.29 $\pm$ 0.13, 1.87 $\pm$ 0.07, 1.85 $\pm$ 0.08, 1.92 $\pm$ 0.05, and 2.41 $\pm$ 0.05, respectively. We find that there is almost no significant power-law index evolution across different energy bands, indicating that the self-organized criticality may quite be likely to exist in GRB systems or, in other words, be intrinsic.

Significantly, similar results are obtained for Swift GRBs in Figures~\ref{fmdff2} - ~\ref{tmdff2}. Overall, the results for Swift GRBs are approximately consistent with those of BATSE GRBs. In Figures~\ref{fmdff2} - ~\ref{FWHMdff2} (a), the best-fitting power-law indexes of differential distributions of peak flux and FWHM for the whole Swift sample are 2.44 $\pm$ 0.07 and 1.87 $\pm$ 0.13, both similar to the values of BATSE GRBs. Similarly, we can see from Figures~\ref{trdff2} -~\ref{tmdff2} (a) that the power-law differential distributions of rise time, decay time, and peak time for the total Swift sample with the indexes of 1.82 $\pm$ 0.13, 1.86 $\pm$ 0.14, and 1.75 $\pm$ 0.17, respectively, which are slightly smaller than those of BATSE GRBs.

Meanwhile, we find that the cumulative distributions of the total Swift sample can also be well described by power-law model with the indexes of 2.50 $\pm$ 0.01, 1.99 $\pm$ 0.01, 1.99 $\pm$ 0.01, 1.99 $\pm$ 0.02, and 1.97 $\pm$ 0.03 for peak flux, FWHM, rise time, decay time, and peak time, respectively, as shown in Figures~\ref{trdff2} -~\ref{tmdff2} (b).
The mean values of these indexes of four individual channels are 2.40 $\pm$ 0.05, 1.81 $\pm$ 0.08, 1.80 $\pm$ 0.07, 1.80 $\pm$ 0.08, and 1.83 $\pm$ 0.07, respectively. Above all, it can be found that there is also no significant power-law index evolution across different energy bands for Swift GRBs. On the other hand, the fact that there is no significant instrumental effect between BATSE and Swift bursts, strengthens the evidences of the self-organized criticality in GRB systems.

\subsection{SOC} \label{sec:results}

For a fractal-diffusive SOC model, \cite{Aschwanden2012,Aschwanden2014} predicted power-law distributions for duration and peak flux with the indexes $\alpha_T = (1+S)\beta/2$ and $\alpha_P = 2-1/S $, where S = 1, 2, 3 is Euclidean space dimensions of SOC system and $\beta $ is the diffusive spreading exponent. According to the prediction, it is easy to obtain the indexes $\alpha_{P}= 1.67$ and $\alpha_{T}= 2 $ for S = 3 and the classical diffusion with $\beta = 1$.

Our results can be used to derive the possible Euclidean space dimension of GRB SOC system. Owing to the small differences between the best-fitting power-law indexes of
differential and cumulative distributions, we choose the cumulative case to examine GRB system. Furthermore, considering invariance of the power-law index on energy,
we adopt the mean value of power-law indexes of four individual channels to check. For example, the mean value of power-law index of FWHM for BATSE GRB is 1.87 $\pm$ 0.07 and can determine the Euclidean space dimension S = 3 according to the prediction of the FD-SOC model with the theoretical index $\alpha_{T}$ = 2. The result is very consistent with those of BATSE GRBs with at least three pulses reported by \cite{Lv2020}. Similarly, the Euclidean space dimension can be obtained using the results of rise time, decay time, and peak time. Although the results are slightly smaller than the theoretical index $\alpha_{T}$ = 2, Euclidean space dimension of GRB SOC system is essentially in agreement with the model prediction for S = 3.
It is worth pointing out that it is difficult to determine which dimension for peak flux  since the value is obviously larger than the theoretical index 1.67 for S = 3. This phenomenon is similar to that reported by \cite{Lv2020}. The steeper index of the peak flux distributions can be explained due to the fainter peak fluxes in small bin size (5ms and 8ms).

\section{Discussions} \label{sec:result3}

\begin{longtable*}{l c c c c c c c}
\caption{The best-fitting parameters with the power-law models for BATSE GRBs.}
 \label{tab1}\\
\hline
Parameters & Energy Band & Satellite & $x_{\rm 0d}$ & $x_{\rm 0c}$ & $\alpha_{\rm d}$ &$\alpha_{\rm c}$& $\chi{_{\rm \nu}^2}$\\
\hline
peak flux & total sample&BATSE&7.34$\pm$0.40&$--$&2.48$\pm$0.02&$--$&2.67\\
peak flux & total sample&BATSE&$--$&14.94$\pm$0.08&$--$&3.05$\pm$0.02&1.28\\
peak flux & Ch1         &BATSE&$--$&6.81$\pm$2.70&$--$&2.17$\pm$0.31&0.22\\
peak flux & Ch2&BATSE&$--$&7.18$\pm$0.38&$--$&2.48$\pm$0.03&0.85\\
peak flux & Ch3&BATSE&$--$&10.88$\pm$0.53&$--$&2.48$\pm$0.02&1.03\\
peak flux & Ch4&BATSE&$--$&6.32$\pm$2.65&$--$&2.02$\pm$0.40&0.11\\
\hline
FWHM  & total sample&BATSE&0.10$\pm$0.03&$--$&1.89$\pm$0.12&$--$&1.36\\
FWHM  & total sample&BATSE&$--$&0.20$\pm$0.00&$--$&2.00$\pm$0.01&2.93\\
FWHM  & Ch1&BATSE&$--$&0.11$\pm$0.05&$--$&1.77$\pm$0.21&0.21\\
FWHM  & Ch2&BATSE&$--$&0.20$\pm$0.01&$--$&1.98$\pm$0.03&1.45\\
FWHM  & Ch3&BATSE&$--$&0.19$\pm$0.01&$--$&1.97$\pm$0.04&1.07\\
FWHM  & Ch4&BATSE&$--$&0.07$\pm$0.03&$--$&1.75$\pm$0.20&0.11\\
\hline
$t_{\rm r}$ & total sample&BATSE&0.04$\pm$0.01&$--$&1.89$\pm$0.12&$--$&2.83\\
$t_{\rm r}$ & total sample&BATSE&$--$&0.08$\pm$0.00&$--$&2.10$\pm$0.01&2.86\\
$t_{\rm r}$ & Ch1&BATSE&$--$&0.08$\pm$0.04&$--$&1.65$\pm$0.29&0.35\\
$t_{\rm r}$ & Ch2&BATSE&$--$&0.08$\pm$0.00&$--$&1.97$\pm$0.04&1.08\\
$t_{\rm r}$ & Ch3&BATSE&$--$&0.08$\pm$0.00&$--$&1.97$\pm$0.04&0.98\\
$t_{\rm r}$ & Ch4&BATSE&$--$&0.02$\pm$0.01&$--$&1.82$\pm$0.16&0.15\\
\hline
$t_{\rm d}$ & total sample&BATSE&0.05$\pm$0.01&$--$&1.92$\pm$0.10&$--$&2.31\\
$t_{\rm d}$ & total sample&BATSE&$--$&0.13$\pm$0.00&$--$&2.20$\pm$0.01&3.48\\
$t_{\rm d}$ & Ch1&BATSE&$--$&0.12$\pm$0.03&$--$&1.90$\pm$0.16&0.48\\
$t_{\rm d}$ & Ch2&BATSE&$--$&0.11$\pm$0.01&$--$&1.98$\pm$0.03&1.62\\
$t_{\rm d}$ & Ch3&BATSE&$--$&0.09$\pm$0.00&$--$&1.98$\pm$0.03&1.15\\
$t_{\rm d}$ & Ch4&BATSE&$--$&0.05$\pm$0.02&$--$&1.80$\pm$0.14&0.12\\
\hline
$t_{\rm m}$ & total sample&BATSE&0.21$\pm$0.02&$--$&2.46$\pm$0.05&$--$&1.78\\
$t_{\rm m}$ & total sample&BATSE&$--$&0.28$\pm$0.01&$--$&2.50$\pm$0.01&3.21\\
$t_{\rm m}$ & Ch1&BATSE&$--$&0.18$\pm$0.03&$--$&2.41$\pm$0.09&0.51\\
$t_{\rm m}$ & Ch2&BATSE&$--$&0.23$\pm$0.01&$--$&2.47$\pm$0.04&1.55\\
$t_{\rm m}$ & Ch3&BATSE&$--$&0.28$\pm$0.02&$--$&2.46$\pm$0.06&0.93\\
$t_{\rm m}$ & Ch4&BATSE&$--$&0.19$\pm$0.05&$--$&2.30$\pm$0.17&1.07\\
\hline
\hline
\end{longtable*}
\begin{longtable*}{l c c c c c c c}
\caption{The best-fitting parameters with the power-law models for Swift GRBs.}
 \label{tab2}\\
\hline
Parameters & Energy Band & Satellite & $x_{\rm 0d}$ & $x_{\rm 0c}$ & $\alpha_{\rm d}$ &$\alpha_{\rm c}$&$\chi{_{\rm \nu}^2}$\\
\hline
\hline
peak flux & total sample&Swift&0.15$\pm$0.02&$--$&2.44$\pm$0.07&$--$&2.06\\
peak flux & total sample&Swift&$--$&0.16$\pm$0.00&$--$&2.49$\pm$0.01&1.73\\
peak flux & Ch1&Swift&$--$&0.11$\pm$0.02&$--$&2.36$\pm$0.12&0.42\\
peak flux & Ch2&Swift&$--$&0.26$\pm$0.03&$--$&2.43$\pm$0.09&0.93\\
peak flux & Ch3&Swift&$--$&0.17$\pm$0.03&$--$&2.43$\pm$0.09&0.41\\
peak flux & Ch4&Swift&$--$&0.13$\pm$0.03&$--$&2.36$\pm$0.16&1.10\\
\hline
FWHM  & total sample&Swift&0.05$\pm$0.02&$--$&1.87$\pm$0.13&$--$&1.29\\
FWHM  & total sample&Swift&$--$&0.08$\pm$0.00&$--$&1.99$\pm$0.01&0.58\\
FWHM  & Ch1&Swift&$--$&0.04$\pm$0.02&$--$&1.73$\pm$0.15&0.06\\
FWHM  & Ch2&Swift&$--$&0.07$\pm$0.02&$--$&1.86$\pm$0.18&0.25\\
FWHM  & Ch3&Swift&$--$&0.08$\pm$0.01&$--$&1.88$\pm$0.12&0.33\\
FWHM  & Ch4&Swift&$--$&0.03$\pm$0.01&$--$&1.75$\pm$0.18&0.12\\
\hline
$t_{\rm r}$ & total sample&Swift&0.01$\pm$0.01&$--$&1.82$\pm$0.13&$--$&1.37\\
$t_{\rm r}$ & total sample&Swift&$--$&0.04$\pm$0.00&$--$&1.99$\pm$0.01&0.89\\
$t_{\rm r}$ & Ch1&Swift&$--$&0.01$\pm$0.01&$--$&1.58$\pm$0.17&0.13\\
$t_{\rm r}$ & Ch2&Swift&$--$&0.03$\pm$0.01&$--$&1.88$\pm$0.12&0.14\\
$t_{\rm r}$ & Ch3&Swift&$--$&0.03$\pm$0.01&$--$&1.94$\pm$0.08&0.79\\
$t_{\rm r}$ & Ch4&Swift&$--$&0.03$\pm$0.01&$--$&1.81$\pm$0.20&0.22\\
\hline
$t_{\rm d}$ & total sample&Swift&0.05$\pm$0.01&$--$&1.86$\pm$0.14&$--$&1.25\\
$t_{\rm d}$ & total sample&Swift&$--$&0.05$\pm$0.00&$--$&1.99$\pm$0.02&0.54\\
$t_{\rm d}$ & Ch1&Swift&$--$&0.04$\pm$0.01&$--$&1.78$\pm$0.19&0.11\\
$t_{\rm d}$ & Ch2&Swift&$--$&0.03$\pm$0.02&$--$&1.68$\pm$0.20&0.27\\
$t_{\rm d}$ & Ch3&Swift&$--$&0.05$\pm$0.01&$--$&1.88$\pm$0.12&0.22\\
$t_{\rm d}$ & Ch4&Swift&$--$&0.02$\pm$0.01&$--$&1.87$\pm$0.13&0.28\\
\hline
$t_{\rm m}$ & total sample&Swift&0.04$\pm$0.02&$--$&1.75$\pm$0.17&$--$&1.43\\
$t_{\rm m}$ & total sample&Swift&$--$&0.07$\pm$0.00&$--$&1.97$\pm$0.03&0.28\\
$t_{\rm m}$ & Ch1&Swift&$--$&0.04$\pm$0.01&$--$&1.79$\pm$0.17&0.11\\
$t_{\rm m}$ & Ch2&Swift&$--$&0.05$\pm$0.01&$--$&1.78$\pm$0.15&0.18\\
$t_{\rm m}$ & Ch3&Swift&$--$&0.05$\pm$0.01&$--$&1.90$\pm$0.11&0.26\\
$t_{\rm m}$ & Ch4&Swift&$--$&0.04$\pm$0.01&$--$&1.83$\pm$0.15&0.20\\
\hline
\hline
\end{longtable*}


After Swift was launched, the bright X-ray flares are detected in nearly half of GRBs \cite[e.g.][]{Burrows2005,Romano2006}. Generally, X-ray flares are characterized as the late central engine activities, through a mechanism similar to that of GRB prompt emissions \cite[e.g.][]{Margutti2010,Chang2021}. At present, some of theoretical models of GRB and X-ray flares involve magnetic reconnection scenario \citep{ZhangB2006,Giannios2006,Kumar2015}. \cite{Dai2006} suggested that the differential rotation of millisecond magnetar after compact star mergers can lead to windup of interior
poloidal magnetic fields to toroidal fields, which are strong enough to
float up and break through the stellar surface. Once penetrating through the surface, the toroidal fields with different polarity may reconnect and give rise to original GRB and multiple X-ray flares.
In the Internal Collision-induced Magnetic Reconnection and Turbulence (ICMART) model proposed by \cite{ZhangB2011}, internal collisions distort the ordered magnetic field lines in the ejecta. Then, GRBs and X-ray flares can be triggered by magnetic reconnection in the distorted magnetic field. The ICMART model can well reproduce the properties of GRB prompt phases and X-ray flares simultaneously.
Our statistical results of the similar statistical framework of SOC system of GRBs and X-ray flares support the model and can impose strong constraints on the same magnetically dominated stochastic process and mechanism of them.

\cite{Wang2013} studied three statistical properties: power-law frequency distributions for energies, durations and waiting times of GRB X-ray flares with known redshift. They suggested that GRB X-ray flares can be explained with the same statistical framework with solar flares and correspond to a one-dimensional SOC system. In our previous work \citep{Yi2016}, we studied the peak times, rise times, decay times, waiting times, and durations of a larger GRB X-ray flare sample and further strengthened the self-organized criticality in GRB X-ray flares. Nevertheless, the Euclidean space dimension S = 3 is
different from the result reported by \cite{Wang2013}. As to GRB prompt phases, it is very exciting that our conclusion is consistent with that reported by \cite{Lv2020} in despite of different GRB samples. It can be naturally explained by the theory that GRB prompt phases and X-ray flares arise from different active stages of internal shocks \cite[e.g.][]{Burrows2005,M2006}. The radial component of magnetic fields decays faster with radius \cite{Dai2006,Wang2013}. Thus, the early GRB prompt phases near to the central engine is in a 3-dimensional form while the X-ray flares are closer to one dimension rather
than three dimensions \citep{Zhang2014,Wang2013,Lazarian2020,Lv2020}. It is worth noting that GRB optical
flares are found to share a similar physical origin to X-ray flares and similar frequency distributions further confirm the SOC nature of GRB system \citep{Yi2017}.

\section{Conclusions} \label{sec:result3}
In this paper, we have systematically studied the differential and cumulative distributions
of short GRBs from the fourth BATSE TTE Catalog and the third Swift GRB Catalog from 2004 December to
2019 July. For the first time, we presented the joint analysis among the different individual energy bands
in both BATSE and Swift sGRBs. Our major results are summarized as follows:

1. We find that GRBs have similar power-law distributions with GRB X-ray flares for peak flux, FWHM, rise time, decay time, and peak time, thus both can be attributed to a SOC process.

2. There is no obvious power-law index evolution across different energy bands.

3. It is found that the results for Swift sGRBs are essentially consistent with those in BATSE ones.

4. According to \cite{Aschwanden2012,Aschwanden2014}, our results are used to derive the possible Euclidean space dimensions of the GRB SOC system and obtain the spatial dimension S = 3.

In this work, our survey is restricted to the BATSE and Swift short GRBs and shed new light on the physical processes of compact binaries mergers. In fact, some long GRBs are found to originate from moderately-magnetized millisecond
pulsars with hyperaccreting accretion disks after the collapses of massive stars \cite[e.g.][]{Dai2006,Tang2019,Xie2022}. Thus, it is worth exploring the SOC behavior in long GRB prompt emissions. On the one hand, the longer duration enhances the chance of analyzing the multi-pulses (more than three) within a single burst. On the other hand, the higher detection rate of long GRBs makes it possible to obtain enough pulses to perform the statistics. Therefore, further search for GRBs, especially those with X-ray flares observed simultaneously from the Fermi, HXMT, GECAM, and SVOM catalogs can help to unveil the real physical mechanism of GRBs in the future.
\section*{Acknowledgements}
We thank the referee for very helpful suggestion and
comments. This work is supported by the National Natural Science Foundation of
China (Grant No. U2038106), and China Manned Spaced Project (CMS-CSST-2021-A12).



\clearpage

\listofchanges
\end{document}